\documentclass{ws-ijmpa}
\def\draftnote{}
\pdfoutput=1
\usepackage{epsfig}
\usepackage{graphicx}
\usepackage{multicol}
\usepackage{subfigure}
\usepackage{amsmath,amssymb}
\numberwithin{equation}{section}
\usepackage{cite}
\let\trimmarks\relax
\def\ppone{HRI/ST/1212}
\def\pptwo{CPHT-RR088.1112}
\newcommand{\ba}{\begin{align}}
\newcommand{\ea}{\end{align}}
\newcommand{\comments}[1]{}
\def\pref#1{(\ref{#1})}

\newcommand\czt{{\mathbb{C}^{3}/ \mathbb{Z}_{3}}}
\newcommand{\nn}{\nonumber}
\newcommand{\be}{\begin{equation}}
\def\bel#1{\begin{equation} \label{#1}}
\newcommand{\ee}{\end{equation}}
\newcommand{\bea}{\begin{eqnarray}}
\newcommand{\eea}{\end{eqnarray}}
\newcommand{\mbb}{\mathbb}
\newcommand{\ti}{\times}
\newcommand{\half}{\frac{1}{2}}
\newcommand{\mc}{\mathcal}

\newcommand{\beqa}{\begin{eqnarray}}
\newcommand{\eeqa}{\end{eqnarray}}

\newcommand{\pt}{{\mathbb{P}}^2}

\newcommand{\un}{{\bf 1}}

\newcommand{\f}{{\bf 5}}
\newcommand{\fb}{{\bf \bar{5}}}
\newcommand{\te}{{\bf 10}}
\newcommand{\teb}{{\bf \bar{10}}}
\newcommand{\op}{\oplus}

\newcommand{\cL}{{\cal L}}
\newcommand{\cF}{{\cal F}}
\newcommand{\Tr}{\mathrm{Tr}}
\newcommand{\Ps}{\mathbb{P}^1}%

\begin{document}

%
\catchline{}{}{}{}{}
%

\title{MODELS OF PARTICLE PHYSICS FROM TYPE IIB STRING THEORY AND F-THEORY: A REVIEW
}

\author{ANSHUMAN MAHARANA}

\address{Harish-Chandra Research Institute  \\
Chhatnag Road, Jhusi \\
Allahabad 211 019, India.\\ 
anshumanmaharana@hri.res.in}

\author{ERAN PALTI}

\address{Centre de Physique Theorique, \\
 Ecole Polytechnique, CNRS,  \\
 F-91128 Palaiseau, France. \\
Eran.Palti@cpht.polytechnique.fr  }

\maketitle


\begin{abstract}

We review particle physics model building in type IIB string theory and F-theory. This is a region in the landscape where in principle many of the key ingredients required for a realistic model of particle physics can be combined successfully. We begin by reviewing moduli stabilisation within this framework and its implications for supersymmetry breaking. We then review model building tools and developments in the weakly coupled type IIB limit, for both local D3-branes at singularities and global models of intersecting D7-branes. Much of recent model building work has been in the strongly coupled regime of F-theory due to the presence of exceptional symmetries which allow for the construction of phenomenologically appealing Grand Unified Theories. We review both local and global F-theory model building starting from the fundamental concepts and tools regarding how the gauge group, matter sector and operators arise, and ranging to detailed phenomenological properties explored in the literature. 


\end{abstract}


\newpage
%
\makeatletter
\newcommand\contentsname{Contents}
\newcommand\@pnumwidth{1.55em}
\newcommand\@tocrmarg{2.55em}
\newcommand\@dotsep{4.5}
\setcounter{tocdepth}{3}
\newcommand\tableofcontents{%
    \section*{\contentsname
        \@mkboth{%
           \contentsname}{\contentsname}}%
    \@starttoc{toc}%
    }
\newcommand\l@section[2]{%
  \ifnum \c@tocdepth >\z@
    \addpenalty\@secpenalty
    \addvspace{1.0em \@plus\p@}%
    \setlength\@tempdima{1.5em}%
    \begingroup
      \parindent \z@ \rightskip \@pnumwidth
      \parfillskip -\@pnumwidth
      \leavevmode \bfseries
      \advance\leftskip\@tempdima
      \hskip -\leftskip
      #1\nobreak\hfil \nobreak\hbox to \@pnumwidth{\hss #2}\par
    \endgroup
  \fi}
\newcommand\l@subsection{\@dottedtocline{2}{1.5em}{2.3em}}
\newcommand\l@subsubsection{\@dottedtocline{3}{3.8em}{3.2em}}
\newcommand\l@paragraph{\@dottedtocline{4}{7.0em}{4.1em}}
\newcommand\l@subparagraph{\@dottedtocline{5}{10em}{5em}}
\newcommand\listoffigures{%
    \section*{\listfigurename
      \@mkboth{\listfigurename}%
              {\listfigurename}}%
    \@starttoc{lof}%
    }
\newcommand\l@figure{\@dottedtocline{1}{1.5em}{2.3em}}
\newcommand\listoftables{%
    \section*{\listtablename
      \@mkboth{%
          \listtablename}{\listtablename}}%
    \@starttoc{lot}%
    }
\let\l@table\l@figure
\makeatother
\tableofcontents
%

%
\markboth{ANSHUMAN MAHARANA and ERAN PALTI}
{Models of Particle Physics from Type IIB String Theory and F-Theory}
\newpage
\section{Introduction}
\label{sec:intro}

String theory remains the most promising avenue towards a quantum theory of gravity currently in theoretical physics. Among its many celebrated properties which make it so appealing in this respect are perturbative finitness, a microscopic theory of black hole entropy, an impressive handling of spacetime singularities, and an in-built realisation of gauge/gravity duality. However the theory would not receive as much attention if it was only a theory of gravity and one of the primary motivations for studying string theory is that it is potentially a unified theory of all the forces and matter in the universe. It is quite remarkable that a theory of gravity should so naturally incorporate the key elements of the other interactions in nature: gauge theories, chiral matter, and spontaneously broken symmetries. As well as these elements of the Standard Model (SM) string theory realises some of the most appealing ideas for extensions of it such as Grand Unified Theories (GUTs) and supersymmetry. The ability of the theory to reproduce realistic low energy physics was realised directly after its self-consistency was demonstrated, as part of the first superstring revolution, within the Heterotic string context \cite{Candelas:1985en}. One of the important insights that drove the second superstring revolution is the concept of dualities, and within string phenomenology the implications were that the Heterotic string is not unique in its phenomenological aptitude. Indeed it is by now appreciated that these key aspects of particle physics are somewhat universal within the full structure that is termed as M-theory. However as string phenomenology advances, and more realistic models are sought, certain regions of the M-theory framework prove more fruitful in realising certain phenomenological properties than others. Although frequently this is more down to our ability to calculate within the theory than its innate properties, pragmatically in constructing ever more realistic models the community is naturally drawn to those regions. One particularly interesting region in the landscape is that of type IIB string theory and its strongly coupled generalisation termed F-theory, and the subject of this review article is the study of models of particle physics within this framework.

Perhaps the most important question in the subject of string phenomenology is which particular vacuum of M-theory is the universe in? Since our current understanding of the theory lacks a top-down vacuum selection principle the best we can do is attempt to answer this question by extracting phenomenological properties of vacua and ruling them out through observations and potentially various aesthetic criteria. Of course this procedure first requires defining a specific vacuum and in this respect vacuum degeneracies, or moduli, are a serious problem. Type IIB and F-theory \cite{Vafa:1996xn} are regions in which the moduli stabilisation problem is well understood enough so as to be able to specify vacua that are potentially phenomenologically realistic and have all the moduli fixed. There has been much work on this part of the theory over the last decade and reviewing this aspect is the subject of section \ref{sec:moduli}. 

One of the important ideas to come out form the study of the vacuum structure of type IIB string theory is the idea of the string landscape \cite{bopol, Susskind:2003kw}. It can be argued that it is possible to obtain sets of vacua where the values of certain observables, in particular the cosmological constant, are extremely finely spaced. This has opened up the possibility of making  use of environmental selection as a guiding principle for an explanation of the size of the cosmological constant. The downside of this is that, even restricting to the type IIB setting, it appears difficult to extract universal phenomenological properties from the landscape. However there are two key facts that make the task of string phenomenology a plausible one even in this setting. The first is that we have a vast array of observational constraints on models of particle physics, and it is clear that as each layer of such constraints is imposed vast regions in the landscape can be ruled out. So much so that it is difficult to argue for a single known model that is compatible with all current observations. The second fact is that not all aspects of models depend on the full vacuum configuration. This is particularly the case in type IIB and F-theory because gauge interactions and matter are localised on submanifolds of the full extra dimensions. This implies that some aspects of the models, such as the chiral spectrum and Yukawa couplings, can depend, to a decent approximation, only on the details of the local geometry within a spatial region in the extra dimensions. Therefore such local aspects of models are in fact universal within a large class of global models realising, at least to some degree, the idea of studying general properties of type IIB and F-theory models.

Perhaps the ultimate realisation of locality in a type IIB setting are models based on D3-branes on singularities \cite{botup}. In this class of models many phenomenological properties are attributed to a single singular point in the extra dimensions. They are the subject of section \ref{sec:bs} in this review. These models highlight another advantage of local models which is that they are much simpler to work with than full global constructions. The key aspect of this simplification is that many of the complications of geometry are associated to the global structure over compact manifolds of high dimensions, and restricting to lower dimensions or disregarding compactness is a significant technical simplification.

Although D3-branes on singularities offer a range of appealing theories, the most sophisticated model building within the type IIB framework often involves D7-branes. Chiral matter arises as a result of
D7-branes passing through  singular loci or the intersection of D7-branes in the presence of world-volume flux and therefore such  brane configurations are the framework within which string phenomenology in the type IIB regime is set. We review the key aspects of intersecting brane  constructions in section \ref{sec:int7iib}.

A very appealing aspect of type IIB string theory is that we have a good understanding of its strongly coupled limit through the structure of F-theory. Within F-theory the dilaton profile over the extra dimensions is accounted for by combining it with the geometry of the extra dimensions into an elliptically fibered Calabi-Yau four-fold, with the complex structure of the torus fibre playing the role of the dilaton. This is important not only because it allows for a deeper understanding of the theory but also because the strongly coupled regime can be genuinely motivated by observations. This follows because within GUT theories Yukawa couplings are associated to exceptional gauge groups which can not be realised in the weakly coupled regime. This and other attractive phenomenological properties of exceptional groups have been a significant driving force behind a resurgence in work on F-theory and its applications to model building. We review developments in constructing full F-theory models in section \ref{sec:7inf}. 

Although our understanding of full F-theory constructions is developing rapidly it is as yet not advanced enough to realise all the phenomenological features that are expected to be possible in F-theory. In this context the simplifications of local models can be utilised and using the local approach phenomenologically sophisticated models have been developed addressing issues such as GUT breaking, the chiral spectrum, gauge coupling unification, Yukawa couplings, proton decay, neutrino masses and many others. We review the tools used in constructing local F-theory models and the applications to phenomenology in section \ref{sec:locf}.

Type IIB string theory has proved to be a fruitful region in the landscape for string phenomenology. In principle the main ingredients for constructing a string vacuum which is consistent with all current observations are present. The aim of this review is to summarise the wide range of tools and approaches that have been developed in order to address the different components that come together to paint the big picture of a realistic vacuum. Although the main ideas are present the actual implementation of them to ever more realistic model building is the subject of contemporary research. Whether a fully realistic vacuum can be constructed remains to be seen, and although even if such a vacuum is found we may not yet be able to answer why that one could be favoured over others, it is certainly a required stepping stone towards a full understanding of the role string theory plays in the universe. Therefore this review is aimed at setting the scene for cutting edge research on model building in the type IIB framework. Our aim is to be comprehensive in covering the main ideas explored and offering at least a brief discussion on the tools and techniques used to explore them. Hence the review will not be structured like a text book with detailed derivations of all the equations from first principles but rather in the form of a collection of results and ideas with a comprehensive bibliography for those wishing to pursue any particular direction in more detail. Of course some introduction to the tools and techniques is indispensable in order to understand the key aspects of the various results and we hope to have provided those to a sufficient level. Finally, note that we will not review developments in cosmology models based on the IIB/F-theory framework. Of course it is difficult to completely separate cosmology from particle physics but we will refrain from referring to papers whose main subject are models of cosmology.

\section{Moduli Fields and Their Stabilisation}
\label{sec:moduli}

  Calabi-Yau compactifications of superstrings do not have a unique ground state. The space of vacua is a finite dimensional manifold -- the moduli space.  In the supergravity regime, the moduli space is the set of continuous deformations of the background fields  which map   vacuum configurations to vacuum configurations. Since the energy associated with such deformations vanishes,  they give rise to massless scalar fields  upon Kaluza Klein reduction. These scalars are called the moduli.

      Massless  moduli are a phenomenological disaster. Firstly,
moduli couple universally to all matter with gravitational strength. This
makes massless moduli  in direct contradiction with fifth force experiments.
 There is also a loss in the ability to make predictions.
   Moduli vevs  determine the vacuum configuration; they are needed as input for computing various quantities  in the visible sector. The gauge couplings, Yukawa couplings, unification scale etc. all depend on the moduli vevs.  Presence of a moduli space by definition implies that these vevs cannot be determined.

 Thus,  realistic phenomenology requires a  potential for moduli
fields. Knowledge of the precise form of the potential  is also necessary; so that
  moduli vevs can be computed by minimising it and used to determine quantities of phenomenological interest.   Masses generated from the potential have to be large enough
to evade bounds from fifth force experiments and cosmology \cite{bcosmo, qcosmo}.

      Flux compactifications: supergravity solutions with background values
   of the NS-NS and RR form fields \cite{ gvw, drs, granap, gkp} are central to moduli stabilisation in type IIB.
   Our discussion  of  the subject shall be brief, as there are already many excellent reviews  \cite{frey03, silver04, grana05, dougkac06, denef07, denef08}.
We begin by describing backgrounds involving the five form, three form fluxes and D3 and O3 planes\footnote{The construction can be generalised to F-theory.  Moduli stabilisation in F-theory  will be discussed in section \ref{sec:locf}.}. Following \cite{gkp} the ten dimensional
metric takes the form
\bel{gkpmet}
  ds^{2} = e^{2A(y)} \eta_{\mu \nu} dx^{\mu} dx^{\nu} + e^{-2A(y)} \tilde{g}_{mn} dy^{m} dy^{n}
\ee
with  $\tilde{g}_{mn}$ a Ricci flat metric on a Calabi-Yau. The five form
\bel{gkpfive}
\tilde{F}_{5} = (1 + *) d (e^{4A(y)}) \wedge dx^{0} \wedge dx^{1} \wedge dx^{2} \wedge dx^{3}.
\ee
The three forms $F_3$ and $H_3$  thread non-trivial three cycles of the Calabi-Yau
(satisfying  Dirac quantisation conditions) with
 the complex combination $G_3 = F_3 - \tau H_3$  imaginary self dual,
\bel{isd}
     *G_{3} = i G_{3}.
\ee
This requires that $G_{3}$ be expressible as a linear combination of (0,3) and (2,1) harmonic forms of the Calabi-Yau.
Here we introduce the axio-dilaton $\tau = C_0 + i e^{-\phi}$ with $C_0$ being the RR scalar and $\phi$ the dilaton of type IIB strng theory.
In addition to the background fields there can be space filling D3 branes and O3 planes which are
point like on the Calabi-Yau.

    The function $A(y)$, which appears in the metric \pref{gkpmet} and five form \pref{gkpfive}
is determined by a Poisson  equation on the Calabi-Yau
\bel{tad}
 -\tilde{\nabla}^{2} e^{-4A(y)} = { {G_{mnp} \tilde{G}^{mnp}} \over {12 \rm{Im{\tau}}} } + 2 \kappa_{10}T_{3} \rho_{3},
\ee
where $\rho_3$ is the  local contribution to D3 charge from the D3 branes and 03 planes,  $\kappa_{10}$ the ten dimensional Newton's constant and $T_{3}$ the D3 brane tension. For  \pref{tad} to have
a solution the integral of the right hand side over the compact manifold has to vanish. The  requirement follows from   D3 charge  tadpole cancellation
$$
 0 = \int d \tilde{F} = \int \bigg( H_{3} \wedge F_{3} + 2 \kappa_{10}^{2} \rho_{3}^{loc}\bigg).
$$
 Since the flux contribution to the right hand side of \pref{tad} is positive semi-definite  non-trivial $G_3$ flux requires a negative contribution from the local sources. This in turn necessitates the presence of  orientifold three planes,  the carriers of negative D3 charge (or wrapped
D7 branes in F-theory).

The orientifold projection reduces the number of supersymmetries  to $N=1$  in four dimensions and truncates the moduli space \cite{Grimm:2004uq,grimm04}. The moduli arise as complex scalars,  lowest components of $N=1$ chiral superfields.  Moduli associated with metric deformations  fall into two classes --   complex structure and Kahler moduli, classified according to the nature of perturbation of the Calabi-Yau  they describe. The parameters that describe complex structure deformations of a Calabi-Yau are complex numbers, hence one can naturally associate a complex scalar with these perturbations. On the other hand, the parameters associated with Kahler deformations are real. These combine with axions that arise from the dimensional reduction of the RR four form to form complex scalars. The number of complex structure and Kahler moduli  is given by the betti numbers\footnote{The orientifold action induces
a $\mathbb{Z}_{2}$ grading on the cohomology of the Calabi-Yau.} $h_{-}^{2,1}$ and $h_{+}^{1,1}$  . The axio-dilaton also survives the projection and appears as a modulus\footnote{ An additional $h_{-}^{1,1}$ scalars arise from
fluctuations of $B_{2}$ and $C_{2}$ in the presence of O7 planes.}.

        For fixed flux quanta, the imaginary self dual condition \pref{isd} can be thought of as equations
 imposed   in addition to the requirement of a constant dilaton and Ricci flatness;
 i.e a set of
equations on the moduli space. In the language of $d=4$ $N=1$   supergravity,  these equations are captured by the superpotential \cite{gvw}
\bel{egvw}
W = \int G \wedge \Omega
\ee
where $\Omega$ is the holomorphic three form; and the Kahler potential \cite{grimm04} (to leading order in the inverse volume expansion)
\bel{kpot}
  K = - 2 \log ({\cal{V}})  -\log \bigg( i \int \Omega \wedge \bar{\Omega} \bigg) -\log(-i(\tau -\bar{\tau})),
\ee
where ${\cal{V}}$ is the   Einstein frame volume of the compactification in string units  (expressed
in terms of the Kahler moduli). The superpotential \pref{egvw} depends  on the complex structure moduli (via the holomorphic three form) and the axio-dilaton but is independent of the Kahler moduli.\footnote{ This follows from the shift symmetry of the axionic components of the Kahler moduli. The superpotential has to be  holomorphic, on the
 other hand the shift symmetry requires that it is a function  of  the sum of the field and its conjugate. The shift symmetry holds to all orders in string perturbation theory, is broken non-perturbatively.} Furthermore, the Kahler potential \pref{kpot}  satisfies the no-scale criterion of \cite{nsone}.  Thus while the complex structure and the axio-dilaton acquire masses
  due to  fluxes  the Kahler moduli remain massless. Computing  the F-terms, one
    finds that the presence  of (0,3)  flux implies that the vacuum is non-supersymmetric. To preserve supersymmetry, the flux has to be purely (2,1). But the cosmological constant and (soft) masses for D3 brane position moduli vanish for all choices of flux quanta -- a consequence of the no-scale structure.        \footnote{See \cite{oliver} for the form of the Kahler potential  in the presence of mobile D3 branes. From the ten dimensional perspective, the vanishing occurs due to the
     ``pseudo-BPS condition" of \cite{gkp}. }
    The mass generated for the complex structure moduli is
$$
  m_{{\rm{cs}}} \sim {M_{pl} \over { {\cal{V}} } }.
$$
In the presence of (0,3) flux, the gravitino mass is also of this magnitude. This is parametrically smaller than the Kaluza-Klein scale in the large radius limit, justifying a description in terms of four dimensional supergravity. Explicit examples of the stabilisation of  complex structure moduli and axio-dilaton
can be found in \cite{triv, freypo, tri, trivedi, font, kumartwo}.

  Before moving on to  Kahler moduli stabilisation we would like to discuss two
aspects of these compactifications which will be  important in later sections.
Firstly, by making suitable choices of flux quanta one can have regions where the warp factor
is extremely small compared to the average value in the compactification.
For a Calabi-Yau with a conifold singularity whose A and B cycles are threaded by
M and K units of flux, the local geometry is given by  the Klebanov-Strassler \cite{ks}
solution.  The warp factor on the minimal area three sphere is exponentially
small in the flux quanta \cite{gkp}
$$
  e^{A_{{\rm{min}}}} = e^{-2 \pi K/ g_{s} M}.
$$
This provides the possibility to generate hierarchies in four dimensional
scales\footnote{ The redshifting of scales in terms of corrections
to the N=1 Kahler potential is still not completely understood.
The effective field theory in the presence of highly warped regions has been
studied in \cite{oliver,gm, andrewm, tor, torone, uni}.} along the lines of Randall and Sundrum \cite{rs}.

    Fluxes lift the continuous degeneracy associated with the complex
 structure moduli. But they introduce a new kind of degeneracy: associated with the
 integers needed to specify the flux quanta.
  These integers are constrained by the D3 charge tadpole cancellation
 condition \pref{tad}.  Since fluxes always make a positive contribution to the tadpole, the number of vacua is controlled by  the value of the negative contribution.
 Large values of the negative contribution can be obtained in F-theory
 compactifications \cite{sethi}. This implies
  an enormous degeneracy, and also the possibility to tune the
  vacuum expectation value of the superpotential \pref{egvw} for phenomenological applications \cite{statone, stattwo, statthree}.
\subsection{Kahler Moduli Stabilisation}

\subsubsection*{KKLT construction}

   The KKLT construction \cite{kklt} provided the first examples of compactifications
with all moduli stabilised.  The simplest examples involve a single Kahler
modulus
$$
   T = \tau  + i \theta
$$
where  $\tau$ is the Einstein frame volume of the four cycle  dual to the Kahler form measured in string units and  $\theta$ the integral
of the RR four form over the cycle. Note that we henceforth use $\tau$ for the Kahler modulus and not the axio-dilaton. The complex structure and axio-dilaton are stabilised by three form
flux; integrated out  leaving a constant
superpotential $W_{0}$.  A non-perturbative superpotential is generated
for the Kahler modulus\footnote{This breaks the non-scale structure.}  by  gaugino condensation on  a stack of N D7 branes.
Thus
$$
  W = W_{0} + A e^{-a T}
$$
where $a ={2 \pi \over N}$  and A  an order one constant. The Kahler potential is given by
\pref{kpot} with
$$
    {\cal{V}} =  \bigg( \frac{\tau + \bar{\tau}}{2} \bigg)^{3/2}
$$
The  potential for T  has a supersymmetric minimum with the value
of $\tau$ at the critical point given by
\bel{kkltmin}
  -A e^{- a \tau} ( 1  + {2 \over 3} a \tau) = W_{0}
\ee
The vacuum energy at the minimum is negative with
\bel{kklmin}
   V_{{\rm min}} \approx - { |W_{0}|^{2} \over { {\cal V}^{2} } }
\ee
Control over $\alpha'$ corrections  requires  $\tau \gg 1$, while to justify
the use of the leading term in the instanton expansion one needs $a \tau > 1$. These conditions can be satisfied if the number of D7 branes N is large and $W_{0} \ll 1$. Generic choices for flux quanta yield $W_0 \sim 1$, hence control over the effective field theory requires fine tuning $W_{0}$.

   Explicit constructions on toric Calabi-Yaus and the generalisation to multiple Kahler moduli was presented in \cite{mrdone}. Examples on toroidal orientifolds  and in F-theory were constructed in \cite{lustmodone, lustmodtwo, mrdtwo}. Variations of the scenario by incorporating $\alpha'$
and $g_{s}$ corrections were discussed in \cite{vjper, berha, hebk, aw}. More recently, it has been proposed that all Kahler
moduli can be stabilised using a single gauge instanton \cite{piyushwa}.

  \subsubsection*{Large Volume Scenario}

       The  Large Volume Scenario (LVS) \cite{theoriginalvs} provides a mechanism to stabilise the overall volume of the compactification at an exponentially large value in the ratio of flux quanta. The key ingredient
 is  the  leading $\alpha'$ correction   from IIB string theory. This is captured in
 the four dimensional effective action by a modification of the first term in the Kahler potential \pref{kpot} \cite{haackalpha}
$$
  - 2 \log( {\cal{V}} ) \to  - 2\log \bigg( {\cal{V}} +   \frac{\xi}{ {g}_s^{3/2} } \bigg)
$$
where $\xi$ is proportional to the Euler number of the Calabi-Yau.  On  incorporating this correction  (along  with a
non-perturbative superpotential for the Kahler moduli), one is led to
an interesting scenario for moduli stabilisation in a broad class of Calabi-Yaus. For Calabi-Yaus
with negative Euler number and at least one Kahler modulus which is the blow up of a point like singularity;  the  moduli are stabilised with the overall volume of the compactification  given by the exponential of the volume of a blow up cycle \cite{cicolitwo, fano}.  The size of the blow up
cycle is given by a ratio of flux quanta. The minimum exists for
 $W_{0} \approx 1$.
The vacuum breaks supersymmetry and has a negative  cosmological constant
$$
   V_{\rm{min}} \approx  - { 1 \over {\cal{V}}^{3} }.
$$
A universal feature is that the volume modulus
acquires the smallest mass with
\bel{mlight}
  m_{{\cal{V}}} \approx { M_{\rm{pl}} \over {{\cal{V}}}^{3/2} }
\ee
The systematics and  some variations of this scenario have been explored in  \cite{lvstwo, bh, cicolione, cicolitwo, misra, fanone, fanocol, fano, westpha, gray}. The mass spectrum of axions and U(1) fields were studied in \cite{Goodsell:2009xc,Cicoli:2011yh,Cicoli:2012sz,Cicoli:2012aq,Higaki:2012ar}. It has been possible to obtain models of TeV scale strings with anisotropic extra dimensions \cite{ani,pr}.
\subsubsection*{DeSitter Vacua}

     The vacua discussed above have a negative cosmological constant. It was
 argued by KKLT \cite{kklt} that DeSitter vacua can be obtained if an anti-brane is included
 in such a compactification with a  Klebanov Strassler throat. It is
 energetically favourable for the anti-brane to be located at 
  the bottom of the throat. The
 $\tau$ dependence of its energy can be obtained from the DBI action
\bel{vdbar}
    E_{\small{\rm{\overline{D3}}}} = { {e^{4A_{\rm{min}}}} \over {\tau }^{2} }.
\ee
 If such a term is added to the low energy  action, the effect on the critical point \pref{kkltmin} is an increase in the vacuum energy with
 a small change in the value of $\tau$.   It is therefore often referred to as the uplift term. A DeSitter solution
 with a small cosmological constant is obtained by
  tuning the minimum value of the warp factor so that the uplift term approximately cancels the
  vacuum energy of the AdS minimum \pref{kklmin}. Furthermore,
  by varying  $W_0$ (achieved by varying the flux quanta) one
  obtains vacua whose cosmological constants are finely spaced --
  a dense discretuum  as described in \cite{bopol}.

      Dynamics of the anti-brane at the bottom of the throat in the probe approximation was studied in
\cite{kpv} (see also \cite{lippert}). The   configuration was found
to be metastable;  connected to  the  Klebanov
Strassler throat with mobile D3 branes by brane-flux annihilation. This supports
the inclusion of $\pref{vdbar}$ in the low energy effective action.  There has been
much recent work \cite{benaone, benatwo, benathree, benafour, zag, zagtwo} to test   the consistency of the entire setup --  search for a supergravity
solution which includes the back reaction of the anti-branes  and has no irrelevant perturbation of the Klebanov Strassler throat excited.

 Another avenue \cite{quevedodone, zwi, paw, quevedodtwo, svenone, dila} is to try to obtain the uplift term in the effective action from more conventional hidden sector field theory dynamics. The major challenge is to generate the uplift term  at a  low scale (so that it does not  destabilise the moduli); and yet be able to tune it so as to cancel the vacuum energy of the AdS minimum.
 Recently, \cite{dila} proposed a mechanism involving dilaton dependent non-perturbative effects.

 \subsection{Soft Supersymmetry Breaking}

   Moduli stabilisation plays a central role in understanding visible sector supersymmetry
breaking. The moduli potential  determines F-terms of the moduli  dynamically. These together with the
matter-moduli couplings determine  the strength of the gravity mediated contribution
to soft masses\footnote{Thus to construct a viable model of gauge mediation, one is faced with
the challenge to ensure that these contributions are well below the TeV scale. On the other hand, gravity mediated contributions usually lead to dangerous flavour violating contributions; see \cite{kacflav, liam, softlvsone, mirror ,bergf, bergft}
for discussion  and possible mechanisms for the suppression of flavour violating terms in the type IIB context.}.  The backgrounds of \cite{gkp} exhibit sequestering; soft-masses vanish even in the
presence of supersymmetry breaking fluxes. Thus non-trivial contributions to soft masses
arise as a consequence of the inclusion of the effects that break the no-scale structure
(the non-perturbative superpotential for the Kahler moduli and $\alpha'$, $g_{s}$ corrections to the Kahler potential).
 Often, various  phenomenologically interesting  characteristics (such as mass hierarchies) of the  soft parameters are  independent
 of the precise manner in which the standard model is embedded into the comactification and depend only on certain broad features.

   In the KKLT setting, the AdS vacuum preserves supersymmetry. The source of
supersymmetry breaking  is uplifting. As discussed earlier,
the state with an anti-brane at the bottom of a warped throat is related to the supersymmetric minimum
by brane flux annihilation - a non-perturbative process. Thus the description of the supersymmetry
breaking vacuum in terms of elementary field excitations  of the theory is expected to be highly complicated\footnote{See \cite{antigary} for a recent discussion.}. This makes the analysis of supersymmetry breaking by conventional four dimensional effective field theory methods extremely difficult. A promising approach is to make use of the AdS/CFT correspondence \cite{liam, oliversusy}.

A model description of the anti-brane in the language of N=1 supergravity was proposed in
\cite{mirage}. This led to ``mirage mediation" \cite{miragechoi, falk, choitwo, heb, thecode } phenomenology,
a scenario with competing contributions from gravity mediation and anomaly mediation.
The gaugino masses are compressed  and unify at a fictitious scale above the GUT scale
called the mirage scale.

     In  LVS, the minimum of the moduli potential breaks supersymmetry with non-vanishing
F-terms for the Kahler moduli. The pattern of supersymmetry breaking depends on whether the visible
sector is localised on the blow-up cycle that supports the non-perturbative effects (the dominant source of
supersymmetry breaking) or is physically separated\footnote{The distinction arises as separation
in the  extra dimensions controls the strength of coupling between the visible sector and the source of supersymmetry breaking, see for example \cite{lvstwo}.}. In the first case,
the phenomenology has been extensively studied \cite{lvstwo,  softlvsone, lvspar, apa, angus}. Apart from the
F-terms of the LVS minimum\footnote{It can be argued that  uplifting to
a solution with approximately vanishing cosmological constant leads to a small correction to the soft terms in
the inverse volume expansion.} a key input is the modular weights of the Kahler metric of matter fields \cite{scalemod, apa}.  Using these, for a semi-realistic standard model sector one finds
\bel{lvssofteq}
   m_{3/2} \approx { M_{\rm{pl}} \over { \cal{V} } } \ \ \ \ {\rm{and}}
    \ \ \ \ \ \ m_{\rm{gaugino}} \approx m_{\rm{scalar}} \approx { M_{\rm{pl}} \over { \cal{V} \log( {\cal{V}}) } }
\ee
 Thus  the hierarchy problem motivates (for $W_0 \approx 1$)  ${\cal{V}} = 10^{15-16}$, i.e
$$
   M_{\rm{string}} \approx  10^{12-13} \rm{GeV}
$$
Such a scenario implies that
 the unification exhibited by the low energy  couplings of the MSSM  at the GUT scale under renormalization group flow is an accident. This can be considered as a shortcoming.  But cosmologically, the  scenario suffers from a  serious problem. The soft masses in \pref{lvssofteq}  are significantly greater than the mass of the lightest modulus \pref{mlight}, TeV scale supersymmetry makes the modulus extremely dangerous from the point of view of the cosmological moduli problem \cite{bcosmo, qcosmo}.

 Supersymmetry breaking for models where the visible sector is physically separated from the  blow-up cycles which support non-perturbative effects (such as the models in which the standard model is realised from branes at singularities) was studied\footnote{This was motivated by the tension between having
 chiral matter fields from wrapped D7 branes  on a cycle and the topological criterion for generating  non-perturbative effects \cite{blum} on the same cycle.   Subsequent discussion on the issue can be found in \cite{fano}.} in \cite{svenbreak}. This led to a scenario with
\bel{lvsven}
   m_{3/2} \approx { M_{\rm{pl}} \over { \cal{V} } } \ \ \ \ {\rm{and}}
    \ \ \ \ \ \ m_{\rm{gaugino}}  \approx { M_{\rm{pl}} \over { {\cal{V}}^2 } }
\ee
The scalar masses depend on yet to be determined $\alpha'$ corrections to the Kahler potential
of matter fields. But  their size can be estimated, scalars are at least as heavy as gauginos.
TeV scale supersymmetry motivates ${\cal{V}} = 10^{6}$, this gives a unification scale closer to the
GUT scale. The scenario also offers a solution to the cosmological moduli problem.  The lightest  modulus is much heavier than the lightest sparticle in the visible sector,  making it  compatible with TeV scale supersymmetry.
Unfortunately, unlike  the previous setting,  soft masses in this scenario are highly sensitive to the precise form of the uplift potential \cite{svenbreak} and loop corrections\cite{redef, shanta}. These effects can alter the mass hierarchies and reinstate the cosmological moduli problem; thus  deserve systematic exploration.

  A complementary approach \cite{grana, grana03, camaraone, camara, softgary,  albion, liamd3 } is to study supersymmetry breaking from the ten dimensional perspective. This involves computing the backreaction of the no-scale breaking effects on 
 the ten-dimensional solution. This leads to backgrounds which violate the ``psuedo-BPS condition"
 of \cite{gkp}; one then uses the DBI action  to determine the
soft parameters.   One advantage is that symmetries associated with
the extra-dimensions are more manifest.

\section{Branes at Singularities}
\label{sec:bs}

  The world volume dynamics of D-branes probing a singularity in a Calabi-Yau
is described by an N=1 supersymmetric gauge theory with chiral matter. This makes branes
at singularities  an attractive arena for model building in particle physics.  All the standard
model degrees of freedom arise from a single point  in the extra-dimensions; the models provide
the quintessential realisation of the idea of local model building. These models are best understood from the type IIB perspective, F-theory methods are not well developed for a singular base.

\subsection{Orbifold Models}

 Given a manifold with a discrete symmetry, one can always construct a new manifold by identifying points under the action of the symmetry. If the action of the symmetry group has fixed points, the new manifold is singular in the neighbourhood of the fixed points. Such singularities are called orbifold singlarities. In spite of their simplicity, it has been possible to obtain various semi-realistic models
from orbifolds.  The models admit description by world sheet methods; thus it is possible to obtain
 detailed information about couplings. The intuition gained is  extremely valuable in the study of more
 complicated models.

 Our interest shall be in orbifolds with D3 branes localized at the singular point and D7 branes  extended in a four dimensional hypersurface of the extra dimensions  and intersecting the  D3 branes at the singularity. The gauge degrees of freedom will arise from 3-3 strings and matter from both 3-3 and 3-7 strings. We begin our discussion with $R^{6} / {\mathbb{Z}}_N$, closely
following \cite{botup}. The spectrum and interactions can be obtained by carrying out a projection on the theory on $M^{3,1} \times R^{6}$.

      The low energy dynamics of N D3 branes in flat space is described by N=4 super Yang-Mills with
gauge group U(N). In light cone NSR formalism, the
gauge bosons  and  scalars arise from the NS sector
$$
  A^{\mu} = \lambda \psi^{\mu} | 0 \rangle_{\rm{NS}},  \ \ \mu = 2,3  \ \ \ \ \ \phi^{m} = \lambda  \psi^{m} | 0 \rangle_{\rm{NS}}, \ \ r = 4 ..9
$$
where $\lambda$ is the Chan-Paton matrix. The fermions arise from the Ramond sector ground state
\bel{orbfermi}
    \lambda |s_{23},s_{45}, s_{67}, s_{89} | 0 \rangle_{\rm{R}}
\ee
where $s_{ij}= \pm 1/2$ is the spin of the state in the $ij$ plane. The GSO projection requires the sum of the spins to be odd. States of opposite four dimensional spin combine to form Weyl fermions
in four dimensions.

   A  $R^{6} / \mathbb{Z}_{N}$ orbifold  is specified by a  $\mathbb{Z}_{N}$ action on  SU(4)  (double cover of rotations in $R^{6}$) and  an action on the Chan-Paton factors.
The SU(4) action can be encoded in a matrix giving the action on the fundamental  representation
\bel{actionorb}
    {\rm{Diag}}( e^{2 \pi i a_1 / N},  e^{2 \pi i a_2 / N}, e^{2 \pi i a_3  / N}, e^{2 \pi i a_4  / N})
\ee
where  $a_{i} \in  \mathbb{Z} \  {\rm{mod}} \ N$,  and $a_1 + a_2 + a_3 + a_4 = 0$. The Chan-Paton action is given by
$$
       \lambda \to \gamma^{-1} \lambda \gamma
$$
with $\gamma$   an $N \times N$ block diagonal matrix
\bel{block}
   \gamma = { \rm{diag}} ( {\mathbb{I}}_{n_{0}} \omega^{0}, {\mathbb{I}}_{n_{1}} \omega^{1} , .....,{\mathbb{I}}_{n_{N-1}} \omega^{N-1} ),
\ee
 where ${\mathbb{I}}_{n_j}$ is the $n_j \times n_j$ unit matrix and $\omega$ the Nth root of unity.

\subsubsection*{Spectrum}

                      The spectrum of the orbifold theory consists of states invariant under the combination of
  the $SU(4)$ and Chan-Paton actions. The vector
 bosons are inert under $SU(4)$; thus the Chan-Paton factors of the vector bosons in the orbifold theory satisfy $\gamma \lambda \gamma^{-1} = \lambda$. The invariant states are the $n_j \times n_j$  diagonal blocks in the block structure of \pref{block}. This gives the gauge group to be $U(n_0) \times U(n_1) \times ...... \times U(n_{N-1})$.  The $SU(4)$ action \pref{actionorb} acts diagonally on the
complexified scalars  via
\bel{actionorbt}
    {\rm{Diag}}( e^{2 \pi i b_1 / N},  e^{2 \pi i b_2 / N}, e^{2 \pi i b_3  / N})
\ee
with $b_1 = a_2 + a_3, b_2 = a_3 + a_1$ and  $b_3 = a_1 + a_2$. Thus  scalar states that survive the orbifold projection satisfy $e^{-2 \pi i b_r / N} \gamma \lambda  \gamma^{-1} =  \lambda$.
These are  the $n_j \times n_{j-b_r}$ off diagonal blocks in the block structure of \pref{block}. The
$(j,j- b_r)$ block transforms in  the fundamental of $U(n_j)$ and anti-fundamental of $U(n_{j-b_{r}})$.
The four Weyl fermions that arise from \pref{orbfermi} transform in the fundamental of $SU(4)$; the invariant states satisfy
$e^{2 \pi i a_r / N} \gamma \lambda  \gamma^{-1} =  \lambda$.
These are the $n_j \times n_{j+ a_{r}}$ blocks, they transform in  the  fundamental of $U(n_j)$ and anti-fundamental of $U(n_{j+a_r})$. In summary, the 3-3 spectrum is
\begin{itemize}
 \item Vector bosons  with gauge group $U(n_{0}) \times U(n_1) \times ....\times U(n_{N-1})$.
 \item Bifundamental scalars with quantum numbers $(n_{i}, \bar{n}_{i-b_{r}})$, where $i = 0 .... (N-1)$ and $r =1,2,3$.
 \item  Bifundamental fermions with quantum numbers $(n_{i}, \bar{n}_{i+a_{s}})$, where $i=0 ....(N-1)$
 and $s=1..4$.
\end{itemize}
Additional massless states charged under the 3-3 gauge groups appear in the
presence of seven branes passing through the origin. Let us consider  a stack of
M D7 branes extended along the hypersurface $z_{3} = 0$ (we use complex coordinates $z_1, z_2, z_3$ on $R^{6}$). As in the case of D3 branes, the action on Chan-Paton factors is specified by a $M \times M$ matrix with diagonal blocks of size $m_0, m_1 ... m_{N-1}$.
This gives a $U(m_0) \times U(m_1) \times  ...... U(m_{N-1})$ world volume  gauge
theory. The spectrum of 3-7 strings can be obtained by following  exactly the  same procedure as above\footnote{See \cite{botup} for details.}, one finds
\begin{itemize}
\item For $b_{3}$ even, bifundamental fermions in  $(n_i, \bar{m}_{i + {1 \over 2} b_3}),
(m_i, \bar{n}_{i + {1 \over 2} b_3})$  and scalars in \\ $(n_i, \bar{m}_{i - {1 \over 2} (b_1+b_2)}),
(m_i, \bar{n}_{i - {1 \over 2} (b_1+b_2)})$.
\item For $b_{3}$ odd, bifundamental fermions in  $(n_i, \bar{m}_{i + {1 \over 2} (b_3-1)}),
(m_i, \bar{n}_{i + {1 \over 2} (b_3+1)})$  and scalars in \\ $(n_i, \bar{m}_{i - {1 \over 2} (b_1+ b_2 +1)}),
(m_i, \bar{n}_{i - {1 \over 2} (b_1 + b_2 -1)})$.
\end{itemize}
 Since the world volume of the
7-branes is non-compact, all 7-7 states are non-dynamical from the perspective of
a four dimensional observer on the three branes. The $U(m_{i})$ symmetries are seen
as  flavour symmetries, the vevs of 7-7 matter fields appear as parameters in the
four dimensional effective action. The fields become dynamical once the system is embedded
in a compactification.

     In general, the spectrum of the four dimensional theory is not free from
anomalies\footnote{The orbifold projection does not guarantee that global consistency conditions such as
anomaly cancellation are automatically satisfied.}. The vanishing of non-abelian $SU(n_i)$ anomalies  is equivalent to the vanishing of the RR tadpoles in the twisted sector and is necessary for the consistency of the theory. Thus, consistency requires relations between the number of D3 and D7 branes, these have
been discussed in detail in \cite{botup}. One also has anomaly cancelation conditions associated with the
D7 gauge groups. Unlike the D3 gauge groups,  massless fields charged under these gauge groups do not
 have to be localised at the singularity. Thus, we will not demand that  these anomalies are canceled in the local model; although the issue has to be addressed in any global embedding.  For typical solutions of  non-abelian anomalies associated with the 3-3 gauge groups,  the spectrum is non-anomalous with respect to only a single linear combination of the diagonal $U(1)$ factors
in  $U(n_i)$. The associated charge is
\bel{hych}
  Q_{\rm{ano-free}} = \sum {{Q_{i}} \over {n_{i}} }
\ee
where $Q_{i}$ are charges with respect to the diagonal $U(1)$ factors in $U(n_{i})$.
Other linear combinations acquire a mass by a Stueckelberg term; their
anomalies are canceled by the four dimensional analogue of the Green-Schwarz mechanism. The
mass generated is of the order of the string scale.

\subsubsection*{Phenomenological Considerations}

     Let us examine the spectrum,  with the goal of obtaining the standard model
or its extensions.
\begin{itemize}
\item {\it{Family Triplication:}}   To account for the generation structure in the standard model one needs triplication in the quantum numbers of  states. For 3-3 states this implies
degeneracy in the values of $a_s$. Since their sum must also vanish, one has
 a bound on the number of families. The maximum number of families is three\footnote{Exceptions are $\mathbb{Z}_2$ and $\mathbb{Z}_4$ singularities where it is possible to obtain four families. Since for these cases $\vec{a} = (1,1,1,- 3) \equiv (1,1,1,1)$.},  attained for
$$
\vec{a} = (1,1,1,- 3).
$$
Later, we shall see that
a similar statement can be made for general toric singularities.
Triplication of 7-3 states can be obtained by introducing
multiple stacks of seven branes; extended along the hypersurfaces $z^i=0$, i=1..3.
\item {\it{Supersymmetry:}} To preserve N=1 supersymmetry in four dimensions the holonomy of the orbifold has to be  $SU(3)$. This requires the vanishing of one of the $a_{s}$. When combined with the
 condition for three families, one is led to a unique abelian orbifold:  $\czt$ with
$$
   \vec{a} = (1,1,1,-3) \equiv (1,1,1,0).
$$

\item {\it{Absence of SU(5)/SO(10)  GUT Models :}} The 3-3 matter states are bifundamentals, while the 7-3
states are either  fundamentals or anti-fundamentals. Thus it is impossible to
obtain GUT models, which require antisymmetric/spinorial representations. At the level of the spectrum, this can be ameliorated for $SU(5)$ by orientifolding\footnote{For work in this direction
see \cite{tris, kumarone, ber, bianchi, mirgut}.}
but as we will see in section \ref{sec:7inf} it is not possible to obtain SU(5) models with
realistic Yukawa couplings in the perturbative IIB setting.

\item {\it{Left Handed Quarks and Hypercharge:}} The left handed quarks are charged under
 both the non-abelian gauge groups of the Standard Models. Thus they have to arise as
 bifundamental 3-3 states. The hypercharge assignments for these are given  by that of the anomaly free
abelian charge \pref{hych}.
\end{itemize}

 Keeping the above considerations in mind,  \cite{botup} explored the possibility of
constructing realistic models in this setting. It was not possible to embed the
standard model in any of the non-supersymmetric orbifolds. On the other hand,
the supersymmetric  $\czt$ orbifold was found to be interesting for
phenomenology.

\subsubsection*{ $\czt$ : Spectrum and Interactions}

\begin{figure}[h!]
\centering{\includegraphics[height=5cm]{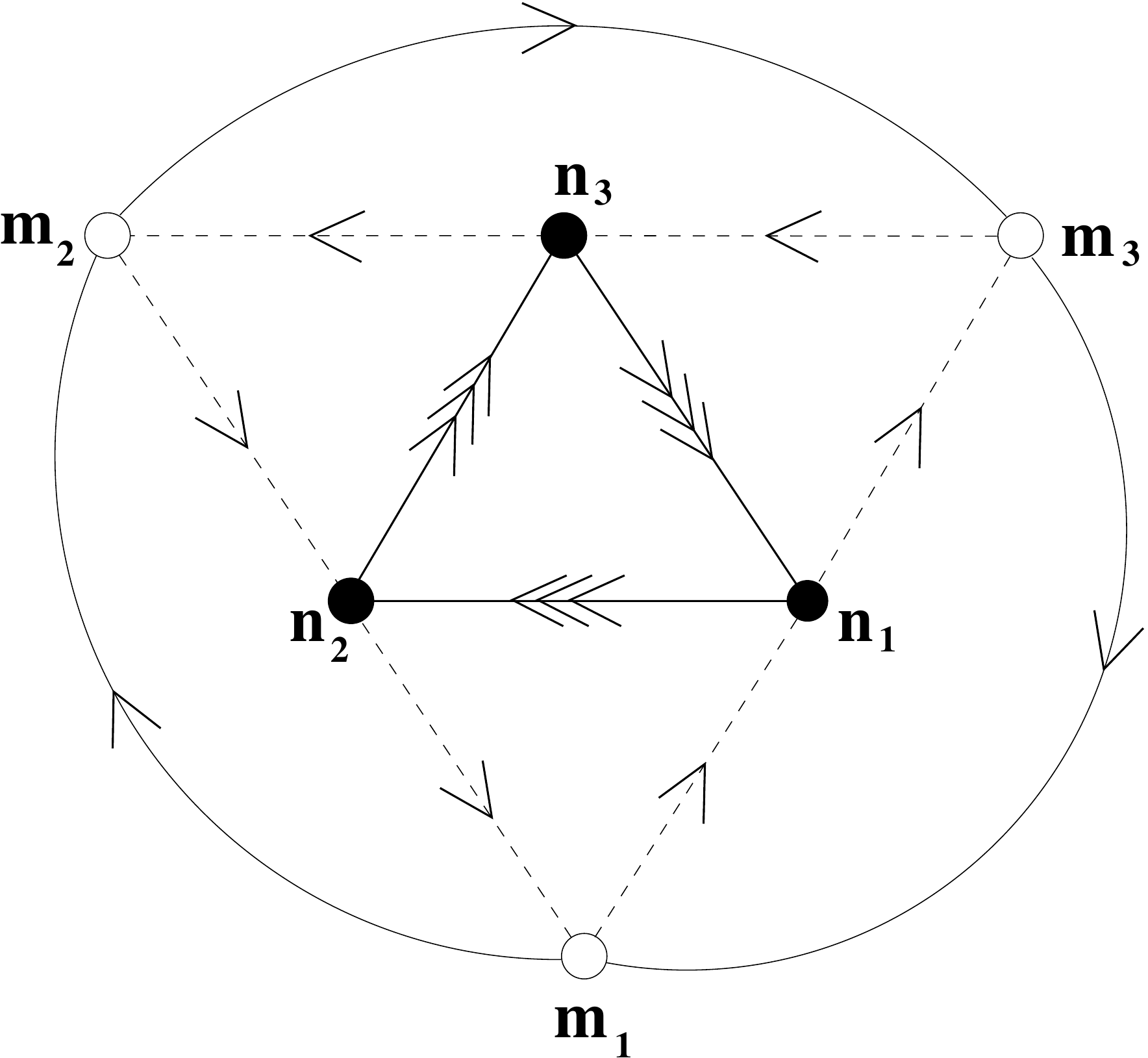}}
\caption{The quiver for D3 and D7 branes probing a $\czt$ singularity.}
\label{dP0}
\end{figure}

    Let  us use our discussion in the previous section to
analyse the spectrum for D3 and D7 branes probing the supersymmetric $\czt$ singularity.
Note that the condition for supersymmetry $(a_4 = 0)$ implies that
the 3-3 Weyl fermions with quantum numbers $(n_{i},\bar{n}_{i+a_{4}})$
are charged under a single gauge group and transform as gauginos.
These combine with the gauge bosons to form vector multiplets. The remaining
fermions have the same quantum numbers as the scalars; they combine to form
chiral multiplets. The spectrum can be summarised using a quiver diagram, as shown in figure \ref{dP0}.
We use white nodes to indicate D3 gauge groups (both the gauge field and the associated gaugino).
Seven branes are indicated by dark nodes. They are characterised by the divisor they wrap and for each such divisor an associated Chan-Paton action. The spectrum is insensitive to the former (since  $b_1 = b_2 = b_3 $ for the orbifold); we do not distinguish between seven branes wrapping different divisors in the quiver diagram. Also, with a slight abuse of notation we use $m_{i}$ to denote the sum of the ranks of the {\it{i}} th identity block of the Chan-Paton actions for the various divisors.
 Matter is indicated by directed  arrows between the nodes;  each arrow corresponds  to a bifundamental chiral multiplet. As mentioned earlier, one sees that there is family triplication for 3-3 states. For three brane gauge groups $U(n_1) \times U(n_2) \times U(n_3)$ the chiral matter
 spectrum is
 \begin{eqnarray}
    && 3 [(n_1, \bar{n}_2, 1 ) +   (1, n_2, \bar{n}_3)   +  (\bar{n} _1,1, {n}_3  ) ]
     + m_1 [ (1, n_2, 1) + (1,1, \bar{n}_3)] + \cr  && m_2 [ (1,1, n_3,) + ( \bar{n}_1,1,1)]  +
     m_3 [ (n_1,1,1) + (1,\bar{n}_2,1)]
\end{eqnarray}
 The condition for cancellation of non-abelian anomalies is
\bel{anocon}
 m_2 = 3(n_3 - n_1) + m_1, \ \ \ \  m_3 = 3(n_3 - n_2) +   m_1,
\ee
 with the constraint $m_{i} \geq 0$. This necessitates the presence of seven branes for unequal gauge groups $n_i$. Given the three brane gauge groups, the
 general solution for $m_{i}$ can be obtained by ordering (without loss of generality) the $n_{i}$ as $n_3 \geq n_{2} \geq n_1$; making an arbitrary choice for $m_1 \geq 0$ and then determining $m_{2,3}$ from \pref{anocon}.

     The interactions can be inferred from the superpotential of N=4 super Yang-Mills.
 The $\czt$ geometry has a manifest $SU(3) \times U(1) $ symmetry - a unitary
transformation of the complex coordinates
$$
     z^{i} \to U^{i}_{\phantom{i} j} z^{j}.
$$
This is reflected as a global symmetry in the interactions of the 3-3 states which interact by a superpotential
$$
    W = \epsilon_{ijk} \rm{Tr} (X_{12}^{i} X_{23}^{j} X_{31}^{k})
$$
where $X^{i}_{12}, Y_{23}^{j}, Z_{31}^{k}$  are  the chiral multiplets with quantum numbers $(n_1,\bar{n}_2,1),
(1,n_2,\bar{n}_3)$ and $(\bar{n}_1,1,n_3)$  respectively.  Each  transforms in the fundamental of the
global $SU(3)$. The presence of seven branes breaks the $SU(3) \times U(1)$ symmetry of the configuration.
This is reflected in the 7-3 interactions. For seven branes warping the divisor $z^3=0$,
the  superpotential is
$$
     W = \epsilon_{ijk} \rm{Tr} (X_{12}^{i} X_{23}^{j} X_{31}^{k}) + {\rm{Tr}}(Y_{\bar1} X_{12} Y_{2})
$$
where $Y_{1}$ and $Y_2$ are the 7-3 fields with quantum numbers $(\bar{n_1},1,1)$ and $(1,n_2,1)$. Note
that the superpotential retains the residual $SU(2) \times U(1)$ symmetry of the background. For
seven branes wrapping sufficiently many divisors, the $SU(3) \times U(1)$ is completely broken.

\subsubsection*{$\czt$: Models}

\subsubsection{Standard Model}

\begin{figure}[h!]
\centering{\includegraphics[height=5cm]{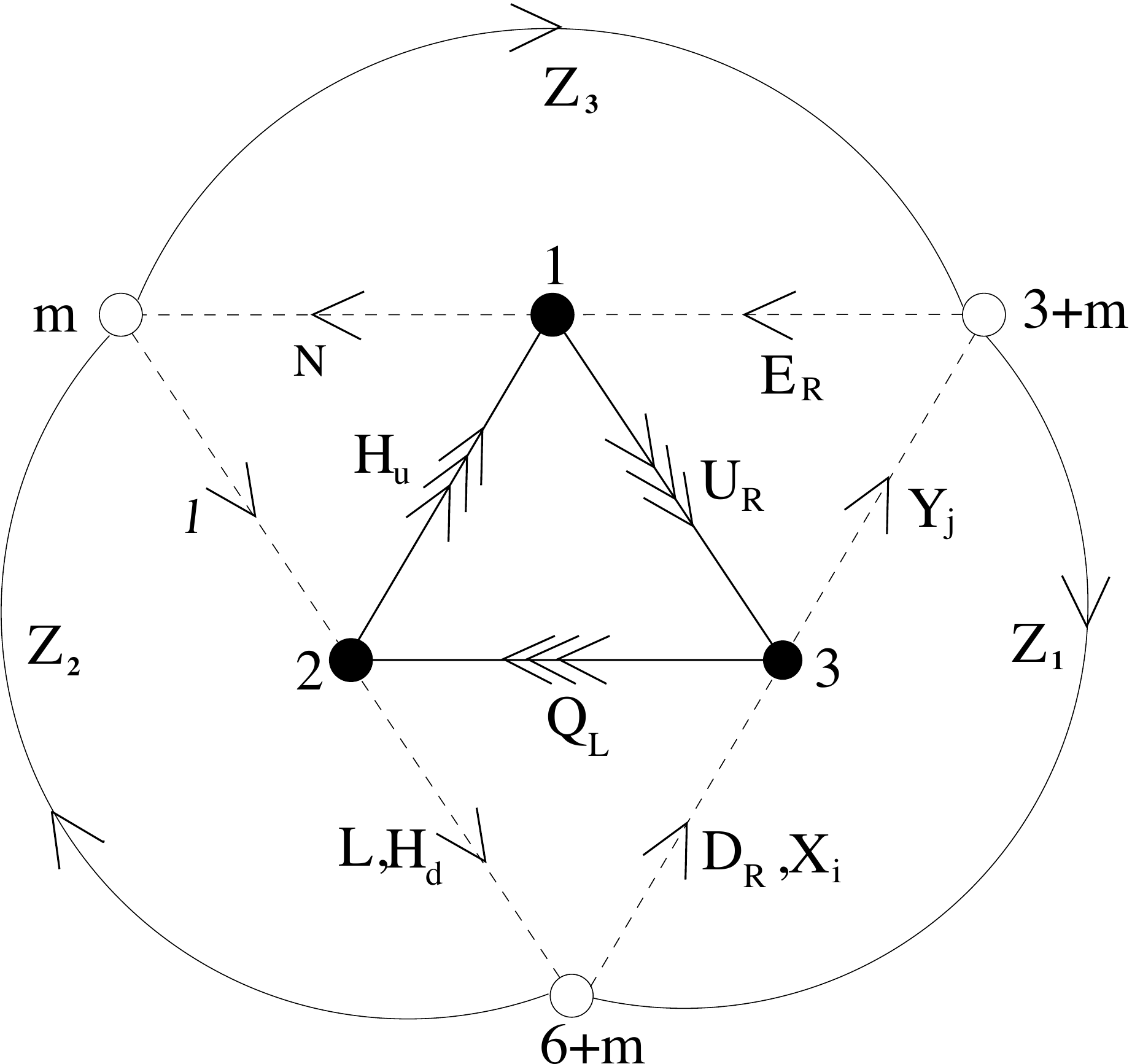}}
\caption{The Standard Model at $\czt$}
\label{dP0SM}
\end{figure}

This was proposed in \cite{botup}. The phenomenological discussion follows
\cite{alda, real}. We use a  quiver diagram in figure
\ref{dP0SM} to show how the Standard Model degrees of freedom arise.

\noindent Some  important features are:

\begin{itemize}
\item{} The total number of D7 branes is related to the free
   parameter $m_{1}$. For the simplest case of $m_1=0$ one has nine seven branes.
\item{} The unique non-anomalous $U(1)$ associated with \pref{hych},  provides
 the  hypercharge.  This has a non-standard normalisation \cite{botup}. Using
 $\rm{Tr} T^2=1/2$ for the $U(n)$ generators,
the hypercharge normalisation is $k_1=11/3$. This gives the Weinberg angle to be $\sin^2\theta_w=1/(1+k_1) =3/14$, close to the experimental value.
\item{} The left handed quarks $Q_L$, right handed up
  quarks $U_R$ and the down Higgs $H_u$ come in three copies and
  couple with a superpotential $\epsilon_{ijk} Q_L^i U_R^j H_u^k$.
  This gives masses to up-quarks.
\item{ }   A general vev of the Higgs fields can be brought to the form (0,0,M) by performing
a global $SU(3)$ rotation. The quark mass matrix then takes the form
\bel{yukprob}
  M_{ij} = \left( \begin{array}{ccc} 0 & M & 0 \\ -M & 0 & 0 \\ 0 & 0 & 0 \end{array} \right),
\ee
This has mass eigenvalues (M,M,0); there are two heavy generations and one light. The mass hierarchies
are highly unrealistic. Later, we will see that this problem can be evaded in singularities closely
related to  $\czt$; the del Pezzo singularities.

\item{} All leptons are 3-7 states; left-handed ones  ($L$) have
  the same origin as the down Higgsses and interact with $Q_L$ and $X$ as
$Q_L L X$.
If the field  $X$ acquires a high mass and is integrated out; the interaction
is irrelevant for low-energy physics.
The $3+m$ right-handed electrons $E_R$ couple to $U_R$ and
  $Y$. There are $m$ extra fields $N, l$ that couple to $H_u$.
Yukawa couplings are not induced for the leptons; we will again find
that this can be ameliorated when one considers models on del Pezzo
singularities.
\item{} The right handed down quarks $D_R$ and the (three) down Higgs $H_d$
are 3-7 states. The 7-3-7 coupling $Q_L D_R H_d$
  provides masses to down quarks.
\item{}    There are $(m+3)$
extra $SU(3)$ vector-like triplets  $X_i, Y_i$ which  can acquire
  a mass if the standard model singlets $Z_{1,2,3}$ get a vev. The spectrum then  reduces to
three copies of: $Q_L, U_R, D_R, L, E_R, H_u, H_d$ which is precisely
the MSSM spectrum;  with two additional Higgs pairs\footnote{For attempts to obtain
just the standard model see \cite{alday}.}. 
\item{}  If the blow up mode is  stabilised at the singularity, the anomalous U(1)s that survive
at low energies as global symmetries can be responsible for the stability of the proton; by forbidding all dangerous $R$-parity
 violating operators  \cite{proton} (see also \cite{alda, anton}).
  If the Fayet-Iliopoulos (FI) parameter  is stabilised at a finite value,
 R-parity violating operators can be generated in the effective action.
 The coefficients of these operators are expected to be suppressed by
powers of the blow-up vev  in string units. The operators would
lead to proton decay via sfermion exchange. The decay rate is
\begin{equation}
    \Gamma \sim  \bigg(   {   |\phi| \over M_{\rm string} } \bigg)^{2(p+q)}    {   m^{5}_{\rm proton} \over 16 \pi^{2} {  M^{4}_{\rm susy}} },
\end{equation}
where $M_{\rm susy}$ is the SUSY breaking scale, $\phi$ the vev of the blow up, and $p$ and $q$ the suppression
powers  of the two MSSM vertices involved in the process. Comparing with the current bounds on
 the proton lifetime, one requires
\begin{equation}
   \bigg({   \langle|\phi| \rangle\over M_{\rm string} } \bigg) ^{(p+q)} < 10^{-27}.
\end{equation}
\item Another mechanism is the presence of discrete symmetries that forbid
proton decay. A concrete example was found in  \cite{alda}  where  a $\mbb{Z}_2$ symmetry
arising  from  the fact that 7-3 states couple in pairs combined with a
residual $\mbb{Z}_2$ from the breaking of the gauge symmetry to yield an effective $R$-parity symmetry.
\item There are no distinct candidates for right-handed neutrinos.
  They could arise as  standard model singlets $Z_{1,2,3}$ or
  other heavy singlets such as KK excitations of moduli fields.

 \end{itemize}

\subsubsection{Left-Right Symmetric Models}

\begin{figure}[h!]
\centering{\includegraphics[height=5cm]{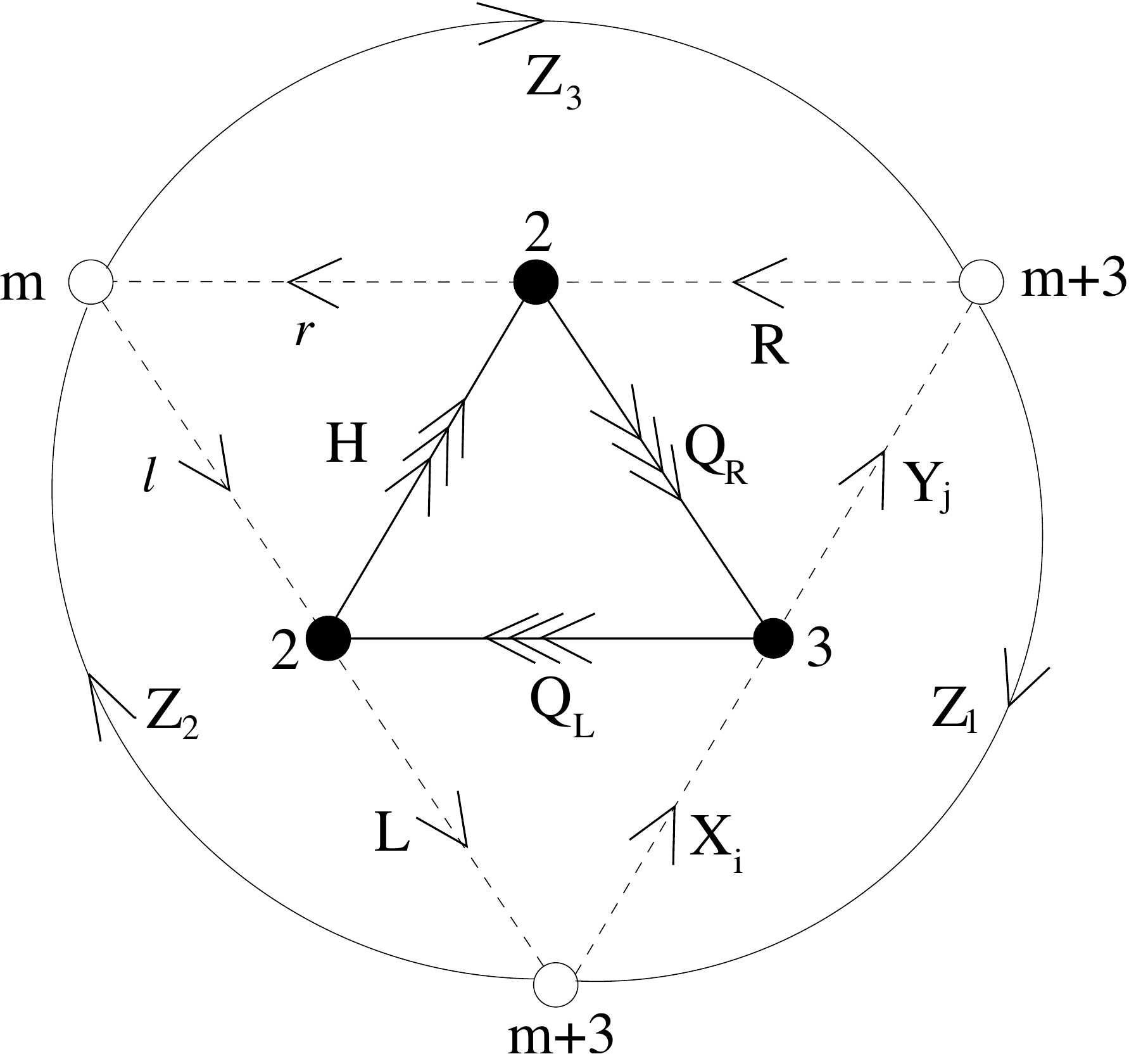}}
\caption{The  Left-Right symmetric model at $\czt$}
\label{dP0LR}
\end{figure}

  It is also possible to obtain the left-right symmetric models with
gauge symmetry $SU(3)_c\times SU(2)_L\times SU(2)_R \times U(1)_{B-L}$
\cite{alda, botup}. Figure \ref{dP0LR} represents the general class of the
models.
We highlight some of the important features: 

\begin{itemize}
\item{} The anomaly free combination \pref{hych} corresponds to
  $U(1)_{B-L}$, with a normalisation factor $k_{B-L}=32/3$. Once the
  breaking to the Standard Model is implemented, the Weinberg angle
  is same as in the Standard Model like construction discussed earlier.
\item{}  Both left-handed and right handed quarks
 $Q_L,U_R, D_R$ arise from 3-3 states. The right handed quarks form an $SU(2)_R$ doublet
$Q_R$.  The Higges $H_u, H_d$  form an $SU(2)_R$
doublet $H$ and come in three families. Unlike the Standard Model case, they are clearly distinguished from leptons.
\item The Yukawa couplings for all quarks arise from the superpotential term
 $\epsilon_{ijk} Q_L^i Q_R^j H^k$.
\item{}
The leptons $L,R$ arise from the 3-7 sector. Again, lepton Yukawas are absent.
The right handed leptons $R$ include  right-handed neutrinos $\nu_R$.
\item{} There are $m+3$ pairs of vector-like triplets $X,Y$ that can acquire a   mass if the left-right singlet $Z_1$ gets a vev.


\item{} If all $Z_{1,2,3}$ get a vacuum expectation value, the model reduces to  the
  supersymmetric version of the left-right model with two extra Higgs.
\item{}
The breaking $SU(2)_R\times U(1)_{B-L} \rightarrow U(1)_Y$ can be achieved by giving the field R a vev.
Hypercharge is related to  $SU(2)_R$ and $Q_{B-L}$ by $Y=T^{3}_R + Q_{B-L}$.
\item{} $U(1)_{B-L}$ forbids proton decay, this
   can survive as a global symmetry if the blow-up mode is
  stabilised at the singular point.
\item{}It was found in \cite{alda} that the
  models lead to gauge coupling unification at the
  intermediate scale $\sim 10^{12} $ GeV with the same level of
  precision as the MSSM.
\item{}  An extension of this model  to a singular F-theory compact model with the left right symmetric model realised by seven D3 
branes and six
D7 branes at a $\mathbb{Z}_3$ singularity was presented in \cite{botup}.    Realisation of the models at $\czt$ in highly warped geometries was discussed in \cite{quw}

\end{itemize}

\subsubsection{Pati-Salam Models}

\begin{figure}[h!]
\centering{\includegraphics[height=5cm]{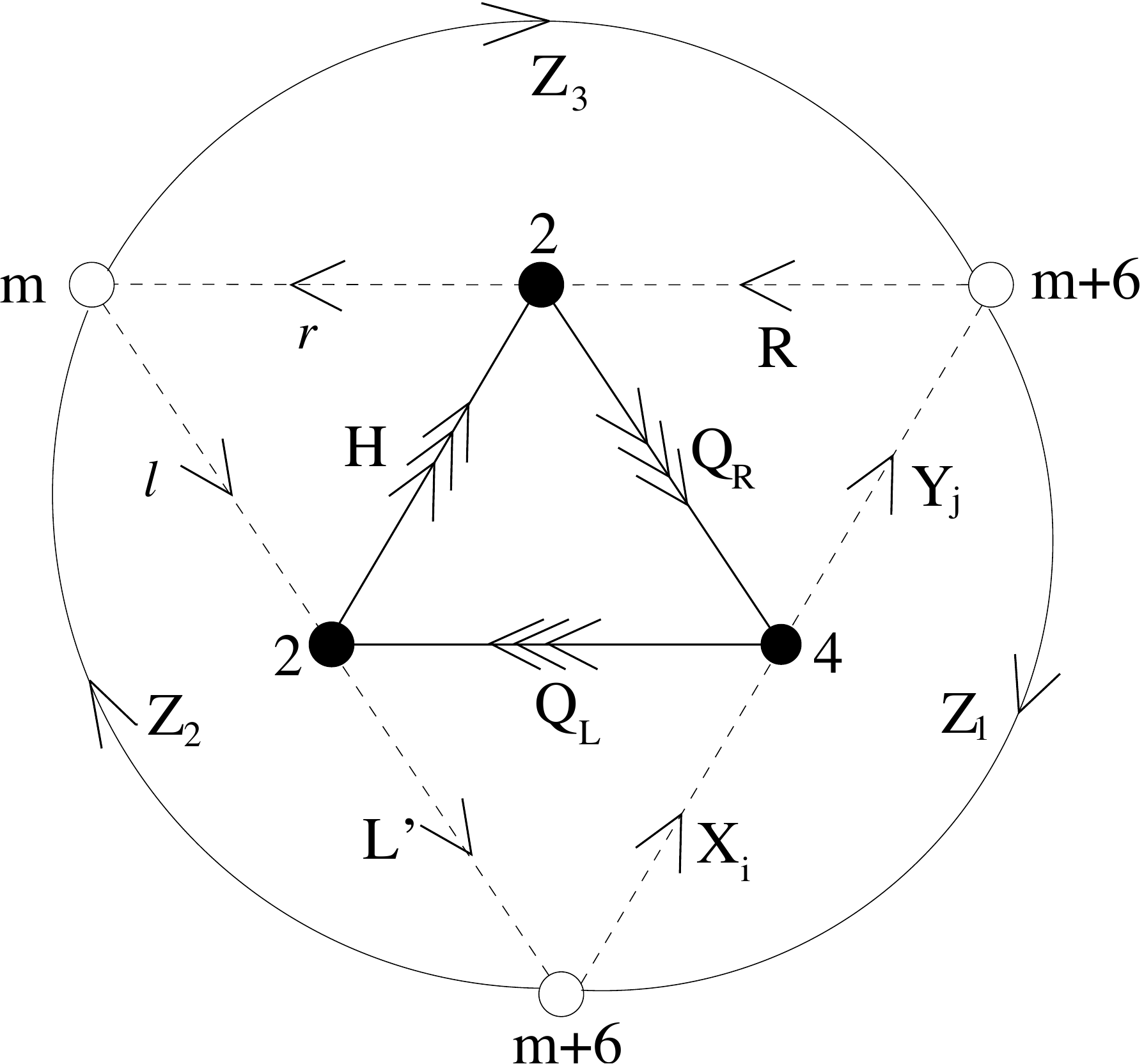}}
\caption{Pati-Salam models
 at a $\czt$ singularity.}
\label{ps}
\end{figure}

Variants of the  Pati-Salam model (figure \ref{ps}) with gauge group
$SU(4)\times SU(2)_L \times SU(2)_R\times U(1)$ were constructed in \cite{gul}
.
The main ingredients of the construction are:

\begin{itemize}
\item{}
The sixteen standard model particles of a single generation, including the right handed neutrinos,
arise  from 3-3 strings.
Yukawa couplings for all quarks and leptons can be generated from the
coupling
$\epsilon_{ijk} Q_L^i Q_R^j H^k$.
\item{} The model reduces to the  original Pati-Salam model along with six
  copies of the left(right) doublets $L' (R)$ if the fields $Z_{1,2,3}$  acquire a vacuum expectation value.
\item{}  Breaking to 
  the standard model can be induced by the right-handed neutrino in $(\bf{\bar{4}},
  \bf{1}, \bf{2})$. Although this would generate masses for some of the quarks
  and the leptons.
\item{}
There are extra doublets of both $SU(2)$'s arising   from the 3-7 sector. There are  also
additional states charged under $SU(4)$-  $X$ and $Y$. These fields can  be used for the
 breaking $SU(4)\times U(1)$ to $SU(3)_c\times
U(1)_{B-L}$. A vev of the field $Z_1$ induces a mass for X,Y.
\end{itemize}

\subsubsection{Trinification Models}
\begin{figure}
\centering{\includegraphics[height=5cm]{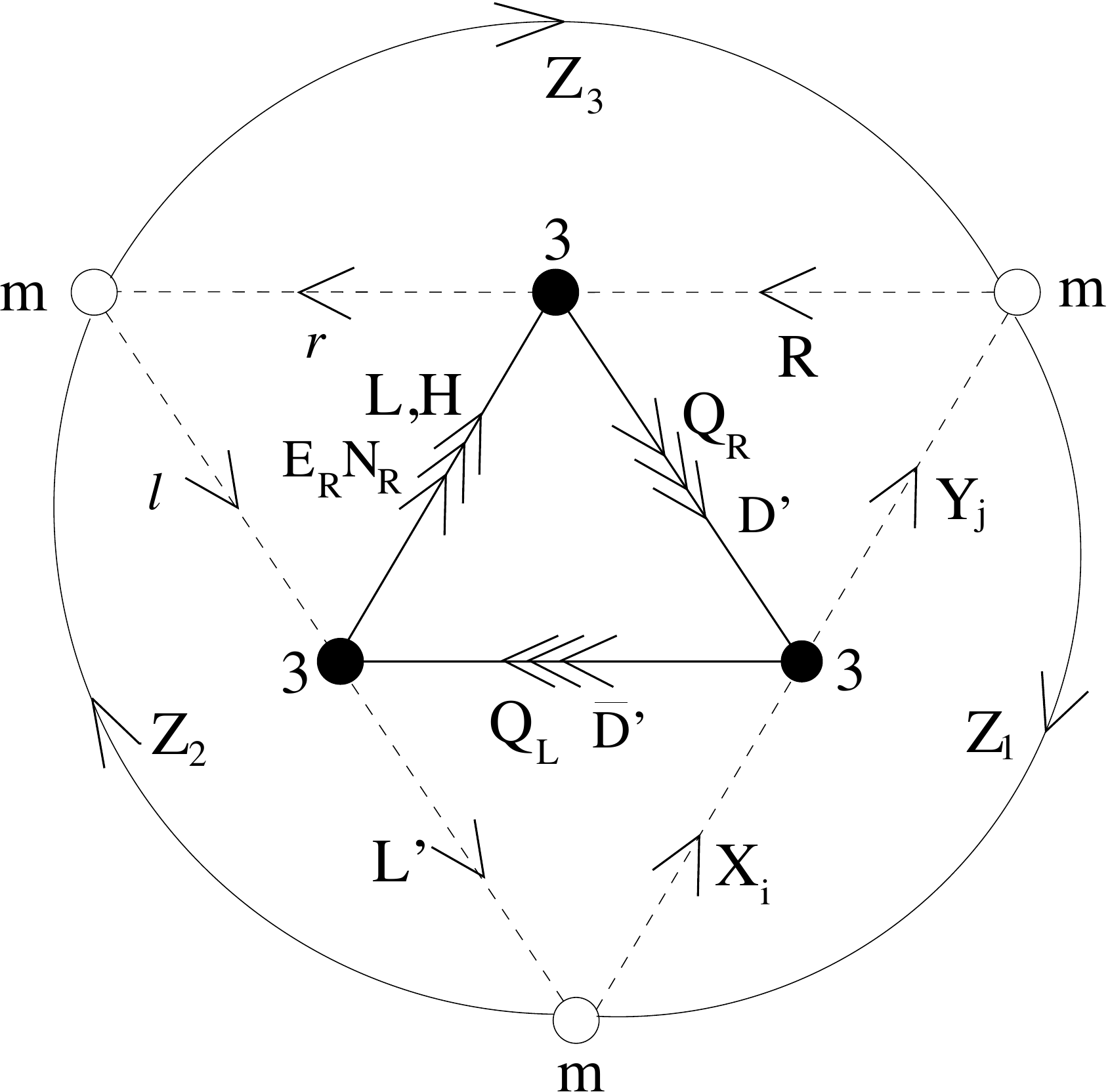}}
\caption{ Trinification Models at $\czt$ }
\label{trin}
\end{figure}

 Trinification models provide another class of interesting extensions; with
 gauge group  $SU(3)^3$. The quantum number of various matter fields
 is shown in  figure \ref{trin}.

\begin{itemize}

\item{}
The anomaly free $U(1)$ in \pref{hych},  is a trivial
overall $U(1)$ that decouples. Thus the gauge group is $SU(3)_c\times
SU(3)_L\times SU(3)_R$; one does not have additional massless $U(1)$'s.
 Hypercharge
has to arises as a $U(1)$ subgroup of $SU(3)_L \times SU(3)_R$.
\item{}
Since all the 3-3 gauge groups have equal ranks; D7 branes are not needed
to cancel the anomalies.
\item The standard model particles, along with additional matter,
 arise from the 27 states in the 3-3 sector:
\be
3 [ \left(\bf{3}, \bf{\bar{3}}, \bf{1}\right) + \left( \bf{1}, \bf{3},
  \bf{\bar{3}} \right) + \left( \bf{\bar{3}}, \bf{1}, \bf{3} \right) ].
\ee
These would map to the adjoint in  an $E_6$ theory. The first nine states
include the left-handed quarks $Q_L$ plus one (exotic)
triplet $\bar{D}'$
of hypercharge $Y=-1/3$. The second nine  consist of the right
handed quarks plus an extra down quark, $D'$. The
leptons and Higgs and two right-handed neutrinos  arise from the last nine.
\item{}Gauge symmetry breaking can be induced by the 3-7 states; this consists  of $m$ 
pairs of $\bf{3}$ and
  $\bf{\bar{3}}$ for each of the $SU(3)$.

\end{itemize}

In summary,  one can construct  various supersymmetric extensions of the standard model 
with three families all containing the matter content of the MSSM and no chiral exotics.
None of the models are fully realistic. One of the major shortcomings is the absence of
lepton Yukawa couplings in the Standard Model and Left-Right symmetric like models.
The vanishing is due to  the anomalous U(1)s, the leptons $L$ and $E_R$
arise from different 7-3 sectors, and the orientation of the arrows
(which indicate the $U(1)$ charge of  the abelian factor in $U(n)=SU(n)\times U(1)$) 
forbids  a coupling between them. For Pati-Salam and Trinification models,
all the Standard model degrees  of freedom arise from 3-3 states thus one has non-trivial Yukawa couplings for both the quarks and leptons.

   However there is a general problem with the Yukawa couplings in the 3-3 sector.
The $SU(3)$ global symmetry requires that the coupling between the matter and Higgs fields
involves the epsilon tensor. This in turn implies that the Yukawa matrix can always be brought
to the form \pref{yukprob} by a global $SU(3)$ rotation. Thus the fermion masses have the form  $(M, M, 0)$, i.e
there are two heavy generations and one light. Another problem is that the CKM matrix associated
with  \pref{yukprob} does not have the required hierarchical structure. All these problems can be
addressed by considering singularities closely related to  $\czt$ -- the del Pezzo singularities.

\subsection{Models on Del Pezzo and general Toric Singularities}
\label{dP1sec}

 Recall that the $\czt$ geometry admits a supersymmetric resolution in which the origin
is blown up to a $\mathbb{P}^2$. The del Pezzo series of surfaces are closely related to
$\mathbb{P}^2$, the {\it{n}}-th del Pezzo surface is   $\mathbb{P}^2$ blown up at {\it{n}} points.
 The reason for unrealistic Yukawas for models on $\czt$, is the large global symmetry - $SU(3) \times U(1)$. These are closely related to the geometric isometries of $\pt$. These isometries are broken in the higher del Pezzo singularities; where the shirking surface is no longer a $\pt$ but a $\pt$ with points blown up. For instance $dP_1$ has a $SU(2) \ti U(1)$ isometry and $dP_2$ a $U(1)$. This makes it natural to consider model building on the higher del Pezzo surfaces, and see if it is possible to retain the attractive features of the models on $dP_{0}$ while addressing flavour issues.
We begin our study with $dP_1$ as this retains some of the family symmetry. The quivers for lower degree del Pezzos can be obtained from higgsing higher del Pezzos,
so the models we now describe are also closely related to the models that can be obtained on $dP_n$,
$n \geq 2$.

\begin{figure}[h!]
\centering{\includegraphics[height=5cm]{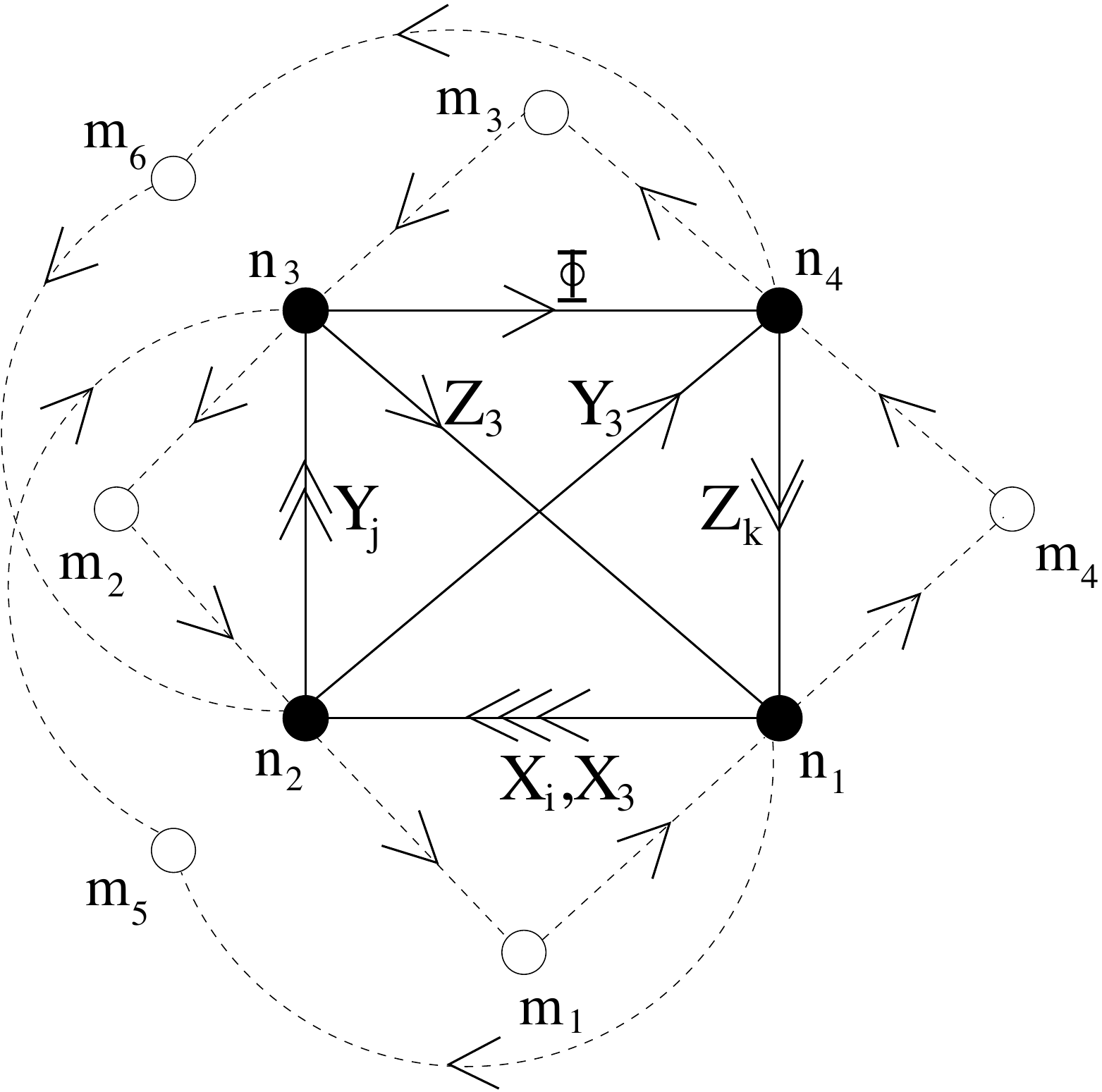}}
\caption{The $dP_1$ quiver}
\label{delPezzo1}
\end{figure}

               The $dP_1$ singularity  is a toric singularity. It can
 be obtained by performing successive blow-ups of the $\mbb{C}^3/\mbb{Z}_3 \ti \mbb{Z}_3$  singularity \cite{feng}. The spectrum of fractional D3 and D7 branes for this model was  computed in \cite{franco}. This is summarised in the quiver diagram  \ref{delPezzo1}.
For every $33$ state $X{i j}$, there exists  a  seven brane giving  a fundamental $Y_i$  and an
antifundamental $Y_j$  with the Yukawa coupling $X_{ij} Y_i Y_j$.

The superpotential for the 33 interactions is
\be
W = \epsilon_{ij}X_i Y_j Z_3 - \epsilon_{ij} X_i Y_3 Z_j + \frac{\Phi}{\Lambda} X_3 \epsilon_{ij} Y_i Z_j,
\label{sp}
\ee
where $\Lambda$ is the  UV cutoff.
There is an $SU(2)$ flavour symmetry under which $X_i$, $Y_i$ and $Z_i$ transform in the fundamental, and also a $U(1)$ flavour symmetry under which
$X_3$ has charge $+1$ and $\Phi$ charge $-1$.

\vspace{0.5cm}

\begin{table}
\tbl{Table showing the continuous flavour symmetry associated with 33 states for various del Pezzo singularities.}
{\begin{tabular}{ | c| c|}
\hline
 singularity  & flavour symmetry  \\
\hline
$dP_{0}$ & SU(3)  \\
\hline
$dP_{1}$ & SU(2) $\times$ U(1) \\
\hline
$dP_{2}$ & $U(1)$ \\
\hline
$dP_{n>2}$ & \rm{none} \\
\hline
\end{tabular}
}
\end{table}

For D3 brane gauge groups of unequal rank, seven branes are necessary for anomaly cancellation.
The number of D7 branes $(m_{i})$ have to satisfy
\bea
\label{oneano}
m_4 &= &n_4+n_3-n_1-n_2+m_1 -m_2+m_3, \nonumber \\
m_5 & = & n_1-2n_2+n_4+m_2-m_3,
\nonumber \\
 m_6 &= & n_4-3n_1+2n_3+m_1-m_2
\eea
Thus for  fixed D3 brane gauge groups  $SU(n_i)$, the models
have three free parameters  $m_{1,2,3}$, with the remaining {\it{m}}-s
fixed by \pref{oneano}. If the field $\Phi$ acquires a vacuum expectation value,
the matter content is that of $\czt$ at low energies. This higgsing is necessary
for models on $dP_1$, as this is the only way to generate family  triplication.
The mass generated for the $Z'$ has to be high enough to evade existing bounds
(see for e.g. \cite{bounds}).

      The reduced family symmetry  $SU(2) \ti U(1)$,  allows for realistic mass hierarchies as opposed to the
$\czt$ case.  Upon diagonalisation of the Yukawa matirx, the mass squared matrix takes the from
$$
 \left( \begin{array}{ccc} M_1^{2} & 0 & 0 \\ 0 & M_2^{2} & 0 \\0 & 0 & 0 \end{array} \right).
$$
with $M_1 \gg M_2 $ for $\frac{\langle\Phi\rangle}{\Lambda}\ll 1$. Instanton
 contributions to Yukawa couplings, which can be relevant to give the lightest generation a mass have been  computed for branes at singularity  in \cite{iu, lerda}.  Radiative corrections can 
also be relevant  for the  lightest generation  to acquire mass \cite{rad}. Another mechanism is
the breaking of isometries as a result of compactification \cite{iso, isot}.

\subsubsection*{Standard and Left-Right Symmetric $dP_1$ Models}

\begin{figure}
\centering
\mbox{\subfigure{\includegraphics[width=3in]{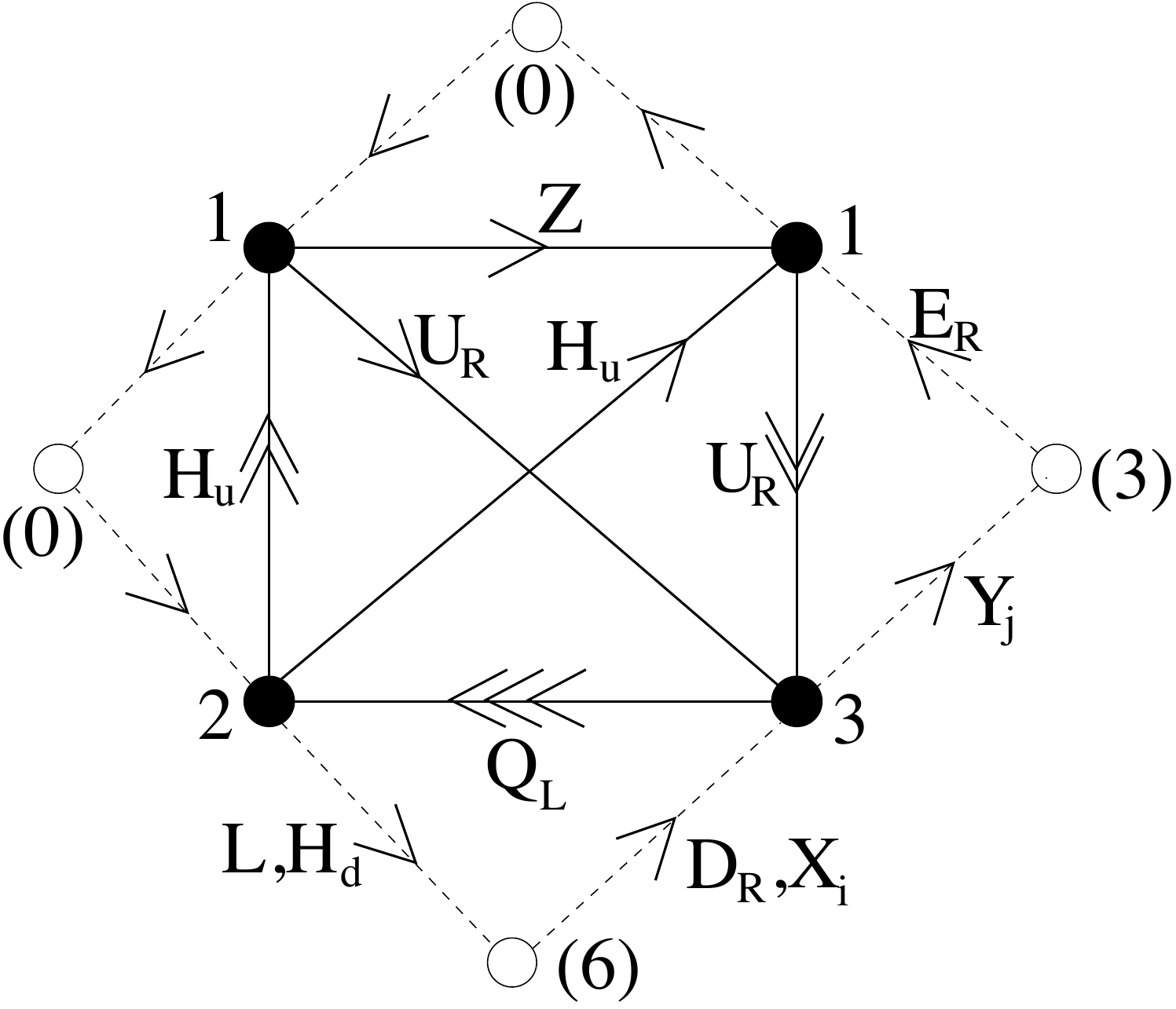}}\quad
\subfigure{\includegraphics[width=3in]{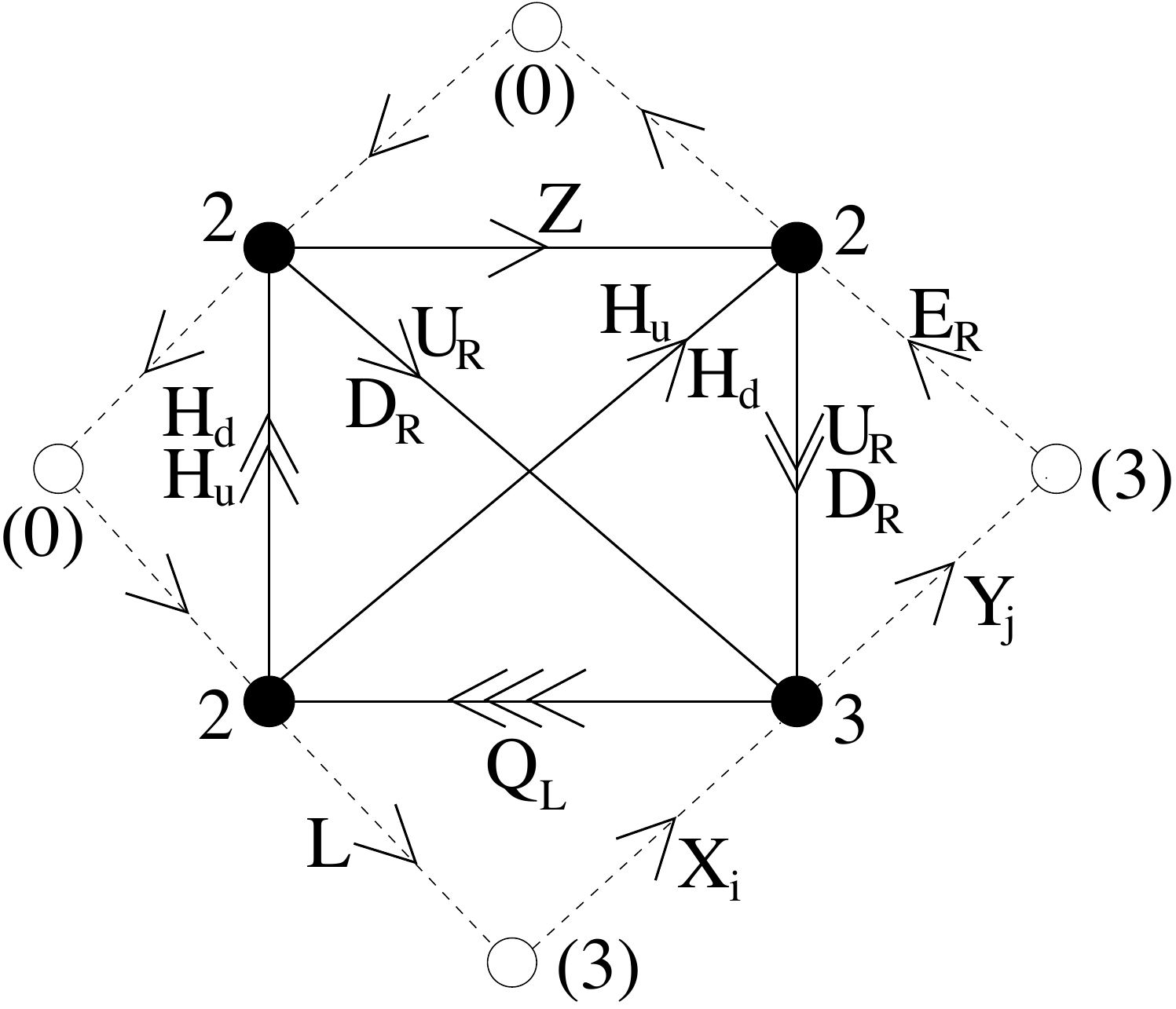} }}
\caption{The Standard Model and Left-Right Symmetric Model at dP1. } \label{dP1mod}
\end{figure}

We now embed the models of the previous section in the $dP_1$ quiver, giving them a more
realistic flavour structure.  Figure  \ref{dP1mod} shows  quivers for an MSSM-like model and  a left-right symmetric model on $dP_1$.

In the non-compact limit, the standard model like construction has (anomaly free)  gauge group $SU(3) \ti SU(2) \ti U(1)_Y \ti U(1)_Z$.
One of these $U(1)$s corresponds to hypercharge and the other to an additional $U(1)_Z$. In the non-compact limit, $U(1)_Z$ is massless. However, if all 2-cycles of $dP_1$ remain 2-cycles of the Calabi-Yau upon compactification; the $U(1)$ acquires a mass by the Green-Schwarz mechanism. 

      Since it is necessary to vev the $dP_1$ quiver down to the $dP_0$ quiver to obtain the standard model gauge group at low energies, let us emphasise the role of $dP_1$ in model building.  The $dP_1$ quiver reduces to the $dP_0$ quiver
upon higgsing; this represents the fact that if the Z field gets a vacuum expectation value and supersymmetry is unbroken; the $dP_1$ theory flows to the $dP_{0}$ theory  in the deep infrared. But if the scale of supersymmetry breaking is well above the scale at which the theory evolves to the $dP_{0}$ theory, the interactions of the model are governed by the $dP_1$ quiver and
not the $dP_0$ quiver. In particular the symmetry relevant for flavour is $SU(2) \times U(1)$ and not $SU(3)$.

\subsection*{Models on general toric singularities and $dP_{2-3}$}

   The constructions on $dP_1$ illustrate that it is possible to obtain semi-realistic models by a judicious
choice of the singularity and brane configuration. This motivates a systematic study of branes at singularities; with the
goal of trying to obtain models which satisfy all phenomenological constraints. One class of singularities for which the  effective field theory is well understood are conical singularities\footnote{See \cite{inaki}
for global embeddings of a large class of such singularities.} in toric Calabi-Yaus. For conical singularities, the metric takes the form
\begin{equation}
ds^2=dr^2+r^2 g_{ij}\, dx^i dx^j\, ,
\end{equation}
where  $g_{ij}$ is the metric on a Sasaki-Einstein space $X$.  In general, one does not have knowledge of the  explicit form of the metric on $X$. But,  the matter content and superpotential of the gauge theory describing D3 branes probing the singularity can be determined from the topology  of $X$. This can be seen via the large volume perspective \cite{doug, dia , freddy, wij, maly}: the matter content and superpotential for D3 branes can be understood
 as arising from D5 branes  wrapping the two-cycles in the non-singular surface $X$. The gauge theory on the D5 branes can be determined  in the intersecting brane picture: D5 branes  wrapping distinct cycles correspond to  gauge groups, chiral matter  arises from intersection of two cycles and triple intersections  give  superpotential terms. The theory
  obtained has gauge groups of equal rank;  unequal ranks can be obtained by including wrapped D7 branes. 

     For toric Calabi-Yaus,  one can use the  powerful techniques of toric geometry to extract the gauge theory  \cite{ he, vafa, sparks} (the first four del Pezzo surfaces are toric).  
A systematic study of generic aspects of model building in toric singularities was carried out in \cite{gul}, primarily making use of the techniques in
\cite{gulo}. The key results of the analysis were:
\begin{itemize}
\item{\it Three families bound:}
 From the general structure of the gauge theories that arise from toric singularities and some phenomenological inputs it was argued that the number of families cannot exceed three.  As we have discussed earlier, a similar result was obtained for $Z_N$ singularities in \cite{botup}.  Given the arbitrariness in the number of families in most string constructions, it is intriguing that the physical value plays an important role in this class of models. Also, it was found that the singularities most attractive from the point of view of model building are
 the del-Pezzo surfaces.
\item{\it Hierarchy of masses:}  Possible forms of the quark mass matrix  were explored. When a physically realistic choice  was made for the quark fields(from the point of view of quantum numbers),  the mass matrix always had a vanishing eigenvalue . Generically it was possible to find hierarchical masses for the other two
eigenstates (except for the zeroth del Pezzo singularity $dP_0$).
\item{\it CKM matrix:}
  The CKM matrix was computed for two  classes of models at  del Pezzo singularities: the first where quarks arise from 3-3 states and the
second where quarks arise from both 3-3 and 3-7 states . It was possible to obtain realistic hierarchical structures in the CKM matrix in both type of models for appropriate values of the additional Higgs vevs.  For the first class of models, the $dP_1$ singularity allowed for the correct form of flavour mixing. In the second class of models  the correct form of mixings  could be obtained for the $dP_2$ and $dP_3$ singularities.

\end{itemize}

   A detailed exploration to obtain a model with the goal of satisfying all phenomenological constraints
was carried out in \cite{dq}. The most promising candidate is a variant of the Pati-Salam model at the $dP_3$ singularity. Some
 interesting features from the point of view of phenomenology are:
\begin{itemize}
\item The Standard model $U(1)_{Y}$ arises as a subgroup of the non-abelian gauge symmetries.
Thus it cannot acquire a mass by the Green-Schwarz mechanism upon compactification.

\item The potential for the heavy higgses allows for hierarchical separation
 in the quark and lepton masses.
\item FCNC contributions are suppressed due to  the additional gauge symmetries
 of $dP_3$ at the high scale which constrain the form of the Kahler potential.

\end{itemize}

A systematic procedure  to glue toric singularities has been developed in \cite{lego},
 allowing for modular model building involving hidden sectors.

\subsubsection*{Models on $dP_{8}$}

     Another interesting class of models have been constructed on cones
over $dP_8$. The geometry is non-toric, but there exists  
 a point in moduli space where the geometry is a non-abelian orbifold
- ${\mathbb{C}} / \Delta_{27}$. The spectrum and basic structure of the
interaction terms at the orbifold point was discussed in \cite{botup}. The minimal consistent
set up  has gauge group $U(6) \times U(3) \times U(1)^9$. The overall $U(1)$ decouples
and two of the other $U(1)$s are anomalous; acquire mass by the Green-Schwarz 
mechanism. Model building in this setting was initiated in  \cite{jaj} . The spectrum 
from the large volume perspective was obtained in \cite{ver}, systematic study of 
the D-terms led to the conclusion  that the ${\mathbb{C}}/ \Delta_{27}$ phase 
does not allow for breaking to the standard model gauge groups; although it
was shown that a Seiberg dual phase is attractive  for model building.  For these models
the hypercharge arises as a linear combination of the abelian factors, it is 
crucial to ensure that it does not become massive upon compatification. This
issue has been discussed in \cite{vert}. 
\subsubsection*{Global Models}
 
 While the  approach  of local model building allows to decouple many issues related
to the standard model degrees of freedom from that of moduli stabilisation, once
an attractive local model is developed it has to be  embedded  in a compact construction.
Early work on this subject includes   \cite{softgary, mirglobal, shiumir, quw, botup}. More recently, a systematic program to carry out global embeddings
of  IIB constructions with D3/D7-branes and O3/O7-planes has been initiated \cite{fano, fanone, fanocol} .
The study exploits the explicit description of compact Calabi-Yaus by means of toric geometry.
Various compact models have been presented, these satisfy global consistency conditions - D7-tadpole, torsion charges and Freed-Witten
anomaly cancellation. This exhibited how  global consistency conditions
can be important for model building for e.g.  tension between moduli stabilisation via non-perturbative effects and chirality (first pointed out in \cite{blum}), tension between moduli stabilisation via non-perturbative effects and
the cancellation of Freed-Witten anomalies. Two classes of models have been studied - the first class \cite{fanone} with visible sector D7-branes in the geometric regime: chiral SU(5) or
MSSM-like models  and the second \cite{fanocol} with visible sector fractional D3-branes at del Pezzo 
singularities. Global model building in the F-theory context will be discussed in  section \ref{sec:int7iib}.

%
\section{Intersecting 7-branes in IIB}
\label{sec:int7iib}

The Standard Model of particle physics is a gauge theory with charged chiral matter and one of the most satisfying aspects of string theory is that such a sector arises very naturally. Gauge theories are supported on D-branes and matter representations are supported on intersections of branes. In this section we study the realisation of realistic matter sectors in type IIB string theory. In particular we limit ourselves to the geometric regime where all geometric curvature scales are much larger than the string scale and so we are able to use field-theory techniques to study the open-string sector. We discuss D3-branes on singularities in section \ref{sec:bs}.\footnote{Although see \cite{Heckman:2010pv} for work on the singular limit in F-theory.} In the geometric regime of O3/O7 compactifications the supersymmetric D-branes of type IIB string theory are D3- and D7-branes. Since in the absence of singularities on the manifold the theory on a D3-brane is non-chiral, realistic chiral theories are associated to the D7-brane sector. In this section we review the basics of how to construct gauge theories with chiral matter in a type IIB setting using D7-branes and O7-planes. Such theories can only involve gauge groups of type $U(N)$, $SO(N)$ and $Sp(N)$ and in particular do not allow for the exceptional gauge groups $E_6$, $E_7$ and $E_8$. The absence of exceptional group structures can be evaded in the strong coupling regime of type IIB string theory which can be described using the framework of F-theory \cite{Schwarz:1995dk,Witten:1995ex,Vafa:1996xn}. Contemporary model building makes significant use of this due to attractive phenomenological properties of exceptional groups as discussed in section \ref{sec:yukawaF}. Therefore the following sections will be concerned with this approach. In section \ref{sec:7inf} we review how 7-branes, and their worldvolume fluxes, are described in F-theory through the geometry of CY four-folds. One of the interesting aspects of studying 7-branes is that they only wrap a submanifold of the full internal space which means that some aspects of the physics associated to them can be studied by only considering the local  geometry around the 7-brane. In section \ref{sec:locf} we review the tools and results that can be obtained from this approach. Finally in section \ref{sec:feno1} we review how the ideas and techniques presented can be utilised for detailed phenomenological applications.

For a review, particularly of type I model bulding see \cite{Blumenhagen:2006ci}, while in this section we follow the discussion of \cite{Blumenhagen:2008zz} which summarises the rules for D7 model building.

Consider a stack of $N_a$ D7 branes wrapping a (four-cycle) divisor $D_a$ in a CY 3-fold $Y$. We denote the divisor wrapped by the O7-plane $D_{O7}$ and the orientifold image of $D_a$ by $D'_a$. The gauge group of such a configuration depends on relative homology classes of of $D_a$ and $D_{O7}$ and is given by $U(N_a)$ in the case where they are not in equivalent homology classes, by $SO(2N_a)$ in the case where the brane stack is on top of the O7 pointwise, and by $Sp(2N_a)$ otherwise.\footnote{Note that the intersection of $O7$-planes with $D7$-branes in IIB is always a double point intersection \cite{Collinucci:2008pf}.}

The gauge coupling strength or, with supersymmetry, gauge kinetic function for a D7-brane is given by the size of the 4-cycle that it wraps, see for example \cite{Jockers:2004yj}. The simplest way to see this is to consider the Chern-Simons term in the D7-action\footnote{For simplicity we have dropped the $\hat{A}$-roof genus terms though it should be noted that these are important for the cancellation of gravitational anomalies, see for example \cite{Plauschinn:2008yd}.} 
\be
S^{CS}_{D7} = 2\pi \int_{M_4\times D_a} \sum_{2p} C_{2p} \Tr\left[e^{\frac{\cF_a}{2\pi}}\right] \;. \label{d7act}
\ee
Here $M_4$ denotes the external space-time, $C_{2p}$ are the closed string RR fields, and $\cF_a$ is the world-volume gauge field strength. The particular term of interest is
\be
S^{CS}_{D7} \supset \frac12 \int_{M_4} \Tr\left[ \cF_a \wedge \cF_a \right] \int_{D_a} \frac{1}{2\pi} C_4\;.
\ee
Within a supersymmetric action this term appears as part of the gauge kinetic function which must be holomorphic in the fields. The relevant superpartner to $C_4$ appears from the complexification with the volume of the 4-cycle \cite{triv, grimm04}
\be
T_i = \int_{\tilde{\omega}_i} \left(e^{-\phi} \left[-J \wedge J + B \wedge B\right] + iC_4\right) \;,
\ee
where $\phi$ is the dilaton, $J$ is the Kahler form on the CY, $B$ is the NSNS 2-form, and $\tilde{\omega}_i$ denotes a basis of 4-cycles in the CY. The gauge kinetic function is therefore the appropriate combination of the $T_i$ according to how $D_a$ is decomposed in terms of the basis $\tilde{\omega}_i$, and so the real part which is the gauge coupling strength is given by the associated volume.

Massless matter in the adjoint representation arises from strings with both ends on the same brane along $D_a$. There is the vector multiplet of the gauge group, $h^{(1,0)}\left(D_a\right)$ chiral multiplets corresponding to Wilson line moduli, and $h^{(2,0)}\left(D_a\right)$ chiral multiplets coming from deformations of the brane. Matter in the bifundamental representation arises from strings stretching between two different D7-branes and this is only chiral in the presence of background world-volume flux. Therefore such a flux is necessary for any realistic model building. For a stack of $N_a$ D7-branes carrying gauge group $U(N_a)$, we can consider a background flux 
\be
\cF_a = T_0 \left(f_a^0 + i^*B\right)+ \sum_i f_a^i T_i \;.
\ee
Here the $T_i$ denote the generators along which the flux is turned on with $T_0$ denoting the diagonal $U(1)_a \subset U(N_a)$. $i^*B$ is the pull-back to the brane of the universal NS-NS anti-symmetric two-form which always appears in combination with the flux as above. We consider for simplicity only Abelian flux, see for example \cite{Blumenhagen:2005pm} for a general discussion of the non-Abelian case in terms of D9-branes. Since the flux is Abelian it corresponds to line bundles 
\be
c_1\left(\cL^0_a \right) = \frac{1}{2\pi} \left(f_a^0+i^*B\right) \in H^2\left(D_a\right)\;,\;\; c_1\left(\cL^i_a \right) = \frac{f^i_a}{2\pi} \in H^2\left(D_a\right) \;.
\ee
The flux must satisfy a quantisation condition \cite{Freed:1999vc,Blumenhagen:2008zz}
\be
T_0 \left( c_1\left(\cL^0_a \right) - i^* B\right)+\sum_i T_i c_1\left(\cL^i_a \right) + \frac12 T_0 c_1\left(K_{D_a} \right) \in H^2\left(D_a,\mathbb{Z}\right) \;,
\ee
where $K_{D_a}$ is the canonical bundle of $D_a$ which is odd for non-{\it Spin} manifolds. In such cases one way to allow for integer fluxes is to turn on a background half-integer $B$-field. The presence of the orientifold implies that the fluxes split into even and odd components; for simplicity we do not describe this in detail but refer to \cite{Blumenhagen:2008zz} for a discussion of this.

\subsection{Tadpoles}
\label{sec:d7tad}
 
The flux is also subject to additional constraints coming from Tadpoles in the closed string sector. The $D7$ tadpole limits the brane stacks themselves
\be
\sum_a N_a \left(\left[D_a\right]+\left[D'_a\right] \right) = 8 \left[D_{O7}\right] \;.
\ee  
While the D5-tadpoles constrain the flux
\be
\sum_a \int_{CY} \omega \wedge \left(\left[D_a\right] \wedge \mathrm{Tr}\cF_a+\left[D'_a\right] \wedge \mathrm{Tr}\cF'_a \right) = 0 \;, \label{d5t}
\ee
where we define
\be
\mathrm{Tr}\cF_a = 2\pi \sum_I \mathrm{Tr}\left[ T_I \right] c_1\left(\cL_a^I\right)\;,\;\; I=0,i \;,
\ee
and (\ref{d5t}) must hold for all elements $\omega \in H_2\left(CY,\mathrm{Z}\right)$.

The D3 tadpole receives contributions also from closed string fluxes $H_3$ and $F_3$, and any O3/D3-branes that are present in the compactifications
\be
N_{D3}+N_{D3'}+ \frac{1}{2(2\pi)^2}\int_{CY} H_3 \wedge F_3 - \sum_a\left(Q^a_{D7}+Q^a_{D7'}\right) = \frac12 N_{O3} + Q_{O7} \;.
\ee
Here 
\be
Q_{O7} = \frac16 \chi\left(D_{O7}\right) = \frac16 \int_{CY} c_2\left(D_{O7}\right) \wedge \left[ D_{O7} \right] \;,
\ee
where $\chi$ is the Euler characteristic. While
\be
Q^a_{D7} = N_a \frac{\chi\left(D_a\right)}{24} + \frac{1}{8\pi^2} \int_{D_a} \mathrm{Tr} \cF^2_a \;,
\ee
with
\be
\frac{1}{8\pi^2} \mathrm{Tr} \cF^2_a = \frac12 \sum_{I,J} \Tr\left[T_IT_J\right]c_1\left(\cL^I_a\right)\wedge c_1\left(\cL^J_a\right) \;.
\ee
In the case where $\left[D_a\right] = \left[D'_a \right]$ but not pointwise, the D3-tadpole expression is modified to include possible pinch points in $D_a$ and the expression is given in \cite{Blumenhagen:2008zz}.

\subsection{Chiral Spectrum}
\label{sec:d7spec}

The presence of flux induces a chiral spectrum of states stretching between D7-branes $D_a$ and $D_b$ and their orientifold images in the bi-fundamental, symmetric and anti-symmetric representations. The spectrum is show in table \ref{tab:d7spec}. 
\begin{table}
\tbl{Table showing the chiral spectrum of states stretching between $N_a$ D7-branes wrapping a divisor $D_a$ with orientifold image $D'_a$ and $N_b$ D7-branes wrapping divisor $D_b$. Here ${\bf N_a}$ denotes the fundamental representation of $SU(N_a)$ and the subscript denotes the diagonal $U(1)$ charge, while ${\bf A.S}$ and ${\bf S}$ denote the anti-symmetric and symmetric representations respectively.}
{\begin{tabular}{|c|c|c|c|}
\hline
Sector & $U\left(N_a\right)$ & $U\left(N_b\right)$ & Chirality \\
\hline
$a - b$ & $\left({\bf \overline{N}}_a\right)_{-1}$ & $\left({\bf N_b}\right)_{+1}$ & $I_{ab}$ \\
\hline
$a' - b$ & $\left({\bf N_a}\right)_{+1}$ & $\left({\bf N_b}\right)_{+1}$ & $I_{a'b}$ \\
\hline
$a' - a$ & $\left({\bf A.S}\right)_{+2}$ & $1$ & $\frac12\left(I_{a'a}+2I_{O7a}\right)$ \\
\hline
$a' - a$ & $\left({\bf S}\right)_{+2}$ & $1$ & $\frac12\left(I_{a'a}-2I_{O7a}\right)$ \\
\hline
\end{tabular}}

\label{tab:d7spec}
\end{table}
The appropriate chiral indices $I_{ab}$ can arise from two types of matter: bulk matter and matter localised on curves. Bulk matter arises when the two D7 stacks wrap the same divisor $D_a=D_b=D$. In this case we have
\be
I_{ab}^{bulk} = - \int_{CY} \left[D\right] \wedge \left[D\right] \wedge \left( c_1(\cL_a) - c_1(\cL_b)\right) \;. \label{iibbulkchir} 
\ee
It is important to note that although this type of matter is denoted bulk matter it is still localised to some extent within the bulk by the flux.\footnote{This follows from general considerations since if the internal wavefunction of a state has additional isometries then correspondingly the state has additional supercharges. So for example a complex one-dimensional isometry in the wavefunction would imply the state is six-dimensional and preserves 8 supercharges and is therefore non-chiral.}

The chiral spectrum of matter localised on curves in the CY where two D7-branes wrapping different divisors intersect is counted by the similar expression
\be
I_{ab}^{curve} = - \int_{CY} \left[D_a\right] \wedge \left[D_b\right] \wedge \left( c_1(\cL_a) - c_1(\cL_b)\right) \;. \label{iibcurflux}
\ee
The relevant index for calculating the symmetric and anti-symmetric representations is given by
\be
I_{O7a} = \int_{CY} \left[D_{O7}\right] \wedge \left[D_a\right] \wedge c_1(\cL_a) \;.
\ee

In the presence of a chiral spectrum the field theory can exhibit gauge and gravitational anomalies. Within a globally consistent setting these anomalies of the massless spectrum are cancelled by the Green-Schwarz mechanism. The tadpoles discussed in section \ref{sec:d7tad} play an important role in this and they can be mapped explicitly to the cancellation of anomalies \cite{Plauschinn:2008yd}. The appropriate terms for the GS mechanism arise from the CS term in the D7 action
\be
\int_{M_4 \times D_a} C_4 \wedge \Tr\left[\cF_a \wedge \cF_a\right] = \int_{M_4} C_2^i \wedge \Tr \left[ F_a \int_{D_a} f_a \right] \wedge i^*\left(\omega_i\right) + \int_{M_4} a_i \Tr\left[F_a \wedge F_a\right] \int_{D_a} i^*\left(\tilde{\omega}^i\right) \;. \label{d7gs} 
\ee
Here we have expanded $C_4$ in a cohomology basis of 4-forms and their 2-from duals $C_4=a_i \tilde{\omega}^i+ C_2^i \omega_i$, wth $i^*$ denoting their pullback to the brane.
The first term of (\ref{d7gs}) is a Stueckelberg mass for the anomalous $U(1)$ gauge field and the second term is an axion coupling. Since the cohomology bases of 4-forms and 2-forms on the CY are not independent but are related by Hodge duality the four-dimensional fields $C_2^i$ and $a_i$ are not independent degrees of freedom but rather are related by the four-dimensional duality between axions and tensors. The precise relations are such that all cubic anomalies involving $U(1)$ factors are cancelled.

In the case where the orientifold odd cohomology of the CY, denoted with index $\alpha$, is non-vanishing there are additional anomalies and correspondingly an additional contribution to the GS anomaly cancellation mechanism. The contributing terms this time are 
\bea
& & \int_{M_4 \times D_a} C_2 \wedge \Tr\left[\cF_a \wedge \cF_a \wedge \cF_a \right] + \int_{M_4 \times D_a} C_6 \wedge \Tr\left[\cF_a\right] \nn \\
&=& \int_{M_4} a^{\alpha} \wedge \Tr \left[ F_a \wedge F_a \int_{D_a} f_a \right] \wedge i^*\left(\omega_{\alpha}\right) + \int_{M_4} C_{2{\alpha}} \wedge \Tr\left[F_a \right] \int_{D_a} i^*\left(\tilde{\omega}^{\alpha}\right) \;.\label{d7gsodd}
\eea

An interesting thing about (\ref{d7gsodd}) is that the  Stueckelberg mass term is independent of the internal flux and in fact can induce a mass of the diagonal $U(1)$ in the $U(N_a)$ even in the absence of any flux. Indeed this will occur in the case where $[D_a] \neq [D'_a]$ since then it is possible to make an orientifold odd combination from the divisor and its image.

\subsection{Supersymmetry}
\label{sec:d7susy}

Since supersymmetry is expected to only be broken well below the string scale we are mostly interested in brane and flux configurations that preserve some supersymmetry. Supersymmetric configurations are also favourable since they are typically stable while non-supersymmetric ones are typically unstable. This is related to the fact that for a supersymmetric vacuum cancellation of RR tadpoles, as discussed in section \ref{sec:d7tad}, implies the cancellation of NSNS tadpoles which in turn are related to vacuum instability.

The conditions for the D7-branes to be supersymmetric themselves is that they must wrap (orientifold even) holomorphic divisors. For the fluxes, being line bundles, we have an induced D-term which must be satisfied. The Fayet-Iliopoulos (FI) term induced by the flux is
\be
\xi_a \sim \int_{D_a} \left(i^* J\right) \wedge c_1\left(\cL_a\right) \;,
\ee
where $J$ is the Kahler form on the CY so that the FI term is a function of the Kahler moduli. Of course, the D-term also recieves contributions from any charged matter fields that arise in the open-string sector. Overall the question of finding a field configuration which solves the D-term is intimately tied to the issue of moduli stabilisation and the most complete studies of this are presented within the context of explicit global models \cite{Blumenhagen:2007sm, fanone ,Balasubramanian:2012wd, fanocol}. A particularly important question to understand in this context is why the D-terms do not induce vaccum expectation values for fields charged under the standard model gauge groups.

\subsection{Yukawa Couplings: A Motivation for F-theory}
\label{sec:yukawaF}

So far we have primarily discussed the spectrum of fields in D7-models, however a realistic model is also expected to recreate the correct interactions between the fields. In particular reproduce the observed pattern of Yukawa couplings at least approximately. Yukawa couplings for matter arising from intersecting D7-branes can be studied by modeling the full set of intersecting branes as an 8-dimensional enhanced gauge theory which is broken to a the actualy gauge group of the intersecting D7-branes by a spatially varying background Higgs vev. We will describe this in some detail in section \ref{sec:fultralocal} in an F-theory context but the complete discussion will equally apply to type IIB D7-branes. We therefore refer the reader interested in more detailed aspects of Yukawa couplings on D7-branes to those sections. However, at this stage, while considering only D7-branes there is a serious problem that arises at much more fundamental level which is the subject of this subsection. 

The problem we are interested in is the presence of a top quark Yukawa coupling for realisations of Grand Unified Theories (GUTs) in type IIB. GUTs are well motivated from the apparent unification of the Standard Model gauge couplings in the MSSM. The minimal such unification is to $SU(5)\supset SU(3)\times SU(2) \times U(1)_Y$. The MSSM spectrum fits into the fundamental ${\bf 5}$ and anti-symmetric ${\bf 10}$ representations of an $SU(5)$ GUT theory broken to the MSSM gauge group 
\bea
{\bf 5} &\rightarrow & \left({\bf 3},{\bf 1}\right)_{-\frac13} + \left({\bf 1},{\bf 2}\right)_{\frac12}  \;, \nn \\
{\bf 10} &\rightarrow & \left({\bf 3},{\bf 2}\right)_{\frac16} + \left({\bf \bar{3}},{\bf 1}\right)_{-\frac23} + \left({\bf 1},{\bf 1}\right)_{1}  \label{su5decomp} \;.
\eea
The up Higgs is embedded into a ${\bf 5}_{H_u}$ multiplet and the top quark Yukawa coupling arises from the GUT coupling
\be
Y_{\mathrm{top}} \subset {\bf 5}_{H_u} {\bf 10}\; {\bf 10} \;.
\ee

In type IIB O7/D7 constructions such an $SU(5)$ would be realised by a stack of 5 D7-branes wrapping divisor $D_a$ and intersecting a single D7 brane wrapping divisor $D_b$. Then according to table \ref{tab:d7spec} matter transforming as the ${\bf 5}$ of $SU(5)$ arises on the intersection of $D_a \cap D_b$ and matter in the anti-symmetric comes from $D_a \cap D_{O7}$. The Abelian gauge group is $U(1)_a\times U(1)_b$, and with respect to it the fields appearing in the Yukawa coupling have the following charges
\be
Y_{\mathrm{top}} \subset {\bf 5}^{(1,-1)}_{H_u} {\bf 10}^{(2,0)}\; {\bf 10}^{(2,0)} \;.
\ee
It is manifest that the coupling is not gauge invariant under the Abelian gauge group and is therefore perturbatively forbidden. This is obviously a serious setback for such model building because the top quark Yukawa coupling is known to be of order 1 at the GUT scale.

A possible solution to this problem was given in \cite{Blumenhagen:2006xt,Ibanez:2006da,Blumenhagen:2007zk,Kiritsis:2009sf} where it was shown that the coupling could be generated non-perturbatively with an $O(1)$ E3-instanton. However such an instanton is non-perturbative in both the string coupling $g_s$ and the 4-cycle volume and therefore should not be of order 1 if the perturbative $g_s$ and $\alpha'$ expansion is to hold.

The requirement of a perturbatively allowed top quark Yukawa coupling for type IIB GUT models necessarily leads to its strong coupling regime and this is described by F-theory. F-theory will be discussed extensively in the upcoming sections and the solution to the top quark Yukawa problem that it offers has played no small part in much of the interest that it has attracted recently. The key aspect of F-theory is that it allows for more states than are present in type IIB and in particular for different realisations of the anti-symmetric SU(5) representations. One particular such realisation comes from the presence of exceptional symmetries and in particular from the adjoint of $E_6$ which decomposes to $SU(5)\times U(1)_a \times U(1)_b$ as
\bea
\bf{78} &\rightarrow& \bf{24}^{(0,0)} \op \un^{(0,0)} \op \un^{(0,0)} \op \un^{(-5,-3)} \op \un^{(5,3)} \nn \\
 & & \op \left(\f \op \fb\right)^{(-3,3)} \op \left(\te \op \teb\right)^{(-1,-3)} \op \left(\te \op \teb\right)^{(4,0)} \;. \label{e62su5}
\eea
The crucial thing to note is that the ${\bf 10}$ representations are charged under 2 separate Abelian symmetries, and this is something which is not possible in perturbative type IIB since they arise from strings stretching between the brane and its orientifold image and so are only charged under the diagonal $U(1)$ of the brane. The F-theory ${\bf 10}$s can instead be thought of as 3-pronged strings which are bound states of F and D strings that are able to form string junctions. With the new states it is manifest from (\ref{e62su5}) that a top quark Yukawa coupling can be gauge invariant under all the symmetries of the theory.
Hence we can deduce that the requirement, following substantial phenomenological motivations, of an order one top quark Yukawa coupling within a GUT model implies that F-theory, and not the weakly coupled type IIB string, is the more appropriate arena for model building. with this motivation in mind we proceed to describe the relevant aspects of the F-theory framework.

\section{7-brane description in F-theory}
\label{sec:7inf}

F-theory \cite{Vafa:1996xn} is not so much a theory as a collection of results regarding type IIB in backgrounds with spatially varying, and possibly strong, string coupling. The idea is that aspects of the background geometry, the 7-branes and the dilaton profile can be described in a combined way by the geometry of an elliptically fibered CY four-fold. The elliptic fiber, which is a torus fibered over every point in the CY base $B$, describes the profile of the dilaton, and 7-branes correspond to degenerations of this fibre. There is no sense in which a 12-dimensional theory is reduced on the four-fold to four dimensions, but it is possible to define F-theory as a certain limit of the dimensional reduction of M-theory on an elliptically fibered Calabi-Yau fourfold to three dimensions. The limit is one where one of the internal dimensions along the torus is shrunk to zero size, and therefore if we T-dualise this to IIB this direction is decompactified to yield four large dimensions. This is the more working definition since this way we know, at least the low energy limit of, the higher dimensional theory as 11-dimensional supergravity. For methodical discussions of this way of approaching F-theory see \cite{Grimm:2010ks,Grimm:2012yq,Cvetic:2012xn} for example. Some of the more detailed aspects of F-theory can only really be understood in this way.

It is important to note that apart from the phenomenological motivations, such as the top quark Yukawa discussed in the previous section, F-theory is important for understanding type IIB vacua in the presence of 7-branes generally. This is because one of the interesting properties of 7-branes is that since they must wrap an internal 4-cycle they live in (complex) co-dimension 1 of the internal space. This means that modes that are sourced by them, in particular the axio-dilaton, propagate in two dimensions and therefore have a logarithmic propagator. Hence the backreaction of 7-branes on the geometry is not localised and in particular we must keep track of the sourced dilaton profile over the full compact geometry. This global constraint on the dilaton profile is precisely captured by the CY condition of F-theory.

In the previous section considering D7-branes in type IIB string theory we were thinking of a background geometry in which we embed some choice of D7-branes and O7-planes with chosen world-volume bundles and appropriate intersections to build models. In F-theory the input of background flux remains but the distinction between 7-branes and the background geometry is blurred. The closest thing we have to such a separation is to think of the IIB background geometry as corresponding to the geometry of the base of the elliptic fibration and the 7-branes as being described by the fibre. With this point of view the analogoue of the discussion on model building in IIB, which did not specify a particular CY but assumed some general background geometry and instead focused on the configuration of D7-branes and O7-planes, is studying the structure of the fibration without specifying the base. In this sense we are learning to redescribe 7-branes in terms of the elliptic fibre and this is the topic of this section. To understand all the aspects of the fibration structure it is important to account for its full global structure over the whole CY four-fold. In this section we will consider such a global approach, but it is important to state that substantial information can still be learned regarding 7-branes in F-theory by not considering the full global structure but only the geometry near the 7-brane and this approach is discussed in section \ref{sec:locf}. 
%
%
\setcounter{footnote}{0}
\def\thefootnote{a\alph{footnote}}
%
There are a number of nice reviews of the formalism relating F-theory and 7-branes which we refer to for any further details required \cite{denef08,Weigand:2010wm,Braun:2010ff,Knapp:2011ip,Leontaris:2012mh,Weigand:2012cx}.

\subsection{7-branes as codimension 1 singularities}
\label{sec:7braco1}

Every elliptically fibered CY four-fold can be represented as a Weierstrass model where we consider the elliptic curve (the torus) to be a hypersurface in the weighted projective space ${\mathbb P}\left[2,3,1\right]$, with respective homogenous coordinates $\left(x,y,z\right)$, and fibre this over the base $B$ with coordinates $u_i$ so that the CY is given by the locus
\be
y^2 = x^3 + f\left(u_i\right)x + g\left(u_i\right) \;, \label{weisform}
\ee
with $f$ and $g$ being holomorphic functions of degree 4 and 6 respectively. The elliptic fibre degenerates whenever the discriminant
\be
\Delta = 27 g^2 + 4 f^3\;,
\ee
vanishes.\footnote{A nice discussion of why (\ref{weisform}) describes a torus fibration and why it degenerates at $\Delta=0$ can be found in \cite{Johnson:2003gi}.} For generic $f$ and $g$, the vanishing of the discriminant defines a complex codimension 1 locus of the base, a 4-cycle, and we should associate this to 7-branes wrapping the 4-cycle. There is a singularity on this locus and the 7-brane configuration is described by the type of singularity. There is an algorithm for determining the type of singularity from the vanishing orders of $f$, $g$ and $\Delta$.\footnote{In the case where the determinant vanishes to linear order while $f$ and $g$ do not, the so-called $I_1$ locus, only the fibre pinches and the CY four-fold is not singular.} Tate's algorithm applies when the elliptic fibration (\ref{weisform}) can be written in Tate form\footnote{Generally this is possible but there are some exceptions \cite{Katz:2011qp}. For the cases of realistic gauge groups $SU(5)$ and $SO(10)$ it is always possible \cite{Katz:2011qp} at least up to higher order corrections in the vanishing locus which may have meromorphic coefficients.}
\be
P_T = -y^2 + x^3 - a_1 x y z + a_2 x^2 z^2 - a_3 y z^3 + a_4 x z^4 + a_6 z^6 = 0\;. \label{tateform}
\ee
The relation of the coefficients in (\ref{tateform}) to those of (\ref{weisform}) is given by
\bea
\beta_2 &=& a_1^2 + 4 a_2 \;, 
\;\;\beta_4 = a_1 a_3 + 2 a_4 \;, 
\;\;\beta_6 = a_3^2 + 4 a_6 \;, \nn \\
\beta_8 &=& \beta_2 a_6 - a_1 a_3 a_4 + a_2 a_3^2 -a_4^2 \;, \nn \\
\Delta &=& -\beta_2^2 \beta_8 - 8 \beta_4^3 - 27\beta_6^2 + 9 \beta_2 \beta_4 \beta_6 \;, \nn \\
f &=& -\frac{1}{48} \left(\beta_2^2 - 24 \beta_4 \right) \;, \;\;
g = -\frac{1}{864} \left(-\beta_2^3 + 36\beta_2 \beta_4 - 216 \beta_6 \right) \;. \label{usequant}
\eea
Tate's algorithm takes into account the refinement of the Kodaira classification of the fibre structure required to account for possible monodromies. In this case it is possible to read off the gauge group induced at a singular point from the vanishing order of $f$, $g$ and $\Delta$ according to table \ref{tab:kodsing} \cite{Bershadsky:1996nh}.\footnote{There actual further refinement in terms of vanishing orders of the $a_i$ is presented in \cite{Bershadsky:1996nh}.}
\begin{table}
\tbl{Table showing the fibre structure, as classified by Kodaira, according to Tate's algorithm.}{\begin{tabular}{|c|c|c|c|c|}
\hline
 ${\rm ord}(f)$ & ${\rm ord}(g)$ & ${\rm ord}(\Delta)$   & fiber type & singularity type \\
\hline
$\geq 0$ & $\geq 0$ & $0$ & smooth &none \\
\hline
$0$ & $0$ & $n$ &  $I_n$  & $A_{n-1}$ \\
\hline
$\geq 1$ & $1$ & $2$& $II$ & none\\
\hline
$1$ & $\geq 2$ & $3$ &  $III$ & $A_1$ \\
\hline
$\geq 2$ & $2$ & $4$ &   $IV$  & $A_2$\\
\hline
$2$ & $\geq 3$ & $n+6$ &  $I_n ^*$ & $D_{n+4}$ \\
\hline
$\geq 2$ & $3$ & $n+6$ &  $I_n ^*$ & $D_{n+4}$ \\
\hline
$\geq 3$ & $4$ & $8$ &  $IV^*$ & $E_6$ \\
\hline
$3$ & $\geq 5$ & $9$ &  $III^*$  & $E_7$\\
\hline
$\geq 4$ & $5$ & $10$ &  $II^*$  & $E_8$\\
\hline
\end{tabular}}
\label{tab:kodsing}
\end{table}
The classification in table \ref{tab:kodsing} is of ADE type (with $A_{n}=SU(n+1)$ and $D_n=SO(2n)$) and the claim of F-theory is that a singularity of specific type describes a stack of 7-branes carrying the associated gauge group. The fibre type $I_1$ describes a single D7 brane. To see how the associated gauge group arises we should go to the M-theory side by compactifying the special F-theory direction of space-time and performing a T-duality on it. Since the direction is compact the T-dual can be of finite size and this allows for a resolution of the singularity. The resolution of an ADE singularity consists of a tree of complex projective spaces $\Ps_i$ that intersect according to the Dynkin diagram of the associated group. Fibering these over the four-cycle on which the 7-brane/singularity have support gives rise to a set of resolution divisors $E_i$. There is a further divisor $E_0$ that corresponds to the elliptic fibre over the four-cycle and all together these divisors intersect according to the affine Dynkin diagram.
\be
\int_{CY_4} E_I \wedge E_J \wedge D_a \wedge D_b = C_{IJ} \int_{CY_4} W \wedge D_a \wedge D_b \;. \label{exdivint}
\ee
Here we define the affine index $E_I=\left\{E_0,E_i\right\}$ and $C_{IJ}$ is the affine Cartan matrix, $D_a$ and $D_b$ are generic divisors in the base $B$ and $W$ is the 4-cycle divisor of the 7-brane which we define by the vanishing of an appropriate holomorphic polynomial 
\be
W \;:\; w=0 \;.
\ee
The generators of the associated gauge group are accounted for as follows: there are M2 branes wrapping suitable combinations of the resolution $\Ps_i$ which give rise to the non-Cartan elements of the gauge group. The Cartan elements come from the two-form duals $w_I$ of the resolution divisors $E_I$ which give rise to gauge fields by dimensional reduction of the M-theory three-form
\be
C_3 = A^I \wedge w_I \;.
\ee

For the case of an $SU(5)$ gauge group on the divisor $W$ the required vanishing orders of the coefficients of (\ref{tateform}) are
\be
a_1 = b_5 \;,\;\; a_2 = b_4 w \;,\;\; a_3 = b_3 w^2 \;,\;\; a_4 = b_2 w^3\;,\;\; a_6 = b_0 w^5 \;, \label{su5sing}
\ee
where the $b_i$ are functions which do not vanish at $w=0$. The explicit resolution of the $SU(5)$ singularity for this Tate model has been studied in detail in \cite{Blumenhagen:2009yv,Grimm:2009yu,Esole:2011sm,Marsano:2011hv,Krause:2011xj,Grimm:2011fx,Krause:2012yh,Marsano:2012yc}.\footnote{Similar analysis was performed for $SO(10)$ in \cite{Esole:2011cn,Tatar:2012tm} and for $E_6$ in \cite{Kuntzler:2012bu}.} The works differ in the form of the resolution in that some use a small resolution while others use a blow-up, currently there is no disagreement in the final results. For the purposes of this section we will follow the blow-up methods of \cite{Krause:2011xj} and refer to that work for full details. A blow-up is a smoothing of the singularity by the introduction of an additional coordinate and an additional scaling relation such that the new scaling implies, through the Stanely-Reisner Ideal (SRI), that the singularity locus is no longer part of the manifold and is replaced by the exceptional divisor associated to the newly introduced coordinate. The resolution of an $SU(5)$ singularity this way requires four blow-ups and the four new coordinates, $e_i$ with $i=1,...,4$, are introduced through the transformation \cite{Krause:2011xj}
\be
\left(x,y,w\right) \rightarrow \left(x e_1 e_4 e_2^2 e_3^2, y e_1 e_4^2 e_2^2 e_3^3, e_0 e_1 e_2 e_3 e_4 \right) \;.
\ee
After this replacement, and dividing out by an overall prefactor, we have that the so called proper transform of the Tate equation is
\be
P_T = -y^2 e_3 e_4 + x^3 e_1 e_2^2 e_3 - a_1 x y z + a_2 x^2 z^2 e_0 e_1 e_2 - a_3 y z^3 e_0^2 e_1 e_4 + a_4 x z^4 e_0^3 e_1^3 e_2 e_4 + a_6 z^6 e_0^5 e_1^3 e_4^2 e_2 = 0\;. \label{tateformresolved}
\ee
The modified SRI after the blow-up is given in \cite{Krause:2011xj} and it can be checked using it that the resulting manifold is smooth, ie. there is no solution to $P_T=dP_T=0$. The divisors associated to the $e_i$ are the $E_i$ of (\ref{exdivint}). Note that the actual resolution $\Ps_I$ can be extracted as the intersection of $e_I \cap P_T \cap D_a \cap D_b$ with $D_a$ and $D_b$ generic divisors in the base that intersect $w$ at a point.

\subsection{Matter curves as codimension 2 singularities}
\label{sec:matglob}

For generic $b_i$ in (\ref{su5sing}) Tate's algorithm implies an $SU(5)$ singularity over $w=0$. However over loci where the $b_i$ themselves vanish the manifold may become more singular and naively applying Tate's algorithm on this restricted locus we would associate an enhanced symmetry to that locus. To be more explicit consider the discriminant for the ansatz (\ref{su5sing}) which reads
\be
\Delta = - w^5 \left[P^4_{10} P_5 + w P^2_{10} \left(8b_4P_5 + b_5 R\right) + {\cal O}\left(w^2\right)  \right] \;,
\ee
where we define
\bea
P_{10} &=& b_5 \;, \\
P_5 &=& b_3^2 b_4 - b_2 b_3 b_5 + b_0 b_5^2 \;, \\ \nn
R &=& -b_3^3-b_2^2b_5 + 4b_0b_4b_5 \;.\label{p5p10}
\eea
Now over loci where the combinations $P_{10}$ or $P_{5}$ vanish we have further degeneration and according to Tate's algorithm a gauge enhancement. Since these constraints restrict to complex codimension 2 they correspond to curves on $w=0$ or $S_{GUT}$ and are referred to as matter curves because on these loci additional massless degrees of freedom localise. Such enhancements of the gauge group over curves is familiar from early work on F-theory \cite{Katz:1996xe} and will closely tie in with the local approach to model building explored in section \ref{sec:locf}. More specifically on the locus $P_{10}=0$ Tate's algorithm predicts an enhancement to $SO(10)$ and on $P_5$ to $SU(6)$. The additional massless degrees of freedom are required to complete the adjoint of $SU(5)$ which is present everywhere to the adjoint of the enhanced gauge group. Or in terms of M-theory they correspond to membranes wrapping the additional $\Ps$s required to resolve the enhanced singularity. Working the other way we can deduce the spectrum of localised modes by decomposing the adjoint of the enhaced gauge group to $SU(5)\times U(1)$ 
\bea
SO(10) \;&:&\; {\bf 45} \rightarrow {\bf 24}_0 \oplus {\bf 10}_{2} \oplus {\bf \overline{10}}_{-2} \oplus {\bf 1}_0 \;,\nn \\
SU(6) \;&:&\; {\bf 35} \rightarrow {\bf 24}_0 \oplus {\bf 5}_{1} \oplus {\bf \overline{5}}_{-1} \oplus {\bf 1}_0 \;. \label{adjmat}
\eea
Note that the additional $U(1)$ may be completely broken in the vacuum. Therefore on $P_{10}$ we have localised modes in the representations ${\bf 10} \oplus {\bf \overline{10}}$ and on $P_5$ a pair ${\bf 5} \oplus {\bf \overline{5}}$. 

The above reasoning relied to application of Tate's algorithm to singularities of co-dimension higher than one but, as emphasised in \cite{Esole:2011cn}, the algorithm strictly only holds for co-dimension one singularities. Because of this it is worth examining in more detail the structure of the fibre over the matter curves. Again this was done a number of different ways in \cite{Blumenhagen:2009yv,Grimm:2009yu,Esole:2011sm,Marsano:2011hv,Krause:2011xj,Grimm:2011fx,Krause:2012yh,Marsano:2012yc}, but we will follow the analysis of \cite{Krause:2011xj}. We are particularly interested in the resolution $\Ps$-tree over the matter curves and to study this explicitly we should study the intersection $e_I \cap P_T$ over the matter curves. Consider for example the case of the first $\Ps$ given by
\be
e_1 \cap \left\{-y^2 e_3 e_4 - a_1 x y z \right\} \;,
\ee
which using the SRI (presented in \cite{Krause:2011xj}) to set to 1 coordinates which can not vanish simultaneously with $e_1$ can be simplified to
\be
e_1 \cap \left\{e_3 e_4 + a_1 x \right\} \;. \label{e1p1}
\ee
Now over the ${\bf 10}$ matter curve $a_1=b_5=P_{10}=0$ the right hand side of (\ref{e1p1}) factorizes and so the $\Ps$ splits into two
\be
\left\{e_1 \cap e_3\right\} \;\mathrm{and}\; \left\{e_1 \cap e_4\right\} \;.
\ee
It can then be checked that additional $\Ps$ is exactly such that the intersections of the $\Ps_I$ forms the Dynkin diagram of $SO(10)$, and so indeed we recover the extra membrane states as expected from the naive application of Tate's algorithm. The same can be shown to hold for the case of $P_{5}$ enhancing to $SU(6)$.
	
\subsection{Yukawa couplings as codimension 3 singularities}
\label{sec:yukglob}

Similar to the enhancements of the singularity over the matter curves it is possible to have further enhacements where matter curves intersect. These rank 2 enhancements in co-dimension 3 correspond not to new massless content but rather to operators coupling the modes on the intersecting matter curves, an important subset of which are Yukawa couplings. The naive application of Tate's algorithm implies that we have the following enhancements and associated operators
\bea
P_{10}=b_4=0 \;&:&\; SU(5) \rightarrow E_6 \implies {\bf 5\;10\;10}\; = \mathrm{Up\;type\;Yukawa}, \nn \\
P_{10}=b_3=0 \;&:&\; SU(5) \rightarrow SO(12) \implies {\bf \overline{5}\;\overline{5}\;10} \; = \mathrm{Down\;type\;Yukawa}, \nn \\
P_5=R=0 \;&:&\; SU(5) \rightarrow SU(7) \implies {\bf 5\;\overline{5}\;1} \; = \mu\mathrm{-term\;type\;operator}. 
\eea
where the operators are deduced following the same procedure of decomposing the adjoint of the enhanced group in terms of $SU(5)$ \footnote{Here the charges under the additional 2 $U(1)$s are given for the ${\bf 10}$, ${\bf 5}$, ${\bf 1}$ in each bracket with the conjugate representation having the opposite charge.}
\bea
\bf{78} &\rightarrow& \bf{24}^{(0,0)} \op \un^{(0,0)} \op \un^{(0,0)} \op \un^{(-5,-3)} \op \un^{(5,3)} \nn \\
 & & \op \left(\f \op \fb\right)^{(-3,3)} \op \left(\te \op \teb\right)^{(-1,-3)} \op \left(\te \op \teb\right)^{(4,0)} \;, \nn \\
\bf{66} &\rightarrow& \bf{24}^{(0,0)} \op \un^{(0,0)} \op \un^{(0,0)} \op \left(\f \op \fb\right)^{(-1,0)} \op \left(\f \op \fb\right)^{(1,1)} \op
\left(\te \op \teb\right)^{(0,1)} \;, \nn \\
\bf{48} &\rightarrow& \bf{24}^{(0,0)} \op \un^{(0,0)} \op \un^{(0,0)} \op \left(\f \op \fb\right)^{(-6,0)} \op \left(\f \op \fb\right)^{(0,6)} \op
\left({\bf 1} \op {\bf \overline{1}}\right)^{(6,-6)} \;.
\eea
Note that the GUT singlets are not localised on matter curves on the GUT divisor but do intersect it at the point of enhancement.

Although again the application of Tate's algorithm is strictly not possible, since in the previous section we showed that indeed the expected matter localises on the matter curves, and the enhancement points are where such matter curves intersect, we expect that indeed the conjectured operator is there even if the fibre structure over the point is not that of the full enhancement group predicted by Tate's algorithm. Indeed as shown in \cite{Esole:2011sm,Marsano:2011hv,Krause:2011xj} this is the case: although the $\Ps$s in the fibre do not form the extended Dynkin diagram of $E_6$ say, the appropriate couplings are present. The difference from the matter curves is that the splitting of the $\Ps$s is such that no new classes of $\Ps$ are introduced but only the multiplicities of existing classes are changed. This is expected for an enhancement where an interaction rather than new matter is localised.

\subsection{U(1)-symmetries and G-flux in F-theory}

Unlike non-Abelian symmetries, which are well understood in terms of the ADE classification of singularities, $U(1)$ symmetries in F-theory are more difficult to identify and construct. However they are crucial for model building in two respects: firstly, turning on flux along the $U(1)$s is how chirality is induced in IIB and a similar mechanism is expected to be used in F-theory. Secondly, $U(1)$s, even if  Stueckelberg massive, can be used a symmetries to control operators in the theory, for example forbidding proton decay. Because $U(1)$s play these crucial roles in model building and yet are not well understood their study is an active subject in the contemporary literature and a full and general picture is yet to emerge. In this section we will review the lines of attack and progress that has been made so far. Note that here we will discuss aspects of $U(1)$s that arise from a global perspective, which as emphasised in \cite{Hayashi:2010zp} for example, is required for a full understanding, particularly regarding forbidding operators. In section \ref{sec:locf} we will study $U(1)$ symmetries from a local perspective which does not hold the full information but nonetheless can yield significant results.

In the weakly coupled type IIB string with O3/O7 planes there are two types of $U(1)$s coming from D7-branes: those that are massless in the absence of world-volume flux and those that are massive even in the absence of flux. The latter gain a mass through the Stueckelberg term in (\ref{d7gsodd}). Since in uplifting to F-theory only the world-volume flux is not geometrised it must be that those massive $U(1)$s in IIB are geometrically massive. In the following we will primarily be concerned with massless $U(1)$s according to this classification since they are better understood but will make some comments on the massive ones at the end.

$U(1)$s that are massless and unbroken in the absence of flux correspond to additional sections of the elliptic fibration, i.e additional to the section defining the base $z=0$ \cite{Morrison:1996na}. An additional section implies a set of holomorphic functions $\left\{x_1,y_1,z_1\right\}$ of the coefficients of the fibration (\ref{tateform}) such that when we set $\left\{x,y,z\right\}=\left\{x_1,y_1,z_1\right\}$ we find $P_T=0$. Equivalently, it specifies a point in the torus over every point in the base. In the special case where $z_1=1$ we have that $P_T\left(x_1,y_1,1\right)=0$ and therefore if we transform coordinates by $x \rightarrow x+x_1$ and $y\rightarrow y+y_1$ we arrive at a form for $P_T$ where $a_6=0$. This is the case which was studied in \cite{Grimm:2010ez}, termed a $U(1)$-restricted Tate Model, where it was shown that the additional section generically intersects itself on the locus $a_3=a_4=0$ thereby enhancing to an $SU(2)$ singularity.\footnote{In terms of the general possibilities for $U(1)$s coming from $E_8$ in F-theory, see table \ref{tab:breake8u1s}, this was a restricted model of type 4+1.} The resolution of the $SU(2)$ singularity then introduces an additional exceptional divisor which is in turn associated to the additional massless $U(1)$. In \cite{Krause:2011xj} further evidence for this construction was given by showing that the matter curves have the expected charges under this $U(1)$. 

An additional section was also studied without an $SU(5)$ singularity in \cite{Morrison:2012ei} where the general form of a 2-section model was presented. In \cite{Braun:2011zm} the additional $U(1)$ was identified by noting that the $U(1)$-restricted Tate model constraint $a_6=0$ implies that, after an appropriate coordinate transformation, the $SU(2)$ singularity can be written in the form of a conifold singularity. A small resolution of this then induced the new divisor associated to the $U(1)$. 

Another approach to the second section is to use duality with the Heterotic string, and this was utilised in \cite{Choi:2012pr} to study additional sections. In particular constraints were deduced on the form of the $w$ dependence of the $b_i$ in (\ref{su5sing}) such that they are consistent with the existence of a $U(1)$.

The most systematic approach so far to additional $U(1)$s, or sections, was presented in \cite{mpw}. It was shown that an appropriate restriction of the coefficients of the Tate model induces up to 4 new sections, for an $SU(5)_{GUT}$ model, which can reproduce the general breaking pattern of $E_8 \rightarrow SU(5)_{GUT}\times \prod_i U(1)_i$. The two possible cases with a single $U(1)$ were studied in detail.

Once a massless $U(1)$ is identified we can consider turning on flux along it. This is described on the resolved M-theory side in terms of the four-form $G$-flux. 
This was initially studied independently of the $SU(5)$ resolution in \cite{Braun:2011zm} and for the case of $SU(5)\times U(1)$ within the $U(1)$-restricted Tate model in \cite{Krause:2011xj,Grimm:2011fx,Krause:2012yh} and more generally in \cite{mpw}. The appropriate flux was identified as taking the form
\be
G_X = f \wedge w_X = f \wedge \left(-S + K + Z + \frac15\left(2,4,6,3\right)_i E_i \right) \;,
\ee
where $f$ is the flux on the base and $w_X$ plays the M-theory analogue of the gauge generator along the $U(1)$. Here $S$ is the resolution divisor associated to the $U(1)$ as described above, $K$ is the anti-canonical bundle on the base, $Z$ is the Poincare dual of the section divisor ($z=0$), and the $E_I$ are the resolution divisors of the $SU(5)$ singularity (see section \ref{sec:7braco1}). 

Geometrically massive $U(1)$s uplifted from IIB were first considered in \cite{Grimm:2010ez,Grimm:2011tb} following ideas from massive $U(1)$s in type II \cite{Grimm:2008ed}.\footnote{Similar ideas in the context of M-theory were studied in \cite{Camara:2011jg,BerasaluceGonzalez:2012vb} where the massive $U(1)$s arose from discrete torsion.} The claim is that these arise in F-theory from two-forms that are not harmonic but rather satisfy differential relations of the type
\be
d w_A = C_A^a \alpha_a \;,
\ee
where here $C_A^a$ are integers and $\alpha_a$ are a basis of 3-forms on the manifold. Reducing the M-theory $C_3$ on the $w_A$ gives a set of $U(1)$s but since they are not harmonic the $U(1)$s are massive. Although these $U(1)$s are string-scale massive intuition from IIB implies they still have a role to play, firstly in IIB such $U(1)$s would still leave behind a global symmetry which constrains the low energy theory. In particular this is the case for the diagonal $U(1)$ which as discussed in section \ref{sec:yukawaF} forbids the top Yukawa coupling. Also, in IIB, it is still possible to turn on flux along such $U(1)$s to induce chirality, though such a flux is subject to D5-tadpole cancellation conditions (\ref{d5t}). Whether, and how, these properties lift to F-theory is still under study.\footnote{Alternative promising approaches, based on $E(-1)$ instantons, to the uplift of the diagonal $U(1)$ in IIB were also studied in \cite{Donagi:2010pd}.} 

Relating to this discussion, in \cite{Krause:2012yh} it was shown that fluxes turned on along the geometrically massive $U(1)$ in IIB when lifted to F-theory can be mapped to a particular allowed $G$-flux, first identified in \cite{Marsano:2011hv}, called the universal spectral cover flux (we will discuss this in section \ref{sec:speccov}). This is consistent with the above proposal for the non-closed 2-forms because still a $G$ can be constructed from an appropriate combination of non-closed forms which itself is closed and harmonic. Indeed this is guaranteed by the uplift of the D5-tadpole condition to F-theory which is automatically part of the geometry.

As discussed it is possible to turn on fluxes in F-theory even in the absence of a massless $U(1)$ associated to the flux, and this is more general than the specific massive $U(1)$s described above. A particularly important class of fluxes which have no associated massless $U(1)$s are the universal spectral cover fluxes. These were identified in \cite{Marsano:2011hv,Krause:2012yh,Grimm:2011fx} for $SU(5)$ models, in \cite{Tatar:2012tm} for $SO(10)$ models and in \cite{Kuntzler:2012bu} for $E_6$ models. This class of flux is interesting because, unlike the $U(1)$ flux, it does not take the form $f \wedge w$ with $f$ being a flux on the base and $w$ being the M-theory two-form playing the analogue of the gauge generator in IIB. Rather it has components involving two resolution divisors, explicitly we have for $SU(5)$
\be
G = E_2 \wedge E_4 + \frac15\left(2,-1,1,-2\right)_i E_i \wedge K\;.
\ee

So far we have discussed only fluxes that are given as wedge products of two 2-forms. Such fluxes are said to be in the primary vertical subspace of 4-forms on the CY \cite{Greene:1993vm}. These fluxes present at all loci in complex structure moduli space. In \cite{Braun:2011zm} it was shown how to construct fluxes that can not be written in this way for an $SU(2)$ model, and such fluxes are expected to be only allowed on loci in complex structure moduli space and therefore fix some moduli. These fluxes were generalised to the case with also $SU(5)$ in \cite{Krause:2012yh}.
 
The construction of the four-form G-flux is only half of the input needed to calculate the chirality induced by the flux since we need to determine over which cycle to integrate the flux. In type IIB the two-form flux would be integrated over the matter curves, or D7-brane intersections (\ref{iibcurflux}). In F-theory the G-flux naturally integrates over four-cycles and these are denoted matter surfaces and constructed as follows \cite{Marsano:2011nn,Marsano:2011hv,Krause:2011xj,Krause:2012yh,Grimm:2011fx}. For a representation on a matter curve ${\bf R}$, we consider the Cartan charges of the weights of the representation which can then be identified with some combination of the resolution $\Ps$s with the same Cartan charges. The Cartan charges of the combination of $\Ps$s are determined by intersecting them with the resolution divisors $E_I$ which correspond to the Cartan generators. Therefore to each (weight of a) representation we can associate a combination of $\Ps$s that are fibered over the matter curve supporting that state. Altogether this therefore defines a 2 complex dimensional surface which is the matter surface associated to that state ${\cal C}_{\bf R}$. The chirality induced by the flux for that representation is given by integrating the flux over the associated matter surface.

Introducing chirality through G-flux implies that the massless spectrum can be anomalous. As discussed in section \ref{sec:d7spec}, in IIB the anomalies are guaranteed to be cancelled through the Green-Schwarz mechanism once D5 and D7 tadpole cancellation is imposed. In F-theory these tadpoles are automatically cancelled and so we expect that anomalies are automatically cancelled as well. This result was shown, and the mechanism in F-theory was studied in detail, in \cite{Cvetic:2012xn}. 

Finally, as in IIB, the G-flux is subject to some important quantisation conditions. These were studied and determined in \cite{Witten:1996md,Freed:1999vc,Marsano:2010ix,Collinucci:2010gz,Marsano:2011nn,Braun:2011zm,Marsano:2011hv,Krause:2011xj,Grimm:2011fx,Krause:2012yh,Braun:2012nk,Collinucci:2012as}.
 
\subsection{Global $SU(5)$ F-theory GUT models}

The formalism discussed in the previous section for global F-theory models was primarily concerned with the structure of the elliptic fibration and not with the of the base of the CY. Of course an explicit global model in F-theory is required to specify also the base manifold and identify $S_{GUT}$ within it. There has been substantial literature exploring the geometry of the base and constructing models on it. However most of it was before the developments which lead to the detailed understanding of four-form flux. The work discussed below constructed the base explicitly but then used ideas from gauge theory, or the spectral cover, to discuss the flux. Therefore, strictly speaking, they are not complete global models of F-theory. Nonetheless they do form the background on which true global model building can be viably developed. 

Recently global models which specify also the flux have been constructed in \cite{Krause:2011xj,Grimm:2011fx,Marsano:2012yc}. The state of the art there are models which exhibit 3 chiral generations, though none have a fully realistic Higgs sector: the models of \cite{Krause:2011xj,Grimm:2011fx} do not consider doublet-triplet splitting and in \cite{Marsano:2012yc} Wilson lines are utilised though with results yielding multiple Higgs fields. The models are all based on the $U(1)$-restricted structure which is a particular breaking of $E_8$ with one 10-matter curve and two 5-matter curves, in particular this does not allow for a $U(1)_{PQ}$ scenario. We will describe in detail more general scenarios from a local model building perspective in section \ref{sec:locf}, but for now we note that global models are yet to reproduce the rich structures that local models have been exhibiting and some work remains to reconcile the two approaches.\footnote{It should be noted that Heterotic line bundle models are closely related to the approach adopted in F-theory of using Abelian fluxes and these do exhibit a rich phenomenological structure, especially with respect to additional $U(1)$ symmetries \cite{Anderson:2011ns,Anderson:2012yf}.}

Returning to models that construct the base and study the flux using the spectral cover. Early work studying general global properties of the base was presented in \cite{Andreas:2009uf,Collinucci:2009uh,Blumenhagen:2009up} and actual models were constructed in \cite{Marsano:2009ym,Marsano:2009wr}, where the base was a flop of a blown-up Fano three-fold, and \cite{Blumenhagen:2009yv} where the base was a del-Pezzo transition of a Fano three-fold. In \cite{Grimm:2009yu} Toric methods were utilised to construct the geometry, and \cite{Chen:2010ts} to build $SO(10)$ models. In \cite{Knapp:2011wk} a scan over half a million base geometries was performed and many models constructed and a database formed. In \cite{Cvetic:2010rq} global models were studied with an emphasis on instanton dynamics. In \cite{Braun:2010hr} global models were studied with respect to supporting possible Wilson lines.

\section{Local models in F-theory}
\label{sec:locf}

One of the important aspects of type IIB and F-theory is that the gauge degrees of freedom are localised on 7-branes which wrap only a submanifold of the full CY. This is particularly interesting in the case of a GUT theory where all the SM gauge groups and matter are localised on a single $SU(5)$ 7-brane wrapping a four-cycle. In such a setup much, but certainly not all, of the information regarding the gauge and matter sector can be obtained by considering only the local geometry of the four-cycle and not the full CY. This is a substantial simplification which is similar, at least in spirit, to branes on singularities \cite{botup} (see section \ref{sec:bs}). The idea of local F-theory models is to use this to study aspects of the models which do not require a fully global understanding as outlined in the previous section. This approach has proved quite fruitful and the phenomenological success of the resulting local models identified F-theory as a potentially attractive region in the landscape for string phenomenology and was therefore also the driving force behind much of the global aspects research as well. Indeed some of the earliest papers on the applications of F-theory for string phenomenology advocated the local approach \cite{Donagi:2008ca,Beasley:2008dc,Beasley:2008kw,Donagi:2008kj}. The advantages and disadvantages of the local approach are the universal ones for bottom-up versus top-down approaches: you gain freedom and simplicity for model building but you lose the UV sector, both the UV physics itself and any constraints on the IR physics that arise from it. There is an additional aspect which is that local models project large regions of the landscape onto a relatively small number of models. This allows us to make general statements about F-theory models but also does not distinguish between different possible global realisations that can lead to the same local models. This may be regarded as an advantage or a disadvantage subjectively.

Sections \ref{sec:localgaugetheory} and \ref{sec:speccov} will describe the formalism and tools used in the local approach with an aim towards the setting the groundwork for the model building phenomenology reviewed in section \ref{sec:feno1}.

Since the 7-brane gauge coupling strength is set by the size of the 4-cycle that it wraps, while the Planck scale is set by the volume of the full CY, it is possible to consider decoupling gravity in local models by sending the volume to infinity. Of course since the planck scale is finite we may require that decoupling is possible at least in principle. Some motivation for this requirement can be found in the small hierarchy between the Planck scale and the GUT scale, the former being set by the CY volume while the latter by the four-cycle \cite{Beasley:2008kw}. Also in the fact that the GUT coupling strength predicted by gauge coupling unification is quite strong and therefore requires a relatively small cycle. The possibility of an in-principle decoupling of gravity constrains the GUT cycle embedding in the CY to be a contractable cycle, more formally with an ample normal bundle. It is important to stress that the requirement for contractability is not essential for the local approach to model building, though it does complement it. The constraints on the geometry implied by requiring contractibility were studied in \cite{Beasley:2008kw,Cordova:2009fg,Andreas:2009uf,Marsano:2009ym,Grimm:2009yu, fanocol} for example. 

It is important to emphasise that although in principle it is possible to decouple gravity and keep a finite coupling interacting gauge theory at tree-level, the calculation of any 1-loop quantities will typically diverge in such circumstances because open string 1-loop amplitudes are equivalent to tree-level closed string amplitudes and since closed strings propagate in the bulk space they are sensitive the decoupling limit.\footnote{This of course is the reason that also anomalies, particularly mixed gauge-gravitational, see for example \cite{Grimm:2012yq}, can not be studied in a local context.} Indeed in \cite{Conlon:2009xf,Conlon:2009kt,Conlon:2009qa} it was shown that gauge coupling running and therefore unification occurs at scales set by the bulk scale, and therefore would diverge in the decoupling limit. Similarly in \cite{Conlon:2010xb} it was shown that Yukawa couplings also exhibit such a divergence.

Just as the gauge and matter sectors are localised to four-cycles, operators in the theory are localised, though in a less drastic and universal sense, to points in the four-cycle. Operator coefficients in F-theory, just as in IIB, can be calculated by solving for the wavefunction profiles of the associated fields in the extra dimensions and integrating their overlap. The localisation of operators manifests in the peak of the wavefunction profiles around the intersection of the matter curves. This means that, at least to some approximation, it is possible to extract information about the wavefunctions by just considering a local patch inside the GUT four-cycle around the point associated to the operator. This was first applied to study flavour physics in \cite{Heckman:2008qa} and subsequently in \cite{Hayashi:2009ge,Bouchard:2009bu,Randall:2009dw,Conlon:2009qq,Heckman:2009de,Font:2009gq,Cecotti:2009zf,Leontaris:2010zd,Cecotti:2010bp,Peccei:2011ng,Aparicio:2011jx,Camara:2011nj,Palti:2012aa,Hayashi:2010zp} for this and more detailed aspects of operators. In section \ref{sec:fultralocal} we will discuss the details of these constructions which are sometimes termed ultra-local since they are valid only on a patch within the four-cycle.

\subsection{The gauge theory}
\label{sec:localgaugetheory}

From the discussions in sections \ref{sec:7braco1}-\ref{sec:yukglob} we know that the type of configurations we are interested in have a gauge group $G_{S}$ on the GUT divisor $S_{GUT}:w=0$, which is then enhanced to a higher rank one on matter curves inside $S_{GUT}$. Further Yukawa couplings can also be understood from a further enhancement over the intersection of matter curves. This data describes key properties of the effective low-energy theory that should describe intersecting 7-branes in F-theory. Already in \cite{Katz:1996xe} it was shown that these type of symmetry enhancement over loci in a gauge theory can be captured by thinking about a gauge theory with the enhanced gauge group over $S_{GUT}$ which is broken down to $G_S$ by a spatially varying vev for an adjoint Higgs field. The Higgs vev vanishes over certain loci and that is where the gauge group therefore enhances. The additional massless modes that cause the enhancement are associated to matter fields under the decomposition of the adjoint as in (\ref{adjmat}). Therefore we expect that at least a coarse description of F-theory models of intersecting 7-branes can be captured by an $N=1$ 8-dimensional gauge theory on $S_{GUT}$ which a spatially varying vev for an adjoint Higgs. It is also known that for supersymmetry this theory must be twisted \cite{Bershadsky:1997zs,Donagi:2008ca,Beasley:2008dc} which means it has a connection or half a unit of background flux. 

An appropriately twisted 8-dimensional gauge theory was constructed in \cite{Donagi:2008ca,Beasley:2008dc} as a local description of F-theory models. It captures the relevant enhancements of symmetries over matter curves and therefore the appropriate localised massless matter which is the analogue of strings stretching between the GUT brane and additional 7-branes. It also captures the the point enhancements and therefore the associated Yukawa couplings. The theory is termed topological in \cite{Beasley:2008dc} which rather should be interpreted as being independent of some of the metric data, specifically the non-holomorphic part which affects the Kahler potential of the theory. The spectrum, matter curves and Yukawa couplings all have purely holomorphic components and this is the coarse data which is appropriately captured. In \cite{Conlon:2009qq} it was shown how the theory can be mapped directly to twisted N=1 8-dimensional super Yang-Mills which in turn also captures the Kahler potential data. Therefore the theories presented in \cite{Donagi:2008ca,Beasley:2008dc} do also capture some of this data, but the key point is that this is only in the approximation for a slowly varying Higgs vev. In terms of intersecting 7-branes this corresponds to shallow intersection angles \cite{Beasley:2008dc}. 

The precise regime in which we can trust the Kahler potential of this theory, and the associated D-term equation, was studied in \cite{Camara:2011nj,Palti:2012aa}. In particular it was argued that in the case where the 7-branes intersecting the GUT brane go into the bulk of the CY the theory breaks down on $S_{GUT}$ in the limit where we decouple gravity through a large CY volume. For a finite Planck scale it is possible to find a regime where the approximation of shallow intersection angles can be made to work at least locally on a patch in $S_{GUT}$. 

The 8-dimensional theory is comprised of a gauge field ${\bf A}$ and an adjoint valued Higgs field ${\bf \Phi}$. These can be decomposed into 4-dimensional $N=1$ multiplets
\bea
{\bf A}_{\bar{m}} &=& \left( A_{\bar{m}}, \psi_{\bar{m}}, G_{\bar{m}}\right) \;, \\
{\bf \Phi}_{mn} &=& \left( \varphi_{mn}, \chi_{mn}, H_{mn} \right) \;, \\
{\bf V} &=& \left( \eta, A_{\mu}, D \right) \;.
\eea
The indices on the fields denote their form-values on $S$. So for example $A_{\bar{m}} \in \bar{\Omega}^1_{S}\otimes\mathrm{ad}(P)$ where $\Omega^p_S$ denotes holomorphic $p$-form on $S$ and $P$ is the principle bundle (in the adjoint representation) associated to the gauge group $G$.
Here ${\bf A}$ and ${\bf \Phi}$ are chiral multiplets with respective F-terms $G$ and $H$. ${\bf V}$ is a vector multiplet with D-term $D$. $A_{\bar{m}}$ and $\varphi_{mn}$ are complex scalars while $\psi_{\bar{m}}$, $\chi_{mn}$, and $\eta$ are fermions.

The action for the effective theory was given in \cite{Beasley:2008dc}. Setting 4-dimensional variations of the fields to zero, the equations of motion that follow are
\bea
& &H - F^{(2,0)} = 0 \;, \label{eom1} \\
& &i\left[ \varphi, \bar{\varphi} \right] + 2\omega \wedge F^{(1,1)} + \star_{S} D = 0 \;, \label{eom2} \\
& &2i\omega \wedge \bar{G} - \bar{\partial}_{A} \varphi = 0 \;, \label{eom3} \\
& &- \partial \bar{H} + 2 \omega \wedge \bar{\partial} D + \bar{G} \wedge \bar{\varphi} - \bar{\chi} \wedge \bar{\psi}- i 2\sqrt{2} \omega \wedge \eta \wedge \psi = 0 \;,  \label{eom4} \\
& &\omega \wedge \partial_{A} \psi + \frac{i}{2} \left[ \bar{\varphi}, \chi \right] = 0 \;,  \label{eom5} \\
& &\bar{\partial}_A \chi - 2i\sqrt{2} \omega \wedge \partial_A \eta - \left[ \varphi, \psi \right] = 0 \;,  \label{eom6} \\
& &\bar{\partial}_{A} \psi - \sqrt{2} \left[ \bar{\varphi}, \eta\right] = 0 \;, \label{eom7} \\
& &-\sqrt{2}\left[\bar{\eta},\bar{\chi}\right] - \bar{\partial}_{A} G - \frac12 \left[\psi,\psi\right]= 0 \;. \label{eom8}
\eea
Here $F$ denotes any gauge flux present and $\omega$ denotes the Kahler for on $S_{GUT}$. For supersymmetric backgrounds we have $D=G=H=0$ and the backgrounds considered will have vanishing gaugino vev $\left<\eta\right>=0$. 

The matter curves are captured by an appropriate background vev for $\varphi$, which according to (\ref{eom3}), must be holomorphic. As discussed, to model intersecting branes we should take the gauge group on $S$ to be a higher rank than $SU(5)$. For example to model an intersection which has a 5-matter curve we should take $G_S=SU(6)$ and give the Higgs an adjoint vev which breaks $SU(6) \rightarrow SU(5)\times U(1)$ generically on $S$ while enhancing back to $SU(6)$ on the holomorphic curves where $\varphi$ vanishes. It is possible to show then that solving the equations of motion for perturbations of $\psi_m$ and $\chi$ about such a background gives solutions for wavefunction profiles that have a gaussian profile which peaks on the vanishing locus of $\varphi$ \cite{Donagi:2008ca,Beasley:2008dc}
\be
\delta\psi_m \sim \delta\chi \sim e^{-\left|\left<\varphi\right>\right|^2} \;. \label{schemwave}
\ee
These are the trapped matter modes. We do not go into further details on the wavefunction profiles here as we will do so in section \ref{sec:fultralocal}.

The gauge theory description is richest if we consider an $E_8$ theory on $S_{GUT}$ which is broken by the Higgs background to $SU(5)\times \Pi_i U(1)_i$.\footnote{For theories that go beyond $E_8$, specifically by considering additional intersecting $D_3$-branes see \cite{Heckman:2010fh,Heckman:2010qv,Heckman:2011sw,Heckman:2011hu,Heckman:2011bb}.} There are multiple possible breaking patterns of $E_8$ that preserve an additional number of $U(1)$s and these are be classified in table \ref{tab:breake8u1s}.
\begin{table}
\tbl{Table showing possible breaking patters of $E_8 \rightarrow SU(5) \times \Pi_i U(1)_i$. Note that the non-Abelian factors, other than $SU(5)$, are broken by the Higgs and are displayed for classification purposes of the remaining $U(1)$s.}
{\begin{tabular}{|l|c|}
\hline
Breaking Pattern & Number of $U(1)$s \\
\hline
$E_8 \rightarrow SU(5) \times S\left[U\left(4\right)\oplus U\left(1\right)\right]$ & 1 \\
\hline
$E_8 \rightarrow SU(5) \times S\left[U\left(3\right)\oplus U\left(2\right)\right]$ & 1 \\
\hline
$E_8 \rightarrow SU(5) \times S\left[U\left(3\right)\oplus U\left(1\right)\oplus U\left(1\right)\right]$ & 2 \\
\hline
$E_8 \rightarrow SU(5) \times S\left[U\left(2\right)\oplus U\left(2\right)\oplus U\left(1\right)\right]$ & 2 \\  
\hline
$E_8 \rightarrow SU(5) \times S\left[U\left(2\right)\oplus U\left(1\right)^3\right]$ & 3 \\
\hline
$E_8 \rightarrow SU(5) \times S\left[U\left(1\right)^5\right]$ & 4 \\
\hline
\end{tabular}}
\label{tab:breake8u1s}
\end{table}
The breaking patterns are realised by Higgs background inside $SU(5)_{\perp}$ in the decomposition $E_8 \rightarrow SU(5)_{GUT}\times SU(5)_{\perp}$ where a Higgs generator taking value just along the Cartan sub-algebra of $SU(5)_{\perp}$ preserves $S\left[ U(1)^5\right]$ and the other cases corresponding to turning on components away from the Cartan.

The representations trapped on matter curves are determined once the Higgs breaking pattern is specified. They arise from the adjoint of $E_8$ which under the decomposition $E_8 \rightarrow SU(5)_{GUT}\times SU(5)_{\perp}$ yields
\be
\bf{248} \rightarrow \left(\bf{24},\bf{1}\right)\op\left(\bf{1},\bf{24}\right)\op\left(\te,\f\right)\op\left(\fb,\te\right)\op\left(\teb,\fb\right)\op\left(\f,\teb\right) \;.
\ee
Therefore the GUT 10-multiplets are in the fundamental representation of $SU(5)_{\perp}$ and the GUT 5-multiplets are in the anti-symmetric of $SU(5)_{\perp}$. A particulary nice way to parameterise the possibilities in table \ref{tab:breake8u1s} is, see \cite{Anderson:2012yf,Heckman:2009mn} for example, to consider the purely Cartan breaking in which case the charges of the GUT representations can be parameterised by five 5-component charge vectors $\left(t_i\right)_j = \delta_{ij}$ where $i,j=1,...,5$. Here we are parameterising four linearly independent $U(1)$s in terms of five $U(1)$s that satisfy one tracelessness constraint which means we should specify the vector $\left(1,1,1,1,1\right)$ to be neutral or in terms of the basis charge vectors the linear relation
\be
\sum_i t_i = 0 \;.
\ee
Using this basis the charges are
\bea
Q\left(\te_I\right) &=& t_i \;, \nn \label{cur1} \\
Q\left(\f_J\right) &=& -t_i - t_j , \;\; i\neq j\;, \nn \label{cur2} \\
Q\left(\bf{1}_K\right) &=& t_i - t_j , \;\; i\neq j\;, \label{cur3}
\eea
where the capital indices label the different matter curve representations and run over the possible combinations of the small indices, which means $I=1,..,5$, $J=1,..,10$, and $K=1,..,10$.

The charges of the representations for the other cases in table \ref{tab:breake8u1s} can be built from these charges by identifying some of the $t_i$s according to the non-Abelian factors in the decomposition of $SU(5)_{\perp}$. The rule is that if compared to the pure Cartan breaking we have $U(1)^n \rightarrow U(n)$ then we identify $n$ of the $t_i$. So for example for $S\left[U(3)\times U(1)^2 \right]$ we take $t_1=t_2=t_3$ and these all parameterise a single $U(1)$, the diagonal one in $U(3)$. The different breaking patters in table \ref{tab:breake8u1s} are sometimes refered to as different monodromies \cite{Hayashi:2009ge,Heckman:2009mn,Cecotti:2010bp}. 

Note that from the charge assignments (\ref{cur1}) we see that if all the matter generations of the SM arise on a single $10$-matter curve then a non-trivial monodromy is required for an up-type Yukawa since to be gauge invariant this must take the form
\be
Q\left({\bf 5 \; 10\; 10}\right) = -t_1 - t_2 + t_1 + t_2 \;,
\ee
so that $t_1$ and $t_2$ should be identified as the single 10-matter curve carrying the SM generations.

The term monodromies arises from considering the local form of the Higgs field where its vev along two $U(1)$s is mapped under a monodromy around a branch cut in its spatial dependence on $S_{GUT}$ \cite{Hayashi:2009ge}. In \cite{Cecotti:2010bp} a more general way to view this, which is more along the lines described here, is to note that branch cuts arise in Higgs profiles when a holomorphic Higgs vev which has off-diagonal components so that it take values in the full $SU(2)$ is diagonalised. The resulting diagonal Higgs profile then has branch cuts that interchange the two diagonal components.  

A further important point made in \cite{Cecotti:2010bp} is that to extract full information regarding the localisation of matter in the presence of a non-Abelian Higgs vev it is not sufficient to consider only the eigenvalues of the Higgs field but rather the full Higgs profile should be specified. This is as opposed to a purely Cartan diagonal Higgs vev which is fully specified by its eigenvalues. Configurations that are not captured by the eigenvalues of the Higgs were termed T-branes. As yet we do not have sophisticated enough tools to describe such backgrounds over the full $S_{GUT}$ in F-theory, as opposed to a local patch, and so for the rest of this review we restrict ourselves to Higgs backgrounds that are fully characterised by their eigenvalues (which may have branch cuts interchanging them).

If we diagonalise the Higgs field over the full $S_{GUT}$ then it is characterised by its eigenvalues which have a spatial dependence and may be intercharged under monodromies around branch cuts on $S_{GUT}$. Each diagonal element corresponds to a charge vector $t_i$ above in that when it vanishes $SU(5)_{GUT}$ combines with that particular $U(1)_i$ to enhance to $SU(6)$ or $SO(10)$. Therefore the Higgs background can be described by promoting the $t_i$ to holomorphic (up to branch cuts) functions on $S_{GUT}$ and matter charged under the $t_i$ is localised on the locus where the appropriate combination of $t_i$, as specified in (\ref{cur3}), vanishes.\footnote{The GUT singlets are not fully localised on $S_{GUT}$ but the projection of their localisation locus to $S_{GUT}$ is given by the vanishing curve of the appropriate $t_i$ combination.} Although the $t_i$ individually may undergo monodromies, it is possible to describe the Higgs bundle by quantities that are invariant under monodormies, these are the Casimirs of the Higgs field $b_i$ and are given by
\bea
b_1 &\equiv& \Tr\left[ \varphi \right] = s_1\left(t_i\right) = \sum_i t_i \;, \nn \\
b_2 &\equiv& -\frac12 \Tr\left[ \varphi^2 \right] = s_2\left(t_i\right) \;, \nn \\
... & & \nn \\
b_5 &\equiv& \mathrm{det}\; \varphi = s_5\left(t_i\right) = \Pi_i t_i\;. \label{bihiggs}
\eea
Here the $s_n$ denote elementary symmetric polynomials of degree $n$. The labelling $b_i$ is no coincidence, these quantities are actually related to the global $b_i$ of section \ref{sec:7inf} (more precisely to the projection of them onto $S_{GUT}$). This is the first explicit connection between the Higgs bundle and the geometric description of F-theory, in section \ref{sec:speccov} we make this mapping more explicit.

So far we have discussed the background Higgs vev which determines the matter curves. If we only have a holomorphic Higgs background the gauge theory preserves four dimensional $N=2$ supersymmetry and therefore the massless spectrum forms $N=2$ vector multiplets and hypermultiplets and so is not chiral. To obtain more realistic models we have to break to $N=1$ supersymmetry and generate chirality, this is achieved by the presence of a background flux for the 8-dimensional gauge field along $S_{GUT}$. The key difference from the IIB fluxes described in section \ref{sec:d7spec} is that we are describing the whole intersecting brane system in terms of one 8-dimensional theory. However since the integrals involving chirality are always performed restricted to the GUT brane we expect that the naive generalisation of the formulae (\ref{iibbulkchir}) and (\ref{iibcurflux}) should also hold in these cases where we replace the intersection of the GUT brane with a $U(1)$ brane with the matter curves. This was studied in detail in terms of the twisted gauge theory in \cite{Donagi:2008ca,Beasley:2008dc} with the expected results that the chirality is counted by the homology
\be
H^i\left({\cal C},F\otimes K_{\cal C}^{\frac12} \right) \;. \label{u1flho}
\ee
Here ${\cal C}$ denotes the matter curve, $F$ the $U(1)$ flux which transforms in the appropriate representation for the localised matter (ie. is weighted by the $U(1)$ charge), and $K_{\cal C}$ is the canonical bundle by which we twist. The net chirality is therefore given by
\be
I = h^0\left({\cal C},F\otimes K_{\cal C}^{\frac12} \right) - h^1\left({\cal C},F\otimes K_{\cal C}^{\frac12} \right) = \int_{\cal C} F \;.
\ee
Although this approach does capture some of the appropriate fluxes it does not capture all the possible fluxes; a more sophisticated approach to fluxes is the use of the spectral cover as described in section \ref{sec:speccov}.

Apart from $U(1)$ fluxes which induce chirality in complete GUT multiplets it is also possible to turn on flux inside $SU(5)_{GUT}$. Indeed this is a particularly attractive way to break the GUT group to the gauge group of the Standard Model which is the commutant with the Hypercharge inside $SU(5)$. Generically such a flux would induce a mass for the Hypercharge gauge boson through the Stueckelberg coupling (\ref{d7gs}). However the fact that the GUT gauge theory is localised on a submanifold of the full CY, and that the massless axions that can give the hypercharge a mass come from the closed-string sector and therefore are counted by the cohomology of the full CY, gives a way to avoid such a mass \cite{Jockers:2004yj,Buican:2006sn,Beasley:2008kw,Donagi:2008kj}. The idea is to turn on hypercharge flux which is homologically non-trivial in $S_{GUT}$ but homologically trivial in the full CY. This implies that the corresponding mass term vanishes since
\be
\int_{D_a} f_Y \wedge i^*\left(\omega_i\right) = 0 \;\;\;\;\;\;\forall\; \omega_i \in H^2\left(CY\right) \;, \label{trivfy}
\ee
where $f_Y$ is the hypercharge flux.

The presence of hypercharge flux on $S_{GUT}$ can alter the massless spectrum of both bulk modes and matter curve modes. Decomposing the adjoint of $SU(5)$ to the Standard Model gauge group gives
\be
{\bf 24} \rightarrow  \left({\bf 1},{\bf 3}\right)_0 \oplus \left({\bf 8},{\bf 1}\right)_0 \oplus \left({\bf 1},{\bf 1}\right)_0\oplus \left({\bf 3},{\bf 2}\right)_{-5/6} \oplus \left({\bf \bar{3}},{\bf 2}\right)_{5/6}		\;,
\ee
and so a bulk hypercharge flux can induce net chirality in the $\left({\bf 3},{\bf 2}\right)_{-5/6}$ states. To avoid such chiral states we should take the hypercharge flux to come from a bundle ${\cal L}_Y$ which satisfies \cite{Beasley:2008kw,Donagi:2008kj}
\be
I\left(S,{\cal L}_Y^{\frac56} \right) = 1 + \frac12 c_1^2\left({\cal L}_Y^{\frac56}\right) = 0 \implies c_1^2\left({\cal L}_Y^{\frac56}\right) = -2 \;.
\ee
Note that this requires that the bundle ${\cal L}_Y$ is such that ${\cal L}_Y^{\frac56}$ is integer quantised.
The matter curve states under the decomposition are given in (\ref{su5decomp}), and are counted by the cohomology arising from tensoring the appropriate hypercharge flux bundle with the $U(1)$ fluxes in (\ref{u1flho}).	Since $N=\mathrm{deg}\left(L_{Y}^{5/6}\right)$ is integer quantised the $U(1)$ bundles must be appropriately chosen such that $M_{10}=\mathrm{deg}\left({\cal L}_Y^{\frac16}\otimes F_{10}\right)$ and $M_5=\mathrm{deg}\left({\cal L}_Y^{-\frac13}\otimes F_{5}\right)$ are quantised where $F_{10}$ denotes the $U(1)$ flux on the 10-matter curves and $F_{5}$ on the 5-matter curves. This implies that the bundle combinations that generate chirality for all the SM representations are integer quantised. Explicitly we have
\bea
n_{(3,1)_{-1/3}} - n_{(\bar{3},1)_{+1/3} } &=& M_{5} \;, \nn \\
n_{(1,2)_{+1/2}} - n_{(1,2)_{-1/2} } &=& M_{5} + N \;, \label{chi5sm}
\eea
for the 5-matter curves and
\bea
n_{(3,2)_{+1/6}} - n_{(\bar{3},2)_{-1/6} } &=& M_{10} \;, \nn \\
n_{(\bar{3},1)_{-2/3}} - n_{(3,1)_{+2/3} } &=& M_{10} - N \;, \nn \\
n_{(1,1)_{+1}} - n_{(1,1)_{-1} } &=& M_{10} + N\;, \label{chi10sm}
\eea
for the 10-matter curves.
We will return to the spectrum induced by hypercharge flux in more detail in section \ref{sec:localtrivialcurves}. 

Finally we note that the gauge theory description is only a part of the full effective theory expected on an F-theory 7-brane, and in particular it does not include the coupling to the closed-string sector and terms higher order in the gauge field strength as appear in the DBI and Chern-Simons terms of a D-brane. See \cite{denef08} for a review on how to recover some of the terms in the 7-brane effective action and \cite{Grimm:2012rg} for work approaching the problem form the M-theory perspective.

\subsection{The spectral cover}
\label{sec:speccov}

In the previous section we reviewed the approach to F-theory models of using an 8-dimensional gauge theory to describe the configuration of intersecting 7-branes. In contrast the global approach to F-theory 7-branes presented in section \ref{sec:7inf} was through a 12-dimensional geometry with singularities. Roughly speaking the gauge generators are lifted to a geometric description in F-theory, so that for example the 2-form flux with generator index in the 8-dimensional theory is lifted to a 4-form flux in a pure geometry background. In this section we will review an approach which in some sense the half-way point between the 8-dimensional gauge theory and the geometric global F-theory models. We will still retain a purely local analysis but will geometrise the gauge theory.

The starting point is to model the gauge theory as a fibration of an $A_4$ singularity over $S_{GUT}$ where the $A_4$ accounts for the $SU(5)_{\perp}$. Generically over $S_{GUT}$ the $A_4$ is fully resolved but over the matter curves or Yukawa points it degenerates thereby enhancing the singularity. Such an $A_4$ singularity is described by the equation (for detailed studies see \cite{1992alg,Beasley:2008dc,Grimm:2011tb})
\be
y^2 = x^2 + \Pi_{i=1}^5 \left(s + t_i\right) \;, \label{a4res}
\ee
where the $t_i$ are functions on $S_{GUT}$. The singularity lies at $x=y=s=0$. Each $t_i$ amounts to resolving\footnote{Strictly speaking the $t_i$ in (\ref{a4res}) parameterise deformations of the singularity rather than resolutions. However for an ADE singularity local mirror symmetry induces a map, which does not act on $S_{GUT}$ but only in the fibre, between resolutions and deformations so that we can refer to them interchangeably \cite{Katz:1997eq}.} one of the 5 2-cycles in the $A_4$ singularity and so these map directly to the $t_i$ in (\ref{cur3}) and expanding out the equation we see that we can write it in terms of the $b_i$ in (\ref{bihiggs})
\be
y^2 -x^2 = b_0 s^5 + b_2 s^3 + b_3 s^2 + b_4 s + b_5  \;, \label{a4resbi}
\ee
which gives an explicit map between the Higgs in the 8-dimensional field theory and the $A_4$ singularity fibration. The $b_i$ in (\ref{a4resbi}) are also related to the $b_i$ in the full global $E_8$ Tate (\ref{su5sing}) in the limit $w \rightarrow 0$. Intuitively this can be understood as the manifestation in the geometry, in the limit of the $SU(5)_{GUT}$ singularity on $w=0$, of the splitting $E_8 \rightarrow SU(5)_{GUT}\times SU(5)_{\perp}$, while for a more concrete map see \cite{Marsano:2011hv}.

From (\ref{a4resbi}) we can construct the spectral cover as follows \cite{Donagi:2009ra} (see also \cite{Hayashi:2008ba} for earlier work). The right-hand side of equation (\ref{a4resbi}) encodes the local geometry
\be
{\cal C}\;:\;b_0 s^5 + b_2 s^3 + b_3 s^2 + b_4 s + b_5 = 0 \;. \label{speccov}
\ee
This equation naturally defines a divisor inside a non-compact CY 3-fold $X$ composed of the total space of $S_{GUT}$ and its canonical bundle $K_S$, with $s$ a coordinate along the latter. At a generic point on $S$ the roots of (\ref{speccov}), which are the eigenvalues of the associated Higgs, define 5 points in the fibre $K_S$ and these trace out a 5-fold cover of $S$ which is called the spectral cover (of the fundamental representation). The divisor ${\cal C}$ is non-compact since at $b_0=0$ we lose two roots that go off to infinity. In order to construct fluxes in the spectral cover which will be the local version of G-flux it is useful to compactify $X$ to $\tilde{X}$ by taking two homogenous coordinates $U$ and $V$ and identifying $s=U/V$ \cite{Donagi:2009ra}. ${\cal C}$ is compactified to $\tilde{C}$ by adding the divisor at infinity corresponding to $V=0$ which we denote $\sigma_{\infty}$. The divisor $U=0$ is denoted $\sigma$ and the homogneity implies $\sigma \cdot \sigma_{\infty}=0$. 

Note that since the $b_i$ appearing in the spectral cover are the pull backs to $S_{GUT}$ of the $b_i$ of (\ref{su5sing}), then the $b_i$ are sections of 
\be
b_i \in \eta - i c_1\left(S\right) \;. \label{bisecspec}
\ee
Here $c_1\left(S\right)$ denotes the first Chern class of the tangent bundle of $S$, which we henceforth denote $c_1$, and $\eta=6 c_1 - t$ where $t$ is the first Chern class of the normal bundle of $S_{GUT}$ (so the dual bundle to the divisor $w=0$).

To construct fluxes we define the associated projections
\bea
\pi \;&:&\; \tilde{X} \rightarrow S_{GUT} \;,\nn \\
p_{\cal \tilde{C}} \;&:&\; {\cal \tilde{C}} \rightarrow S_{GUT} \;.
\eea
Then fluxes are line bundle that live in the spectral cover and satisfy the tracelessness constraint \cite{Donagi:2009ra}
\be
c_1 \left( p_{\cal \tilde{C}}^* {\cal L}\right) = p_{\cal \tilde{C}}^* c_1 \left(  {\cal L}\right) - \half p_{\cal \tilde{C}}^* r = 0 \;, \label{tracspec}
\ee
where $r$ is the ramification divisor
\be
r=p_{\cal \tilde{C}}^* c_1\left(S_{GUT}\right) - c_1\left({\cal \tilde{C}} \right)\;.
\ee
If we decompose 
\be
c_1\left({\cal L}\right) = \half r + \gamma \;,
\ee
then (\ref{tracspec}) implies 
\be
p_{{\cal \tilde{C}}*} \gamma  = 0 \;. \label{tacespec2}
\ee
Therefore fluxes in the generic spectral cover model correspond to two-forms that satisfy (\ref{tacespec2}). There is a unique such flux, which is called the universal spectral cover flux
\be
\gamma = 5 \left[{\cal \tilde{C}}_{10}\right] - p_{\cal \tilde{C}}^* p_{{\cal \tilde{C}}*} \left[{\cal C}_{10}\right] \;,
\ee
where ${\cal C}_{10}$ is the class of 10-matter curve in the spectral cover $U=b_5=0$. 	

The chirality in the ${\bf 10}$	sector induced by the universal flux can be explicitly computed to be \cite{Donagi:2009ra}
\be
n_{\bf 10} - n_{\bf \bar{10}} = - \int_{S_{GUT}}\eta \wedge \left(\eta - 5 c_1 \right) \;.
\ee
The net chirality of the ${\bf 5}$ representations is equal to that of the ${\bf 10}$ which guarantees anomaly cancellation.

The discussion so far applies to the generic spectral cover by which we mean that the equation (\ref{speccov}) does not factorise. From (\ref{a4res}) we see that in the case where we think of a resolved $SU(5)_{\perp}$ singularity fibered over $S_{GUT}$ it seems the $b_i$ are such that (\ref{speccov}) factorises into 5 factors. However, as discussed in section \ref{sec:localgaugetheory}, the $SU(5)_{\perp}$ singularity has a Weyl group action which interchanges the $t_i$ so as to preserve the $b_i$ and generally the fibration over $S_{GUT}$ can act with this group which gives rise to monodromies in F-theory \cite{Hayashi:2009ge,Heckman:2009mn,Cecotti:2010bp}.

This maps directly to the product structure of (\ref{speccov}), where we see that under no identification of the $t_i$, (\ref{speccov}) factorises into 5 factors. Each factor corresponds to a $U(1)$ with a tracelessness constraint $b_1=0$ leaving the 4 Cartan $U(1)$s as linearly independent. As we identify the $t_i$ (\ref{speccov}) decomposes into fewer factors implying fewer $U(1)$s and finally if the fibration uses the full Weyl group there is no splitting at all and no $U(1)$s. Hence a spectral cover model which corresponds to the presence of a $U(1)$ symmetry is called a split spectral cover, and the splitting factorisation corresponds to the degree of the group theory factor in table \ref{tab:breake8u1s}. So for example the case of a local breaking $E_8 \rightarrow SU(5)_{GUT}\times S\left[U(4)\times U(1)\right]$ is described by a $4+1$ split spectral cover \cite{Marsano:2009gv}
\be
{\cal C}\;:\;\left(c_0 U + c_1 V \right) \left(U^4 d_0 + U^3 V d_1 + U^2 V^2 d_2 + U V^3 d_3 + V^4 d_4 \right) \;. \label{ssc41}
\ee
In the form (\ref{ssc41}) we have to impose an additional tracelessness constraint on the coefficients since expanding it out we find
\be
b_1 = c_0 d_1 + d_0 c_1 = 0 \;. \label{tracconst}
\ee
In writing a split spectral cover the constraint (\ref{tracconst}) must be satisfied automatically. In this case a solution ansazt is
\be
d_0 = \alpha c_0 \;,\;\; d_1 = -\alpha c_1 \;,
\ee
with $\alpha$ some arbitrary section.

In the presence of a single $U(1)$ symmetry, so that after identifying the $t_i$ related by monodromies there remain two classes $t_1$ and $t_2$, the matter curves also split according to the classification (\ref{cur3}). Different curve components are distinguished by the $U(1)$ charges of the representations they carry. In terms of the split spectral cover we see this as the split of $P_{10}$ and $P_{5}$, the 10-matter and 5-matter curves as defined in (\ref{p5p10})
\bea
P_{10} &=& b_5 = c_1 d_4 \;,\;  \label{41splitb5} \\ \nn
P_5 &=& b_3^2 b_4 - b_2 b_3 b_5 + b_0 b_5^2 = \left( c_1^2 d_2 + c_0 c_1 d_3 + c_0^2 d_4\right) \left(c_1 d_2 d_3 + c_0 d_3^2 + \alpha c_1^2 d_4 \right) \;.
\eea
We should think of the splitting of the matter curves as the decomposition of a single curve in the spectral cover into the components inside each factor of the spectral cover. 

Similarly to such a splitting of the matter curves there are additional fluxes possible due to the split and these correspond to fluxes along the additional $U(1)$. They are constructed in a similar way as for the non-factored case. 
Consider a split of (\ref{speccov}) into factors labelled by $i$ which are of order $n_i$ so that
\be
\sum_i n_i = 5 \;.
\ee
Associated to each factor there is a map
\be
p_i : {\cal C}^i \rightarrow S_{GUT} \;.
\ee
Now the fluxes are defined to live on each factor, so that 
\be
c_1\left({\cal L}_i\right) \in H^{(1,1)}\left({\cal C}^i,\mathbb{Z} \right)\;.
\ee
However the tracelessness constraint only applies to their sum
\be
\sum_i c_1\left(p_{i*} {\cal L}_i \right) = 0 \;.
\ee
Now as before to the map $p_i$ we can associate a ramification divisor, factor it out, and therefore work with the associated $\gamma_i$ which satisfy
\be
\sum_i p_{i*} \gamma_i  = 0 \;. \label{tracsplicont}
\ee
There are a number of fluxes that can be constructed such that they are traceless and most of them correspond to non-Abelian bundles. See \cite{Marsano:2012yc,Marsano:2009wr} for some detailed constructions. It is informative to see how the most direct analogue to an Abelian $U(1)$ gauge flux on $S_{GUT}$ is constructed. Given any 2-form on $S_{GUT}$ $\rho$, this flux is nothing but the flux with geometric component $\rho$ and whose gauge generator index is simply lifted to the spectral cover in the canonical way. To do this first we denote the charges of the ${\bf 10}$'s, or equivalently the $t_i$'s, under this $U(1)$ by $q_i$. Note that due to the tracelessness constraint these must satisfy
\be
\sum_i n_i q_i = 0 \;.
\ee
Then the spectral cover flux is constructed as
\be
\hat{\rho} = \sum_i q_i\; p^*_{i} \rho \;.
\ee
The idea here being that we have $p_{i*} p^*_j \rho = n_i \delta_{ij} \rho$ which ensures that the tracelessness constraint (\ref{tracsplicont}) is satisfied.

This flux acts just like a normal gauge flux over $S_{GUT}$ would do: namely to calculate the induced chirality $M_{\bf R}$ associated to some representation ${\bf R}$ the intersection of $\hat{\rho}$ with the matter curves in the spectral cover simply amount to the intersection in $S_{GUT}$ weighted by the charge under the $U(1)$
\be
M_{\bf R} = \hat{\rho} \cdot  {\cal C}_{\bf R} = q_{\bf R} \; \rho \cdot \pi_*\left( {\cal C}_{\bf R}\right) \;, \label{chisimflux}
\ee
where $q_{\bf R}$ is the charge of the representation ${\bf R}$ under the $U(1)$.

Having discussed the spectral cover construction at some length it is important to emphasise that it is still only a local and incomplete description of F-theory models. In particular a split spectral cover does not guarantee that there is a $U(1)$ present in four-dimensions after all the compact space has been integrated over. Similarly there are global properties such as the D3-tadpole and the spectrum of GUT singlets that can not be studied in full generality using the spectral cover. 

The example given in this section of a 4+1 split spectral cover model was first studied in \cite{Marsano:2009gv}. Following this, and the original work \cite{Donagi:2009ra}, the spectral cover was used extensively in local model building. A 4+1 split was used in \cite{Blumenhagen:2009yv,Marsano:2012yc,Choi:2012pr}, a 3+2 in \cite{Marsano:2009wr,Choi:2012pr} and higher split models in \cite{Dudas:2009hu,Choi:2010su,King:2010mq,Dudas:2010zb,Leontaris:2010zd,Marsano:2010sq,Ludeling:2011en,Dolan:2011iu,Palti:2012aa}. Spectral cover models of $SO(10)$ GUTs were studied in \cite{Chen:2010ts,Antoniadis:2012yk}, the related flipped $SU(5)$ in \cite{Chen:2010tp,Chung:2010bn,Kuflik:2010dg,Choi:2011ua,Callaghan:2011jj,Dolan:2011aq}, and directly the SM gauge group in \cite{Choi:2010nf,Choi:2011te}. Spectral covers based on $E_6$ rather than $E_8$ gauge groups were studied in \cite{Chen:2010tg}. While the connection of the spectral cover to fully global models was studied in \cite{Hayashi:2010zp,Grimm:2010ez,Marsano:2010ix,Marsano:2011hv,Krause:2011xj,Kuntzler:2012bu}.

As mentioned in section \ref{sec:localgaugetheory} the Higgs bundle description is incomplete whenever the Higgs background is not fully characterised by its eigenvalues \cite{Cecotti:2010bp}. This limitation transfers over to the spectral cover construction which does not fully capture the most general geometric setting even locally. In particular it is possible to have spatially varying background vevs for the GUT singlet recombination or gluing moduli. This is the geometric analogue of turning on spatially varying non-Cartan components in the Higgs background, and such geometries have been studied in \cite{Donagi:2011jy,Donagi:2011dv,Krause:2012yh,Marsano:2012bf}.

\subsection{Globally trivial curves and hypercharge flux}
\label{sec:localtrivialcurves}

As discussed in section \ref{sec:localgaugetheory} in the context of hypercharge flux GUT breaking, an important aspect of F-theory GUTs is the existence of curves on $S_{GUT}$ that are non-trivial in the homology of $S_{GUT}$ but trivial in the global homology of the CY. In this section we are interested in a particular aspect of this sector which is the net chirality that can be induced by hypercharge flux in the spectrum of the massless modes localised on matter curves. It was first noted in \cite{Marsano:2009gv} that already from the local approach of the spectral cover there are strong constraints on the possible chiral spectrum induced by hypercharge flux. This was subsequently expanded upon in \cite{Marsano:2009wr,Dudas:2009hu,Choi:2010su,King:2010mq,Dudas:2010zb,Leontaris:2010zd,Marsano:2010sq,Ludeling:2011en,Dolan:2011iu}.

The most direct result is the absence of any net chirality induced by hypercharge flux in the absence of any splitting in the spectral cover \cite{Donagi:2009ra}. This follows because the homology classes of the $b_i$ in (\ref{bisecspec}) imply that the matter curves are in the following homology classes 
\bea
\left[P_{10}\right]&=&\left[b_5\right] = \left[\eta - 5 c_1\right] \;,\;  \label{totmathom} \\ \nn
\left[P_{5}\right]&=&\left[b_3^2 b_4 - b_2 b_3 b_5 + b_0 b_5^2\right] = \left[\eta^3 - 10 c_1 \right]\;.
\eea
Now the important point is that the homology classes of $\eta$ and $c_1$ are both globally non-trivial since they correspond to the first Chern classes of the tangent and normal bundles of the GUT divisor. Therefore, from (\ref{trivfy}), the hypercharge flux must restrict trivially to them and hence no net chirality can be induced by it on the matter curve spectrum
\be
\left[F_Y\right] \cdot \left[P_{10}\right] =  \left[F_Y\right] \cdot \left[P_{5}\right] =  0 \;. \label{nonetchi}
\ee 
Here we denote the hypercharge flux $F_Y$ and the inner product is the intersection number which calculates the net chirality. Note that there can still be vector-like massless modes induced by the hypercharge flux.

The situation changes when the spectral cover is split and the representations on the matter curves are distinguished by an additional $U(1)$ symmetry. Now (\ref{nonetchi}) implies that the total chirality induced by the hypercharge flux on the matter curves must vanish, but there can be some net chirality on each class, as distinguished by the $U(1)$ charge, of matter curves. This is because the homology class of the split components are only determined up to an unknown relative factor by the $b_i$, explicitly using the example split (\ref{41splitb5}) we have that
\be
\left[c_1\right]+ \left[d_4\right] = \left[\eta - 5 c_1 \right]\;,
\ee
and so the two 10-matter curves can have equal and opposite globally trivial components which the hypercharge flux could restrict to. It is important to note that this additional freedom does not guarantee that the 10-matter curves do have a globally trivial component but is simply compatible with it.

A non-trivial restriction of the hypercharge flux to a matter curve modifies the spectrum as in (\ref{chi5sm}) and (\ref{chi10sm}). It was shown in \cite{Dudas:2010zb} for a large set of spectral cover models that the possible net restriction of the hypercharge flux to the matter curves always satisfies the relation, using the notation of (\ref{chi5sm}) and (\ref{chi10sm}), 
\be
\sum_{{\cal C}_{10}^i} Q_{10}^i N_{10}^i + \sum_{{\cal C}_{5}^j} Q_{5}^j N_{5}^j= 0 \;. \label{anomNmix}  
\ee
Here ${\cal C}_{10}^i$ and ${\cal C}_{5}^j$ denote the 10 and 5-matter curves respectively, $Q_{10}^i$ denotes the charges of the representations under any $U(1)$ present due to splitting, and $N_{10}^i$ denotes the restriction of the hypercharge flux to that curve. It is worth writing in this notation the constraint that the total restriction vanishes (\ref{nonetchi})
\be
\sum_{{\cal C}_{10}^i} N_{10}^i = \sum_{{\cal C}_{5}^j} N_{5}^j = 0 \;. \label{nochiN}  
\ee
It was shown in \cite{Dolan:2011iu} that, at least for spectral cover models with at most 2 $U(1)$ symmetries, i.e that split into at most 3 factors, the constraints (\ref{anomNmix}) and (\ref{nochiN}) are the only ones on the possible restriction of the hypercharge flux to the matter curves that arise from the type considerations discussed.

In \cite{Marsano:2010sq} it was argued that the constraints (\ref{anomNmix}) and (\ref{nochiN}) can be understood from four-dimensional anomaly cancellation. The important point is that because the hypercharge flux is globally trivial, and so satisfies (\ref{trivfy}), it can not participate in the Green-Schwarz anomaly cancellation mechanism discussed in section \ref{sec:d7spec} since the coupling in (\ref{d7gs}) vanishes.\footnote{Although the particular terms shown in section \ref{sec:d7spec} were for the weakly coupled IIB limit it is reasonable to expect that similar reasoning applies to their F-theoretic uplift, as studied in \cite{Cvetic:2012xn}.} Therefore the hypercharge flux can not modify the anomalies of the massless spectrum and the constraints (\ref{anomNmix}) and (\ref{nochiN}) ensure that this is so for the anomalies involving just the SM gauge groups
\be
{\cal A}_{SU(3)^2-SU(3)}\;,\;{\cal A}_{SU(2)^2-SU(2)}\;,\;{\cal A}_{U(1)_Y-U(1)_Y}\;... \;,
\ee
and for the mixed anomalies involving the $U(1)$ symmetry
\be
{\cal A}_{SU(3)^2 - U(1)} \;,\; {\cal A}_{SU(2)^2 - U(1)} \;,\; {\cal A}_{U(1)^2_Y - U(1)} \;. \label{mixanom}
\ee
Note that ensuring the hypercharge does not modify the anomalies does not imply that the anomalies must vanish but rather that they should be proportional to the anomalies before the hypercharge flux is turned on, ie. at the GUT level, so that 
\bea
{\cal A}_{SU(3)^2-SU(3)} &\propto& {\cal A}_{SU(2)^2-SU(2)} \propto ... \propto {\cal A}_{SU(5)^2-SU(5)} = 0 \;, \nn \\
{\cal A}_{SU(3)^2 - U(1)} &\propto& {\cal A}_{SU(2)^2 - U(1)} \propto {\cal A}_{U(1)^2_Y - U(1)} \propto {\cal A}_{SU(5)^2 - U(1)} \;.
\eea
The fact that these subtle anomaly cancellation conditions are automatically ensured in the spectral cover approach is rather satisfying. However it was shown in \cite{Palti:2012dd} that there are additional anomaly constraints which are not automatically satisfied and these correspond to the anomaly
\be
{\cal A}_{U(1)_Y - U(1)^2} \;. \label{y2anom}
\ee
Within a possible Green-Schwarz counterterm to this anomaly the hypercharge flux appears directly through a Stueckelberg mass coupling and therefore, in order to remain massless, such a counterterm must vanish. Another way to see that it must vanish is that according to the discussion above it must be proportional to the GUT anomaly, but the latter vanishes automatically because of the non-Abelian GUT group ${\cal A}_{SU(5) - U(1)^2} = 0$. Therefore for globally trivial hypercharge flux we would expect that the local geometry should impose this constraint on the spectrum. However, unlike the anomalies (\ref{mixanom}) this does not come out from all currently known constraints on the geometry. Therefore we must deduce that there are additional constraints on the local geometry of F-theory GUT models that have a (possibly anomalous and massive) $U(1)$ symmetry.

The anomaly (\ref{y2anom}) implies that an additional constraint on the net hypercharge restriction should be \cite{Palti:2012dd}
\be
3 \sum_{{\cal C}_{10}^i} \left(Q_{10}^i\right)^2 N_{10}^i + \sum_{{\cal C}_{5}^j} \left(Q_{5}^j\right)^2 N_{5}^j = 0 \;. \label{anomN}
\ee
This additional constraint was shown in \cite{Palti:2012dd} to be quite restrictive so that only two possibilities of the breaking of $E_8$ from the table \ref{tab:breake8u1s} could support a net chirality in at least some of the representations while satisfying all the anomaly constraints. These were particular models based on a $S\left[U\left(3\right)\oplus U\left(2\right)\right]$ and  $S\left[U\left(2\right)\oplus U\left(2\right)\oplus U\left(1\right)\right]$ breaking. In terms of the spectral cover these are 3+2 and 2+2+1 splits, but the conclusion applies also to more general models not fully captured by the spectral cover (see section \ref{sec:speccov} for a discussion) since it just relies on the possible representations that are present. Note that the fact that the spectral cover approach does not automatically satisfy the constraint (\ref{anomN}), even though it does (\ref{anomNmix}), is not so surprising because the triviality aspect of the curves is only determined globally. A geometric understanding of the constraint (\ref{anomN}) on the globally trivial components of the matter curves is as yet missing.

Finally, it is interesting to note that the constrained homology of the matter curves (\ref{totmathom}) can be used to place constraints on their intersection structure and therefore the associated Yukawa couplings \cite{Cordova:2009fg,Hayashi:2009bt,Hayashi:2011aa}. One particularly important result is that in the absence of any splitting the number of $E_6$ points of enhancement is always even.

\subsection{Wavefunctions and ultra local models}
\label{sec:fultralocal}

So far we have reviewed the so called semi-local approach to F-theory models where we consider the local geometry of $S_{GUT}$ within the CY. One interesting aspect of these models is the localisation of modes to matter curves inside of $S_{GUT}$. Similarly, since cubic couplings, and in particular Yukawas, between fields localised on matter curves are associated to intersections of the curves they are localised at points in $S_{GUT}$. In the effective field-theory description used in section \ref{sec:localgaugetheory} such points corresponds to enhancement of the gauge symmetry. Around these points the localisation is manifest in the effective gauge theory through the profile of the matter fields wavefunctions along $S_{GUT}$. The localisation of operators to points or small patches in $S_{GUT}$ implies that we can study them to a decent approximation by considering the gauge theory within just a patch in $S_{GUT}$ around such an enhancement point of interest. Since locally the patch is just flat space the analysis simplifies considerably allowing for a fairly explicit study of the local wavefunction profiles. The approach of considering the gauge theory on a local patch is sometimes termed ultra local.

Studying Yukawa, and other couplings in a local patch was done in \cite{Heckman:2008qa,Hayashi:2009ge,Bouchard:2009bu,Randall:2009dw,Conlon:2009qq,Heckman:2009de,Font:2009gq,Cecotti:2009zf,Leontaris:2010zd,Cecotti:2010bp,Oikonomou:2011ba,Oikonomou:2011kd,Peccei:2011ng,Aparicio:2011jx,Camara:2011nj,Palti:2012aa,Hayashi:2010zp}.\footnote{In \cite{Hayashi:2009bt,Kawano:2011aa} wavefunctions were studied within a more global setting within $S_{GUT}$. See also \cite{Conlon:2008qi} for an analysis of wavefunctions on simple matter curve geometries and \cite{Marchesano:2008rg,Camara:2009xy,Marchesano:2010bs} for related studies of wavefunctions.} In this section we review the local theory around a point of enhancement and some of the applications. We primarily follow the discussion presented in \cite{Palti:2012aa}.

The setting of the ultra local theory is within the gauge theory described in section \ref{sec:localgaugetheory} where we consider a higher rank gauge theory over $S_{GUT}$ which is broken to $SU(5)_{GUT}$ by a spatially varying Higgs profile. The higher gauge group here is defined according to the type of coupling we are interested in studying and therefore the local enhancement group. The most general such configuration is around a point of $E_8$ enhancement \cite{Heckman:2009mn} and in describing the local theory we will consider this and lower enhancement groups, such as $E_6$ which is associated to a up-type Yukawa coupling and $SO(12)$ associated to a down-type coupling.

We are interested in the equations of motion for fluctuations of the fields about a background with a Higgs vev and non-vanishing gauge field flux along $S$. Locally, the leading order contributions from the Higgs and gauge fields are linear in the local co-ordinates $z_1$ and $z_2$ and so take the form
\be
\langle\varphi_H\rangle=M_{K} R\, m^a_{i} z_i \, Q_a dz_1\wedge dz_2+\ldots\ ,\label{phi}
\ee
\be
\langle A\rangle =-M_{K}\,\textrm{Im}( M^a_{ij} z_id\bar z_j)Q_a + \ldots\ , \label{a} 
\ee
where the dots denote higher order terms in the two local complex coordinates $z_1$ and $z_2$. Here the $m^a_i$ and $M^a_{ij}$ are complex constants which, in the case of $E_8$ enhancement, denote the vevs along the 4 $U(1)$ factors with associated generators $Q_a$. Actually it is more convenient to turn on the vev within $S(U(1)^5)$ so that $a=1,..,5$ but we have to impose an additional tracelessness constraint
\be
\sum_{a=1}^5 m^a_i = \sum_{a=1}^5 M^a_{ij} = 0 \;.
\ee
We will also allow for flux $M^Y_{ij}$ along the Hypercharge $U(1)$ inside $SU(5)_{GUT}$ with generator $Q_Y$, and denote it by the index value $a=6$ so that summing over the $a$ index includes the hypercharge. Note that the local expansion of the Higgs and the flux begins with a linear term in the $z_i$ and there is no constant term. For the Higgs background this amounts to defining the enhancement point to be at the origin. For the gauge field a constant term locally is pure gauge and so can be gauged away.

The mass scale $M_{K}$ is a {\it local} mass scale which involves the cutoff scale of the theory $M_*$ and scales with a homogeneous rescaling of the local metric by a length scale $R_{\parallel}$ as 
\be
M_K = \frac{M_*}{R_{\parallel}} \;.
\ee 
The dimensionless scale $R$ is associated to the {\it local} normal length scale $R_{\perp}$ to $S$ and scales as 
\be
R \equiv R_{\parallel}R_{\perp} \;.
\ee
The scaling of the Higgs with $R_{\perp}$ follows from the pullback of the normal metric component to the world-volume of the 7-brane as discussed in \cite{Camara:2011nj}.

In this background the equations of motion (\ref{eom1}-\ref{eom8}) simplify considerably and can be written in the compact form
\begin{equation}
\mathbb{D^-}\Psi=0 \label{fterm} \;,
\end{equation}
with,
\begin{equation}
\mathbb{D^{\pm}}=\begin{pmatrix}0& D_1^{\pm}& D_2^{\pm} & D_3^{\pm}\\
-D_1^{\pm}& 0& -D_3^{\mp}& D_2^{\mp}\\
-D_2^{\pm}& D_3^{\mp}&0&-D_1^{\mp}\\
-D_3^{\pm}& -D_2^{\mp}&D_1^{\mp}&0 \end{pmatrix}\ , \qquad \Psi=\begin{pmatrix}\eta\\ \psi_{\bar 1}\\ \psi_{\bar 2}\\ \chi\end{pmatrix} \;,
\label{z2}
\end{equation}
and
\begin{align}
D_i^-&\equiv \partial_i-\frac{1}{2}q_a\bar{M}^a_{ji}\bar{z}_j& D_i^{+}&\equiv \bar \partial_i+\frac{1}{2}q_aM^a_{ji}z_j\, \qquad i=1,2\label{gaugecov}\\
D_3^-&\equiv - R \, q_a \bar{m}^a_{i} \bar{z}_i & D_3^+&\equiv R \, q_a m^a_{i} z_i \;.
\end{align}
Here $q_a$ denote the charges of the fields under the generator $Q_a$. Since the charges are always contracted with the Higgs or fluxes it is convenient to introduce the notation
\be
M_{ij} \equiv q_a M_{ij}^a \;,\;\; m_i \equiv q_a m_i^a \;. \label{efffluxhiggs}
\ee
These are the equations of motion for massless fields, but in the ultra local theory we also know the equations that determine the wavefunctions for massive fields
 \cite{Marchesano:2010bs,Aparicio:2011jx,Camara:2011nj}
\be
\mathbb{D}^+\mathbb{D}^-\Psi=\left|m_{\lambda}\right|^2 \Psi \label{masseom} \;,
\ee
with associated four-dimensional mass $M_K m_{\lambda}$.

This completes the summary of the ultra local theory in the presence of gauge flux background which is in the Cartan subalgebra of $SU(5)_{\perp}$. Wavefunction solutions to this theory in generality were studied in \cite{Palti:2012aa} which can account for up to a rank 4 local enhancement from $SU(5)$ to $E_8$, for the case of rank 2 enhancement (particularly $SO(12)$) in \cite{Heckman:2008qa,Font:2009gq,Conlon:2009qq,Leontaris:2010zd,Camara:2011nj} and toy models based on a local $U(3)$ enhancement in \cite{Cecotti:2009zf,Aparicio:2011jx}. The key properties of the wavefunctions are localisation onto the vanishing loci of the Higgs background (\ref{schemwave}). Further there is an arbitrary holomorphic prefactor, the different possible independent holomorphic polynomials corresponding to Landau-level degeneracies. Finally in the presence of a flux background the wavefunction peaks at a single point along the matter curves. Such a turning point is associated to the wavefunction of every chiral state. The flux and Higgs backgrounds around this peak point must be consistent with the chirality of the state which for the background (\ref{phi}-\ref{a}) implies that, for a given state with charges $q^a$ peaking at $z_i=0$ we have \cite{Palti:2012aa}
\be
\chi_{\mathrm{local}}\left(q^a\right) = -\mathrm{Re}\left[\left(M_{12}+\bar{M}_{21}\right)\bar{m}_1 m_2 + M_{11} \left(\left|m_1\right|^2-\left|m_2\right|^2\right)\right] > 0 \;. \label{locchi}
\ee

It is worth noting that the idea of a point of full enhancement to $E_8$ is attractive in terms of a very rich ultra-local theory and in some sense is the ultimate realisation of the local approach discussed in this section. Independently of this motivation there are some phenomenological reasons for favouring such a point. In \cite{Heckman:2009mn} it was argued that if flavour physics is attributed to wavefunction profiles then the special diagonal structure of the CKM matrix, which corresponds to a strong correlation between the up and down Yukawa couplings, implies that the local geometry controlling the wavefunction basis around both the Yukawa points should be closely correlated. Hence the points should be in close geometric proximity motivating a coalescence into a point of $E_7$ enhancement. Requiring further a neutrino Dirac mass coupling leads to a motivation for a point of $E_8$.

The backgrounds described are in some ways the simplest local configurations and more complicated settings are likely to be required to achieve fully realistic phenomenology. In particular a Higgs background not fully in the cartan is likely to be required to account for the large top quark Yukawa coupling through monodromies. Such backgrounds were studied in \cite{Cecotti:2010bp} (see also \cite{Chiou:2011js}) and termed T-branes and in particular their relation to local monodromies introduced in \cite{Hayashi:2009ge} elucidated. 

Monodromies were discussed from the perspective of the full $S_{GUT}$ in section \ref{sec:localgaugetheory} and in terms of the spectral cover in section \ref{sec:speccov}. Recall that a monodromy acts as a subgroup of the Weyl group of the $SU(5)_{\perp}$ singularity fibered over $S_{GUT}$. In terms of the Higgs background the Weyl group permutes the eigenvalues $t_i$. So a $\mathbb{Z}_2$ monodromy or example would interchange $t_1$ and $t_2$ as we move around $S_{GUT}$. 
As emphasised in \cite{Cecotti:2010bp}, the most general way to think about this is in terms of a non-diagonal Higgs background, so for example as a matrix in $SU(5)_{\perp}$ we can consider the following
\begin{equation}
\varphi_H=\begin{pmatrix} 
  0& 1 & 0 & 0 & 0\\
  z_1 & 0 & 0 & 0 & 0\\
  0& 0 & t_3 & 0 & 0\\
  0& 0 & 0 & t_4 & 0\\
  0& 0 & 0 & 0 & t_5 \\
 \end{pmatrix} \;,
\label{z2mon}
\end{equation}
where $t_3$, $t_4$ amd $t_5$ are some unspecified functions of the $z_i$. We can diagonalise this local holomorphic Higgs background
\begin{equation}
\varphi_H=\begin{pmatrix} 
  \sqrt{z_1} & 0 & 0 & 0 & 0\\
  0 & -\sqrt{z_1} & 0 & 0 & 0\\
  0& 0 & t_3 & 0 & 0\\
  0& 0 & 0 & t_4 & 0\\
  0& 0 & 0 & 0 & t_5 \\
 \end{pmatrix} \;,
\label{z2mondiag}
\end{equation}
and now as we circle around $z_1=0$ the eigenvalues $t_1$ and $t_2$ are interchanged. It was shown in \cite{Hayashi:2009ge,Cecotti:2010bp} that such a background acts in terms of the matter wavefunctions as if $t_1$ and $t_2$ were identified.

A local background with a non-diagonal Higgs, or monodromy, modifies the D-term equation (\ref{eom2}) since $\left[ \varphi, \bar{\varphi} \right]\neq 0$, and therefore the local form of the wavefunctions. In particular it implies a non-vanishing flux background which localises on the branch cut locus \cite{Cecotti:2010bp}.

One of the most important applications of the ultra local theory is to Yukawa couplings. Yukawa couplings arise from triple overlaps of the wavefunctions of the participating fields \cite{Cremades:2004wa}. This follows from dimensional reduction of the the relevant operator in the 8-dimensional theory
\be
W_Y = \int_S \mathrm{Tr\;} \left[ {\bf A} \wedge {\bf A} \wedge {\bf \Phi} \right]\;. \label{supyuk}
\ee
As discussed in section \ref{sec:localgaugetheory} the matter fields arise from perturbations of the 8-dimensional fields. The wavefunctions take the form of an exponential localisation (\ref{schemwave}) and an arbitrary holomorphic prefactor which counts solutions to the Dirac equation in the presence of flux which are the Landau levels. Since the wavefunctions are localised within the local patch so is their overlap, which means that effectively we can perform the integral over $S$ as an integral over ${\mathbb C}^2$.

Wavefunctions for the charged fields can be decomposed as
\be
{\bf A}_{\bar{1}} = \phi_{4D}^I \otimes \psi_{\bar{1}}^I \;,\;\; {\bf A}_{\bar{2}} = \phi_{4D}^I \otimes \psi_{\bar{2}}^I  \;,\;\; {\bf \Phi}_{12} = \phi_{4D}^I \otimes \chi^I \;.
\ee
Here $\phi_{4D}$ are four-dimensional (super-)fields which do not depend on the coordinates on $S_{GUT}$ while the internal profiles are given by $\psi_i$ and $\chi$. The index $I$ runs over all the representations present in the decomposition of the adjoint representation of the full enhanced gauge group $G$ associated ot the Yukawa point (typically $E_6$ for an up-type and $SO(12)$ for down-type) under the remaining gauge group, which would be of the form $SU(5)_{GUT} \times U(1)^2$, after turning on the Higgs vev and fluxes. The generator structure of the internal wavefunctions is such that the trace in (\ref{supyuk}) simply leads to a selection rule stating that the charges of the three fields under the $U(1)$s should sum to zero. After accounting for this selection rule the relevant four-dimensional cubic coupling is given by
\bea
Y^{\left(I,J,K\right)(i,j,k)} &=& \frac16 G_{IJK} \int_S \left[ \psi^{I,i}_{\bar{1}} \psi^{J,j}_{\bar{2}} \chi^{K,k} + \psi^{I,i}_{\bar{2}} \psi^{K,k}_{\bar{1}} \chi^{J,j} \right. \nn \\
& & \quad - \chi^{I,i} \psi^{K,k}_{\bar{1}} \psi^{J,j}_{\bar{2}} - \psi^{I,i}_{\bar{1}} \psi^{K,k}_{\bar{2}} \chi^{J,j}  \nn \\
& & \quad\left. - \psi^{I,i}_{\bar{2}} \chi^{K,k} \psi^{J,j}_{\bar{1}} + \chi^{I,i} \psi^{K,k}_{\bar{2}} \psi^{J,j}_{\bar{1}} \right] \;.
\label{yukexpterms}
\eea
We have split off from the indices $\left\{I,J,K\right\}$ the generation indices $\left\{i,j,k\right\}$ so that generation independent quantities are manifestly so. Hence a four-dimensional field is specified by fixing $I$ and $i$. The indices $i$, $j$ and $k$, denote the possible Landau level degeneracies of the states, i.e. the generation number in the case of multiple generations coming from flux. Finally the factor $G_{IJK}$ accounts for the $U(1)$ selection rules so that for appropriately normalised generators it gives 1 if the coupling is gauge invariant and vanishes if it is not.

An important result is that in the case where all the generations arise from a single matter curve, the Yukawa matrix (\ref{yukexpterms}) about a single point is always rank 1 in generation space \cite{Heckman:2008qa,Font:2009gq,Cecotti:2009zf,Conlon:2009qq}. This was shown in full generality in \cite{Cecotti:2009zf} by going to an appropriate holomorphic gauge. The rank 1 structure is promising in terms of flavour physics at leading order because the third generation is so heavy, however it must be broken eventually to generate the other quark and lepton masses and mixing. In \cite{Cecotti:2009zf} it was shown that a background closed-string flux can appropriately deform the rank 1 structure by inducing a non-commutative deformation of the 8-dimensional gauge theory. The Yukawas for the lighter generations arise at higher orders in a small parameter associated to the non-commutative deformation and can naturally induce realistic flavour hierarchies realising the idea of \cite{Heckman:2008qa}. The relevant flux in the type IIB limit is of type $G_{1,2}$ and in terms of M-theory G-flux this lifts to $G_{1,3}$ flux. It is important to note that this type of flux is known to not lead to supersymmetric Minkowski space solutions of M-theory on CY four-folds to 3-dimensions \cite{Becker:1996gj} and therefore neither in F-theory.

In \cite{Marchesano:2009rz,madyuk2} it was shown that the required non-commutative deformation can also be induced by non-perturbative effects such as instantons and gaugino condensation on additional 7-brane sectors. It was shown that an equivilant formulation of the structure of the Yukawa couplings in the non-commutative theory, with the Yukawas coming from the non-commutative generalisation of the operator (\ref{supyuk}), can be made by considering a canonical gauge theory but with additional higher dimension operators
\be
W^{np} \in \epsilon \int_{S} \theta_n \mathrm{STr\;} \left[ \Phi^n F \wedge F \right] \;.
\ee
Here STr denotes the supertrace, $\theta_n$ are some holomorphic functions given in \cite{madyuk2}, and $\epsilon$ parameterises the size of the non-perturbative effects. The case with $n=1$ was originally used in \cite{Aparicio:2011jx} within a toy model based on a $U(3)$ gauge group. However for the more realistic $SO(12)$ and $E_6$ cases the trace vanishes and so the operators with $n=0,2$ are the leading terms. Note that the particular operator $n=0$ requires that the instanton intersects one of the 7-branes which in turn intersects the GUT brane over a matter curve. This was studied in detail in \cite{Aparicio:2011jx,madyuk2} and the resulting rank 3 Yukawa couplings were analysed. It was shown that realistic flavour structures could be induced and a quantitative analysis was presented giving the CKM elements and quark mass ratios in terms of parameters attributed to the non-perturbative effects. It was also proposed that the observed non-unification of the masses of the quarks and leptons that come from the 5-matter GUT mutliplet at the GUT scale can be attributed in such models of flavour to the deformation of the wavefunction profiles by hypercharge flux.

In \cite{liamd3} it was shown that the two approaches of non-perturbative effects and background H-flux are, at least in the type IIB limit, equivilant in the sense that the backreaction of the non-perturbative effects sources the relevant flux. This is compatible with the observation that the flux can not lead to supersymmetric Minkowski solutions because it is known that the non-perturbative effects deform the 10-dimensional solution away from CY and lead to supersymmetric AdS solutions \cite{kklt,Koerber:2007xk}.

Finally we note that since the cubic coupling (\ref{supyuk}) involves the full 8-dimensional field, it also determines the the coupling of massless four-dimensional modes to massive four-dimensional modes that arise as KK modes or Landau levels of the 8-dimensional fields. Such a coupling is important when studying proton decay for example since this is how the heavy triplets, following doublet-triplet splitting, couple to the MSSM fields. This and other applications of the coupling to heavy modes were studied in \cite{Camara:2011nj,Palti:2012aa,Ibanez:2012zg}.

\subsection{F-theory local model building phenomenology}
\label{sec:feno1}

Having reviewed the technical tools behind local models in F-theory in this section we review the application of these techniques to phenomenology. Much of the model building in F-theory is motivated by the possibilities of finding new string motivated solutions to classic phenomenological problems in the SM and GUT theories. The results of this body of work have been rather fruitful and this success has driven much of the interest in F-theory over the last years. Many of the seeds of the ideas studied featured already in the early works \cite{Donagi:2008ca,Beasley:2008kw,Donagi:2008kj}, but these were developed and new ideas introduced in a large body of work that followed. In this section we review these developments. There are some themes that thread many of the papers and this section is grouped under such themes, while ideas that do not fall into these rough classification are reviewed in section \ref{sec:miscphen}.

\subsubsection{Controlling operators with $U(1)$-symmetries}
\label{sec:contu1}

Since F-theory models are typically based on an underlying exceptional symmetry additional $U(1)$ symmetries which remain after breaking to the GUT group are ubiquitous in model building. In this section we review one of the immediate application of these which is to forbid certain dangerous operators in the theory. 

One of the classic constraints on GUT theories comes from dimension 4 proton decay operators that arise after GUT breaking from the matter parity violating operator $\fb_M \fb_M \te_M$ in the superpotential where the subscript $M$ denotes that this is a matter representation of the SM rather than a Higgs. Local models which utilise an additional $U(1)$ to forbid such a dimension 4 proton decay operator were proposed in \cite{Jiang:2008yf,Heckman:2009mn,Marsano:2009gv,Bouchard:2009bu,Marsano:2009wr,Dudas:2009hu,King:2010mq,Kuflik:2010dg,Dudas:2010zb,Ludeling:2011en,Callaghan:2011jj,Callaghan:2012rv,Dolan:2011iu,Dolan:2011aq,Donagi:2011dv}. It is important to note however that the use of $U(1)$ symmetries to forbid operators which are very tightly constrained, such as dimension 4 proton decay, can not be fully addressed within a local approach since the symmetries could be broken by bulk effects. In \cite{Hayashi:2010zp} this was studied and estimates for proton decay operators induced by bulk effects were deduced. This motivated the study of global models which ensure an unbroken $U(1)$ over the full CY \cite{Grimm:2010ez,mpw}. 

Since the discrete subgroup of matter parity is also sufficient to forbid such operators, studies of realisation of discrete symmetries in F-theory are also relevant for this purpose. In \cite{Hayashi:2009bt,Hayashi:2010zp,Hayashi:2011aa,Antoniadis:2012yk} realisations of matter parity as a geometric symmetry were  proposed while in \cite{Burrows:2010wz,Ludeling:2011en,Callaghan:2011jj,BerasaluceGonzalez:2011wy} discrete symmetries arising from breaking of the additional $U(1)$s were studied.

Another use of $U(1)$ symmetries is to generate realistic flavour structures. This can be realised by considering different generations of the SM to arise from different matter curves so that they have different $U(1)$ charges. By breaking the $U(1)$ symmetries spontaneously through the GUT singlets (\ref{cur3}) at a scale below the GUT scale Yukawa couplings can be induced by higher dimension operators with naturally small coefficients realising the Froggatt-Nielsen mechanism. Such models were studied in \cite{Dudas:2009hu,King:2010mq,Leontaris:2010zd,Ludeling:2011en,Callaghan:2011jj} where realistic quark and lepton masses and mixings were obtained.

Proton decay can also proceed through dimension 5 operators that are induced after integrating heavy modes out, typically the triplet partners of the Higgs doublets. In terms of GUT fields such superpotential operators take the form ${\bf \bar{5}}_M \te_M \te_M \te_M$ and are suppressed by the mass scale of the heavy modes that were integrated out. Forbidding such operators using $U(1)$ symmetries is connected to the superpotential Higgs mass term $\mu \f_{H_u} \fb_{H_d}$ because the neutrality of the Yukawa couplings implies that the charge $Q$ under any such $U(1)$ of the operators satisfies
\be
Q\left(\f_{H_u} \fb_{H_d}\right) = -Q\left( \fb_M \te_M \te_M \te_M \right) \;.
\ee
Therefore the up and down Higgs should not be vector-like with respect to any $U(1)$ that forbids dimension 5 proton decay. Such $U(1)$s are denoted Pecci-Quinn (PQ).\footnote{Note that this holds in the vacuum where the Yukawa couplings are fixed to their observed large values, it may be that in a model where the Yukawas themselves are generated through higher dimension operators dimension 5 proton decay can be forbidden by a non-PQ $U(1)$ but this must eventually be broken to generate the Yukawas and so effectively the breaking scale of such a $U(1)$ just amounts to playing with the ratio of the Higgs vevs: $\tan \beta$.}
Models utilising a $U(1)_{PQ}$ symmetry to suppress dimension 5 proton decay were studied in \cite{Heckman:2009mn,Bouchard:2009bu,Marsano:2009wr,Dudas:2009hu,King:2010mq,Kuflik:2010dg,Dudas:2010zb,Ludeling:2011en,Callaghan:2011jj,Callaghan:2012rv,Dolan:2011iu,Dolan:2011aq}. In all the models the $U(1)_{PQ}$ was spontaneously broken in order to give a mass to some fields; in \cite{Heckman:2009mn} this was tied to supersymmetry breaking while in the other works in was related to the lifting of exotic fields that were present after GUT breaking by hypercharge flux. Therefore in all models dimension 5 proton decay operators are not exactly forbidden but only suppressed and so the question of whether there is enough suppression to avoid experimental constraints becomes important. The constraints on the higher dimension operator vary according to which generations are involved in the operator, with the strongest constraint in the case with the largest number of light generations (see \cite{Barbier:2004ez} for a review of the constraints). Therefore a full understanding of the constraints on such operators requires a theory of flavour. In four-dimensional GUTs this is well understood because the flavour structure of the proton decay operators is related to the Yukawa couplings since the Higgs doublets and massive triplets share the same couplings. However in F-theory, depending on the method of doublet-triplet splitting this connection to the Yukawa couplings may not be present and the flavour structure of the dimension 5 proton decay operators needs to be determined independently. Initial investigations of this were performed in \cite{Camara:2011nj,Palti:2012aa} for the case of doublet-triplet splitting by hypercharge flux using wavefunctions overlap analysis (see also \cite{Ibanez:2012zg} for dimension 6 operators) with the results that the flavour structure may be compatible with estimates from four-dimensional GUT theories but in some cases can differ by a few orders of magnitude. In the case where large differences occurred however there was additional suppression of the proton decay operator compared to four-dimension GUTs. Therefore it is likely that the suppression level induced by a spontaneously broken $U(1)_{PQ}$ used in the above references is sufficient to be compatible with proton decay experimental constraints, but a more accurate determination of the flavour structure would certainly be useful in this respect.

Another matter-parity violating superpotential operator which must be very strongly suppressed is $\beta \f_{H_u} \fb_M$ as it leads to large neutrino masses from mixing with the Higgs. Forbidding such an operator using $U(1)$ symmetries is closely tied to the theory of neutrinos since if the right handed neutrinos are neutral under the appropriate $U(1)$ then a Dirac mass for them would be forbidden. For this and other reasons this is typically a difficult operator to forbid, but models which achieve this using $U(1)$ can be found in \cite{Marsano:2009gv,Marsano:2009wr,King:2010mq,Dudas:2010zb,Ludeling:2011en,Callaghan:2011jj,Callaghan:2012rv,Dolan:2011iu,Dolan:2011aq,Donagi:2011dv}.

\subsubsection{Doublet-triplet splitting with Hypercharge flux}
\label{sec:dtsp}

One of the most appealing applications of hypercharge flux is to give a mass to the triplet partners of the Higgs doublets thereby presenting a solution to the well known problem of GUT doublet-triplet splitting \cite{Beasley:2008kw,Donagi:2008kj}. The spectrum that can be induced by hypercharge flux was discussed in sections \ref{sec:localgaugetheory} and \ref{sec:localtrivialcurves}. In this section we discuss some constraints on the spectrum that arise in the presence of a $U(1)_{PQ}$ in which case the hypercharge flux must induce some net chirality on the Higgs curves.\footnote{In the absence of such a symmetry the Higgses are vector-like under all symmetries and hypercharge flux can be used for doublet-triplet splitting without inducing any net chirality.} The non-trivial net restriction to the Higgs curves requires for doublet-triplet splitting has implication for the matter spectrum also away from the Higgs curves. This was first noted in \cite{Marsano:2009gv} using the spectral cover techniques and subsequently explored in \cite{Marsano:2009wr,Dudas:2009hu,Dudas:2010zb,Marsano:2010sq,Dolan:2011iu} culminating in the constraints on the net restriction of the hypercharge flux to the matter curves summarised in section \ref{sec:localtrivialcurves}.\footnote{Although, as emphasised in \cite{Cecotti:2010bp} the derivations using the spectral cover techniques of the constraints may be evaded in the most general case, the understanding of the constraints in terms of four-dimensional anomaly cancellation shows that they must be satisfied in any model where the hypercharge is fully embedded in $SU(5)_{GUT}$ and is required to remain massless.} One of the constraints that (\ref{anomNmix}) imposes is that in the presence of a $U(1)_{PQ}$ doublet-triplet splitting implies a non-trivial restriction of the hypercharge to non-Higgs matter curves and therefore the presence of exotics in the spectrum \cite{Dudas:2010zb}. The additional constraint (\ref{anomN}) sharpens this problem and in particular implying a non-trivial restriction to the 10-matter curves as well as ruling out doublet-triplet splitting with a $U(1)_{PQ}$ completely in some cases. 

The exotics implied by the presence of a $U(1)_{PQ}$ symmetry are only massless in the limit where the symmetry is unbroken. Because, as implied by (\ref{nochiN}), the exotics are vector-like with respect to the SM symmetry groups breaking the symmetry spontaneously can induce a mass for the exotics. The fact that the exotics do not form complete GUT multiplets and therefore typically ruin gauge coupling unification motivates breaking the $U(1)_{PQ}$ at a high scale, ie. giving the charged GUT singlets a large vev, so that the exotics are lifted to a high scale and do not affect the gauge coupling running significantly before the GUT scale. On the other hand the $U(1)_{PQ}$ is useful for controlling operators such as those which lead to dimension 5 proton decay (see section \ref{sec:susybf} for other uses of a $U(1)_{PQ}$). The tension between these two objectives was studied quantitatively in \cite{Dudas:2010zb,Leontaris:2010zd,Marsano:2010sq,Ludeling:2011en,Dolan:2011iu,Palti:2012aa} and in particular in \cite{Dolan:2011aq}. Again these studies are subject to the uncertainty in the flavour structure of proton decay operators discussed in section \ref{sec:contu1}. It is also important to note that the models studied did not take into account the constraint (\ref{anomN}) and so some modification to those theories may be required.

Note that the study of exotics in the aforementioned works was performed assuming that the SM families arise from complete GUT mutliplets. Some additional freedom can be gained by letting the generations arise from different GUT mutliplets. This idea was studied in \cite{Font:2008id} with a different motivation of generating a top Yukawa coupling in the absence of monodromies.

Due to some of the difficulties associated to GUT breaking and doublet-triplet splitting with hypercharge flux it is interesting to consider alternative approaches using Wilson lines. Within an F-theory context this was explored in \cite{Braun:2010hr,Marsano:2012yc}.

\subsubsection{Gauge coupling unification}

Perhaps the strongest experimental evidence for grand unification is the unification of the MSSM gauge couplings at $10^{16}$ Gev. At 1-loop this is to the accuracy of 0.5\% while at 2-loops it is at 3\%. The unification of the couplings in F-theory GUTs depends on the form of GUT breaking. The possibility of breaking by Wilson lines (studied in \cite{Braun:2010hr,Marsano:2012yc}) analogous to the heterotic string maintains the unification of the gauge couplings at the GUT scale up to threshold corrections. An important difference from heterotic models is that the GUT coupling can be independent from the coupling of gravity and so there is no tension between the two, though on the other hand there is also no generic prediction of their relation. One of the most interesting aspects of F-theory models is GUT breaking through the use of hypercharge flux which is not available in the heterotic string.\footnote{This is the case for the definition of the hypercharge generator as completely embedded inside of $SU(5)_{GUT}$. Realisations of hypercharge as a linear combination of a $U(1)$ inside $SU(5)_{GUT}$ and additional $U(1)$ factors can allow for the use of hypercharge flux in the heterotic string, though the canonical normalisation of the generator and therefore the prediction of gauge coupling unification is then lost \cite{Blumenhagen:2006ux}.} This is because, as discussed in section \ref{sec:localgaugetheory}, the mechanism of hypercharge flux GUT breaking relies on the flux being nontrivial in the homology of $S_{GUT}$ but globally trivial \cite{Jockers:2004yj,Buican:2006sn,Beasley:2008kw,Donagi:2008kj}. There is no such separation in the homologies of the submanifold that the gauge field propagates on and the full extra dimensions in the heterotic string.

Gauge coupling unification in the presence of hypercharge flux in F-theory was studied in 
\cite{Donagi:2008kj,Beasley:2008kw,Blumenhagen:2008aw,Grimm:2012rg,Conlon:2009xf,Conlon:2009kt,Conlon:2009qa,Dudas:2010zb,Leontaris:2009wi,Donkin:2010ta,Pawelczyk:2010xh,Jelinski:2011xe,Leontaris:2011tw,Dolan:2011aq}. Generally, there are two ways to study threshold corrections to gauge coupling unification at 1-loop in string theory: through the 1-loop open string channel, or gauge theory, or through the tree-level closed string channel, or supergravity. The latter method was applied in \cite{Blumenhagen:2008aw} to show that the gauge couplings are split at tree-level from a correction in the type IIB limit coming from the $F^4$ term in the D7 action (\ref{d7act}) in the background of hypercharge flux. The tree-level holomorphic gauge couplings for the $SU(3)$, $SU(2)$ and $U(1)_Y$ factors are given by \cite{Blumenhagen:2008aw}
\bea
\label{couplings}
f_{SU(3)} & = & T - \half S \int_{S_{GUT}} \mc{F}_{U(1)}^2 \;, \\
f_{SU(2)} & = & T - \half S \int_{S_{GUT}} \left[ \mc{F}_{U(1)}^2 + \mc{F}_Y^2 + 2 \mc{F}_{U(1)} \wedge \mc{F}_Y \right] \;, \nonumber \\
\frac{3}{5}f_{U(1)_Y} & = & T - \half S \int_{S_{GUT}}  \left[  \mc{F}_{U(1)}^2 + \frac{3}{5} \left( \mc{F}_Y^2 + 2 \mc{F}_{U(1)} \wedge \mc{F}_Y \right) \right] 
\eea
Here $S$ is the axio-dilaton superfield $S=e^{-\varphi} + iC_0$, $T$ the Kahler modulus associated to $S_{GUT}$, $F_{U(1)}$ is the flux turned on along the diagonal $U(1)$ generator of $U(5)$, and $F_{Y}$ is the hypercharge flux. Note that as shown in section \ref{sec:localgaugetheory} in order to avoid exotics in the $\left({\bf 3},{\bf 2}\right)_{-5/6}$ representations we must take $\int F_Y^2 = -2$. The gauge couplings are therefore non-universal unless we turn on $F_{U(1)}$ with $\mc{F}_{U(1)} \wedge \mc{F}_Y =-1$, which this implies that $F_{U(1)}$ must have a globally trivial component \cite{Blumenhagen:2008aw}. 

The uplift of this non-universality to general F-theory backgrounds is not fully understood as yet since we do not understand the uplift of the $F^4$ term, though see \cite{Grimm:2012rg} for progress in this direction. Nonetheless a similar correction to the gauge couplings is expected in the F-theory setup. One way to motivate this is to identify this particular correction from the supergravity persepctive as the correction from the gauge theory perspective of integrating out the pair of $\left({\bf 3},{\bf 2}\right)_{-5/6}$ that have been made massive by the hypercharge flux \cite{Donagi:2008kj,Dolan:2011aq}. Since this latter view point can be understood also in the effective gauge theory describing the F-theory setup the same consequences should apply.

Another use of the dual approaches to calculating threshold corrections is to relate them to closed string tadpoles. In \cite{Conlon:2009xf,Conlon:2009kt,Conlon:2009qa} it was shown that hypercharge flux induces a local tadpole in the closed string sector which from the gauge theory perspective implies that the gauge couplings diverge due to threshold corrections. Since the hypercharge flux is globally trivial the local tadpole is also globally trivial and so the associated divergence is regulated, but only at a scale associated to the global compactification radius. It was therefore argued that since the gauge couplings run until this global radius scale, rather than the local string scale where heavy modes first appear, gauge coupling unification in F-theory models utilising hypercharge flux should appear to occur at a scale above the string scale. Practically, as rough order-of-magnitude estimates, for a GUT scale around $10^{16}$ Gev this would imply a string scale around $10^{15}$ Gev. 

Gauge threshold corrections computations directly from the field theory side were performed in \cite{Dudas:2010zb,Leontaris:2009wi,Donkin:2010ta,Pawelczyk:2010xh,Jelinski:2011xe,Leontaris:2011tw,Dolan:2011aq}. One important aspect of these is that, as discussed in section \ref{sec:dtsp}, many F-theory models typically have exotics in the spectrum which do not form complete GUT multiplets and therefore induce non-universal running for the gauge couplings. The exotics can be given a mass through a vev for GUT singlets, and how strong their effect is on gauge coupling unification depends on their proximity to the GUT scale. This was studied for explicit models in \cite{Dudas:2010zb,Dolan:2011aq}.

\subsubsection{Supersymmetry breaking}
\label{sec:susybf}

Supersymmetry breaking is not a phenomenological aspect that can be studied completely locally because there is always a gravity mediated contribution which is sensitive to global aspects. However it is possible to assume that this is subdominant to another source of supersymmetry breaking mediation which can be studied locally within an F-theory context. When this approximation is valid is generally very model dependent and within an F-theory context was studied in \cite{Blumenhagen:2009gk}. Of course even if the main mediator of supersymmetry breaking is gravitational it is possible to use results from IIB and some approximations to make general statements about the F-theory setup and this has been studied in \cite{apa,Blumenhagen:2009gk,Aparicio:2012iw,Ibanez:2012zg}. Motivated by the Higgs mass results there are also models which consider a very high scale of supersymmetry breaking \cite{Ibanez:2012zg,Hebecker:2012qp}.

Within the approximation that the primary source of supersymmetry breaking is the local gauge sector, supersymmetry breaking in F-theory GUTs was studied in \cite{Heckman:2008es,Marsano:2008jq,Heckman:2008qt,Heckman:2008rb,Heckman:2009bi,Heckman:2009mn,Heckman:2010xz,Marsano:2009wr,Dolan:2011aq}. The models are based on gauge mediation with \cite{Marsano:2008jq,Heckman:2008qt,Heckman:2008rb,Heckman:2009bi} using messengers in the $\f\oplus \fb$ representations, \cite{Heckman:2009mn,Heckman:2010xz} in the $\te \oplus \teb$ representations, and in \cite{Dolan:2011aq} the messengers did not form complete GUT multiplets. All of the models utilised a $U(1)_{PQ}$ symmetry to control the $\mu-B\mu$ problem of gauge mediation where the $\mu$ term induced through the Guidice-Masiero mechanism in the Kahler potential
\be
K \supset X^{\dagger} H_u H_d \;,
\ee
is allowed by the $U(1)_{PQ}$, with $X$ the supersymmetry breaking field which is charged under the $U(1)_{PQ}$. While the associated $B\mu$ term is forbidden
\be
K \supset X^{\dagger}X H_u H_d \;.
\ee
A natural consequence of the presence of a $U(1)_{PQ}$ is an additional source of soft masses for the fields coming from integrating out the associated Stueckelberg massive gauge field \cite{ArkaniHamed:1998nu,Heckman:2008qt}. The interesting thing about such soft masses is that they depend only on the charges of the fields that gain a mass under the $U(1)_{PQ}$. Therefore if the generations share the same charge such masses are automatically flavour diagonal thereby avoiding strict experimental constraints on flavour changing currents. In \cite{Heckman:2010xz} the phenomenological implications of this additional mass term were studied in detail. However it is important to note that, since the mass of a Stueckelberg massive gauge field is at the string scale, a mechanism is still required to understand the suppression of flavour changing soft masses induced by other heavy fields compared with the $U(1)_{PQ}$ induced masses. In particular calculations of such masses coming from integrating out heavy chiral mutliplets were performed in \cite{Camara:2011nj} with the results that generally no additional suppression is expected though there is the possibility of such a relative suppression at regions of parameter space.

\subsubsection{Further topics}
\label{sec:miscphen}

In this section we review phenomenological aspects of F-theory models which do not fall into the themes outlined above. Although flavour physics does constitute a central theme in F-theory phenomenology the wavefunction approach to flavour physics was reviewed in section \ref{sec:fultralocal} and the Froggatt-Nielsen models in section \ref{sec:contu1}. It should be noted though that the wavefunction discussion is of course classical and it is important to take into account quantum corrections to the Yukawa couplings from threshold effects, and these have been studied to some extent in \cite{Conlon:2010xb,Leontaris:2011pu}. Additional to  these approachs a model of flavour was also developed in \cite{Donagi:2011dv} where the flavour structure is attributed to the different dimensions over which the matter representation can be localised in the presence of gluing morphisms that correspond to brane recombination modes \cite{Donagi:2011jy,Donagi:2011dv,Marsano:2012bf}.

F-theory GUTs have proved quite fruitful in the arena of Neutrino physics. A key aspect of neutrino model building is the nature of right-handed neutrinos and in F-theory there are two classes of candidates: there are GUT singlets which are neutral under all the gauge symmetries, and so can be termed moduli, and there are GUT singlets that are charged under additional $U(1)$ symmetries. Note however that the two types of candidates can be related once the $U(1)$ symmetries are spontaneously broken. The moduli candidates have been studied in \cite{Tatar:2009jk,Bouchard:2009bu,Heckman:2009mn} and the charged ones in \cite{Marsano:2009gv,Marsano:2009wr,Heckman:2009mn,Dudas:2009hu,Leontaris:2010zd,Callaghan:2011jj,Callaghan:2012rv}. The models had realistic neutrino masses were obtained which arose either through the traditional see-saw mechanism or through the mechanism proposed in \cite{ArkaniHamed:2000bq} where a Dirac mass in Kahler potential naturally leads to correct neutrino mass scale after electroweak symmetry breaking. It is interesting to note that neutrino physics is also an area where experimental predictions were made in F-theory models which have since been confirmed. This refers to the neutrino mixing angle $\theta_{13}$ which until recently was experimentally consistent with zero and in many bottom-up models assumed to vanish through some symmetry. The relevant mixing angle was predicted in the models of \cite{Bouchard:2009bu,Dudas:2009hu} to be non-vanishing and indeed was measured to be so \cite{An:2012eh}.

So far in F-theory we have considered configurations which amount to intersecting 7-branes. An interesting extension of these theories is by adding probe D3-branes intersecting the GUT 7-brane. In particular D3-branes probing exceptional $E_6$, $E_7$, $E_8$ points of enhancement give rise to additional sectors which are strongly coupled $N=2$ superconformal field theories \cite{Heckman:2010fh,Heckman:2010qv,Heckman:2011hu}. Such strongly coupled sectors can couple to the visible sector and modify physics including gauge coupling unification and electroweak symmetry breaking. The phenomenology of these theories is studied in \cite{Heckman:2011bb,Heckman:2012nt} and is particularly interesting given the current experimental probes of the Higgs couplings.

An interesting and often studied scenario in beyond the SM physics is the possiblity that gauginos are Dirac rather than Majorana. In \cite{Davies:2012vu} it was shown how such a scenario could be realised in F-theory.

\comments{A relatively under-developed area of F-theory phenomenology is cosmology. However some work on dark matter candidates was done in \cite{Heckman:2008jy,Heckman:2009mn}, on inflation in \cite{Hebecker:2011hk,Hebecker:2012aw}, and on baryogenesis in \cite{Heckman:2011sw}.}

Although more inspired by F-theory than directly derived from it the no-scale flipped-SU(5) model proposed in \cite{Jiang:2008yf} has been studied throughly and the detailed resulting phenomenology seems quite promising \cite{Jiang:2009za,Li:2009fq,Li:2010mra,Li:2010dp,Li:2010rz,Li:2010uu,Li:2011hya}.

Although most of the work in F-theory is on $SU(5)$ and $SO(10)$ GUTs there are studies on $SU(6)$ \cite{Chung:2009ib,Choi:2010gx} and $E_6$ GUTs \cite{Chen:2010tg,Callaghan:2011jj,Donagi:2011dv,Kuntzler:2012bu,Callaghan:2012rv}. Also it is possible to consider a direct embedding of the SM gauge groups more analogous to intersecting brane models \cite{Chen:2009me,Choi:2010su,Choi:2010nf}.

It is clear that there is an important element missing from the F-theory phenomenology reviewed which is moduli stabilisation. Although a well developed subject in the type IIB limit, see section \ref{sec:moduli}, the uplift to F-theory of the stabilisation of the Kahler moduli and the open-string moduli (which become geometric in F-theory) is less well understood. The effective action approach of \cite{Grimm:2010ks}, and related flux superpotential \cite{Grimm:2009ef}, is a substantial step towards this aim. Also in \cite{Braun:2008pz,Valandro:2008zg,Braun:2009bh} some aspects of moduli stabilsation from the M-theory persepctive were studied. It is clear from the key role that instantons play in type IIB that a similar important role in moduli stabilisation is relevant for F-theory. Therefore studies of instantons in F-theory are important in this respect \cite{Heckman:2008es,Marsano:2008py,Blumenhagen:2010ja,Cvetic:2010rq,Cvetic:2010ky,Donagi:2010pd,Cvetic:2011gp,Bianchi:2011qh,Grimm:2011dj,Kerstan:2012cy,Cvetic:2012ts}. A particularly interesting result of \cite{Cvetic:2011gp} is that a condition for the instanton prefactor Pfaffian to vanish in F-theory is the existence of a point of $E_8$ enhancement. This is could present an attractive way to dynamically motivate the existence of such a point and the resulting phenomenology.

\section{Concluding remarks}

We reviewed the construction of particle physics models within the weakly coupled type IIB string and its strongly coupled extension of F-theory. There are many aspects that combine to form a realistic string vacuum and one of the most important is the stabilisation of all the moduli. This is an aspect which is particularly well understood within the type IIB framework and we began by reviewing how full stabilisation of the moduli can be realised. The rest of the review was dedicated to how the key elements of low energy particle physics; gauge and global symmetries, a chiral matter spectrum, and interaction operators such as Yukawa couplings arise and the actual models that can be constructed combining these elements. The first class of models we discussed were D3-branes at singularities where semi-realistic matter spectra could be constructed as well as some control over more detailed aspects such as quark and lepton masses and mixing. We then summarised the key aspects of models based on intersecting and fluxed D7-branes in IIB. 

Motivated by the realisation of GUTs we argued that within such a unified framework an order one top quark Yukawa coupling requires the presence of exceptional symmetries that can only be realised within the strongly coupled regime of type IIB string theory termed F-theory. F-theory model building was reviewed in two sections, the first concerned global aspects of the framework such as the structure of the elliptic fiber over various loci in the CY four-fold, how gauge symmetries, both Abelian and non-Abelian, can be generated, and how to account for world-volume flux and the chirality it induces in the spectrum. The second section reviewed local model building which is based on a gauge theory description of the GUT 7-brane. We showed that a wide spectrum of phenomenological issues can be addressed in this framework, many of which offer novel stringy solutions to familiar puzzles from field-theory GUTs. For example a geometric understanding of quark and lepton flavour physics and doublet-triplet splitting using background flux. Additional U(1) symmetries beyond those of the SM played a central role in this arena of model building and, although strongly constrained by anomaly cancellation, could offer substantial control over operators in the theory. 

Although the successes of particle physics model building so far are impressive, both in the accuracy to which known aspects of the SM could be recovered as well as the range of puzzles that this top-down framework offers attractive solutions to, the ultimate goal of string phenomenology: an understanding of how the particular universe we inhabit emerges from string theory, still requires substantial advances. It is of course difficult to guess how revolutionary the insights required and how long it would take to make these advances, but some seem more immediately reachable than others. For example just a few of the more immediate aspects to address are how to realise the local model building phenomenology of F-theory in a fully global context, particularly pressing aspects being the realisation of globally trivial hypercharge flux as well as additional U(1) symmetries. Another global aspect is how the rich moduli stabilisation models of IIB are lifted to F-theory? In the opposite regime there remains much to understand about ultra local models based on wavefunction analysis in local patches, for example an analysis of more general Higgs backgrounds with non-Cartan elements.

In terms of model building targets we have not discussed cosmology at all in this review. There is a substantial body of literature on the subject of cosmology in the type IIB framework (see \cite{Baumann:2009ni,Cicoli:2011zz} for reviews), though it is fair to say that a fully realistic model of inflation is yet to emerge completely naturally from the theory, and inflationary model building remains of high priority. In terms of particle physics it is crucial to respond to the findings of the LHC. Two particularly important aspects of this are updating our model building targets away from the MSSM as other theories become more attractive with results for the Higgs mass and couplings, for example theories with additional singlets or massive gauge symmetries that can change the tree-level Higgs mass. Relating to this aspect of LHC limits on supersymmetry it is important to look for models which accommodate soft mass patterns that maintain the naturalness motivation for supersymmetry while evading LHC limits. One of the primary motivations for supersymmetry remain gauge coupling unification and understanding the details of this in F-theory, particularly in the presence of hypercharge flux, is clearly important. 

Perhaps the most serious long term issue facing string phenomenology is that of the landscape: which vacuum of string theory does our universe correspond to and why that particular one?  Two of the most important aspects, in relation to this question, that have been reviewed here are the modular, or local, approach to model building and the idea of an underlying $E_8$ symmetry controlling the full structure of the theory. The local approach allows us to decouple some of the particle physics questions from the full closed-string background. This idea is particularly appealing when we consider that the landscape, and anthropic reasoning, have proved most successful in the closed-string or gravitational sector when applied to the cosmological constant. On the other hand in particle physics the principles of symmetries, unification and naturalness have so far proved reliable guides. Therefore the appealing idea of a closed-string landscape with an open-string sector which is understood from more traditional routes fits well with the modular approach to model building. Within the particle physics sector itself the idea that $E_8$ should be the structure underlying the gauge, matter and interactions of the model, even within a type II setting, restricts the possible spectrum of models that can be studied to a manageable set. For example in F-theory GUTs with hypercharge flux breaking, even before we impose any possible additional constraints coming from global realisations, anomaly cancellation by itself severely restricts any possible additional U(1) symmetries. Therefore one of the crucial questions for string phenomenology is: is there an underlying $E_8$ controlling unification in string theory?

Of course even if, by some criteria, we could convincingly argue for a particular realisation of our universe in string theory, it is difficult to see with our present understanding of the theory\footnote{ For of criticisms approaches typically used to determine vacua of string theory see for example \cite{fiveten}.} how this could emerge as the unique consistent solution to string theory and therefore is not strictly a prediction of the theory. Ideally a theory should make distinct falsifiable predictions, however, the primary criteria by which a theory of the universe should be judged is its ability to reproduce known observations. And by this criteria string models, in incorporating quantum physics, gravity and particle physics, are already the leading theories known. Further, as we have seen the requirement to reproduce the present day observations is highly constraining. These drive us to distinct regions of the landscape; each region providing a phenomenological scenario. Within a given phenomenological scenario it is certainly possible to extract predictions.  In this light, the absence of a unique string theory prediction at present should certainly not be taken as a fundamental obstacle. The ambitious goal of string phenomenology should be pursued intensively: it is difficult to imagine that fundamental physics could advance indefinitely without facing this task sooner or later.

\section*{Acknowledgements}
We would like to thank Shanta de Alwis, Marcus Berg, Cliff Burgess, Pablo Camara, Joseph Conlon, Michele Cicoli,  Matt Dolan, Emilian Dudas, Andrew Frey, Steve Giddings, Thomas Grimm, Michael Haack, Ami Hanany, Hirotaka Hayashi, Janet Hung, Luis Ibanez, Max Kerstan, Sven Krippendorf,  Steve Kom, Matt Lippert,  Fernando Marchesano, Joe Marsano, Liam McAllister, Christoph Mayrhofer, Joe Polchinski, Sakura Schafer-Nameki, Fernando Quevedo, Gary Shiu, Mahdi Torabian, Sandip Trivedi, Angel Uranga, Roberto Valandro, Timo Weigand and Ivonne Zavala for discussions and collaborations that have helped shape our understanding of the subject. The work of EP is supported by a Marie Curie Intra European Fellowship within the 7th European Community Framework Programme.

\bibliography{notes}{}
\bibliographystyle{unsrt}
\end{document}